\documentclass[a4paper,11pt]{article}
\pdfoutput=1 % if your are submitting a pdflatex (i.e. if you have
\usepackage{dcolumn}
\usepackage{bm}
\usepackage{amsthm}
\usepackage{setspace}
\usepackage[normalem]{ulem}
\usepackage{relsize}
\usepackage{cancel}
\usepackage{verbatim}
\usepackage{enumitem}
\usepackage{mathtools}
\usepackage{empheq}
\usepackage{dsfont}
\usepackage{amsmath}
\usepackage{slashed}

\usepackage{graphicx}
\usepackage{caption}
\usepackage{subcaption}

\usepackage{jheppub}

\usepackage{amsthm}

\newcommand{\calL}{\mathcal{L}}

\newcolumntype{I}[1]{>{\centering\arraybackslash$}m{#1}<{$}}

\title{Holomorphic Structure and Quantum Critical Points in Supersymmetric Lifshitz Field Theories}

\author[a]{Igal Arav,}
\author[b]{Yaron Oz}
\author[b]{and Avia Raviv-Moshe}
\affiliation[a]{Blackett Laboratory, 
  Imperial College\\ London, SW7 2AZ, U.K.}
\affiliation[b]{Raymond and Beverly Sackler School of Physics and Astronomy, Tel-Aviv University, 55 Haim Levanon street, Tel-Aviv, 69978, Israel}
\emailAdd{i.arav@imperial.ac.uk}
\emailAdd{yaronoz@post.tau.ac.il}
\emailAdd{aviaravi@mail.tau.ac.il}

\abstract{We construct supersymmetric Lifshitz field theories with four real supercharges in a general number of space dimensions.
The theories consist of complex bosons and fermions and exhibit a holomorphic structure and non-renormalization properties of
the superpotential. We study the theories in a diverse number of space dimensions and for various choices of
marginal interactions. We show that there are lines of quantum critical points with an exact Lifshitz scale invariance and a dynamical critical exponent that depends on the coupling constants. 
}

\begin{document}
\maketitle
\flushbottom

\section{Introduction}

Quantum field theories that are not Lorentz invariant have been studied extensively in recent years.
Bounds on the violation of Lorentz symmetry have been set at high energy, while at low energy one finds that Lorentz violations appear in various condensed matter systems of interest that exhibit quantum criticality. 
Materials such as high $T_c$ superconductors and heavy fermion compounds have a metallic phase whose properties cannot be explained within the standard Landau-Fermi liquid theory \cite{Coleman:2005,Sachdev:2011cs,Gegenwart:2008np, Sachdev:2011cup,Ardonne:2003wa,Grinstein:1981}. In these systems one observes quantities that exhibit a universal behavior, such as resistivity that  is a linear function of the temperature \cite{Gurvitch:1987prl,Trovarelli:2000prl,Bruin:2013s},
which is believed to be the consequence of quantum criticality. These systems possess a Lifshitz scaling symmetry around the quantum critical point \cite{Grinstein:1981,Hornreich:1975zz}. 

Lifshitz scaling is an anisotropic scale symmetry of time and space:
\begin{equation}\label{Intro:LifshitzScaling}
t \rightarrow \lambda^{-z} t \qquad  x^i\rightarrow \lambda^{-1} x^i \qquad i=1,\dots,d \ ,
\end{equation}
where $d$ is the number of space dimensions and $z$ is the dynamical critical exponent. When $z\neq 1$, it measures the anisotropy between space and time. The generators of the Lifshitz algebra in $d+1$ spacetime dimensions are time translation $H$, space translations $P_i$, scale transformation $D$ and spatial rotations $M_{ij}$. The commutation relations read: 
\begin{align}
\begin{split}
&[D, P_i] = i  P_i  , \quad [D,H] = i  z H , \quad   [M_{ij}, M_{kl}] = i \delta_{kj} M_{il}+ \dots  ,
\\
&  
[M_{ij}, P_k] = - i \delta_{ki}P_{j} +\dots ,    ~\quad \qquad  [M_{ij}, H] = 0  \ .
\end{split}
\end{align}
There are no Casimir operators that are polynomial in the generators of the Lifshitz algebra and, therefore,  no obvious quantum numbers to label its irreducible representations, if exist.

Relativistic supersymmetry is a unique extension of spacetime Poincare symmetry algebra, where the anticommutator of the fermionic 
generators $\{Q,Q^\dagger\}$ yields the bosonic spacetime translations. Supersymmetry has been for many years the leading candidate for an extension of the Standard Model of particle physics and there is an ongoing extensive high energy experimental search for it.
At low energy, emergent supersymmetry is potentially a property of some strongly coupled condensed matter system which 
is yet to be realized experimentally.
Relativistic supersymmetric field theories exhibit a rich and calculable holomorphic quantum structure. 
When certain quantities, such as the effective action, have a holomorphic dependence on the
quantum fields and coupling constants, it is possible to get restrictions on the flow of
these quantities under renormalization. 
Indeed, non-renormalization
theorems are common in relativistic theories with a sufficient amount of supersymmetry
(see  e.g. \cite{Seiberg:1993vc,Grisaru:1979wc}).

Supersymmetry of Lifshitz field theories have been studied in e.g. \cite{Orlando:2009az,Dijkgraaf:2009gr,Chapman:2015wha,Gomes:2014tua,Gomes:2015cia,Xue:2010ih,Redigolo:2011bv,Meyer:2017zfg,Gallegos:2018jyg,Auzzi:2019kdd}.
The aim of this work is to construct supersymmetric Lifshitz quantum field theories that exhibit a holomorphic structure
and study the implications.
In addition to the relevance for the study of non-relativistic field theories, this  may also shed light on which properties of relativistic holomorphic supersymmetry follow from the relativistic symmetry and which ones from the holomorphic structure. 
We will consider a supersymmetric Lifshitz algebra where the 
anticommutator of the fermionic 
generators yields the Hamiltonian, that is the bosonic generator of time translation $\left\{Q , Q^\dagger \right\} \sim  H$.
We will refer to such structure as time domain non-relativistic supersymmetry.

We will construct time domain supersymmetric Lifshitz field theories with four real supercharges in a general number of space dimensions.
The theories consist of complex bosons and fermions and exhibit a holomorphic structure and non-renormalization properties of
the superpotential reminiscent of the relativistic $\mathcal{N}=1$ Wess-Zumino model in four dimensions. 
 We will study the theories in a diverse number of space dimensions and for various choices of
marginal interactions and show that they include lines of quantum critical points with an exact Lifshitz scale invariance and a dynamical critical exponent that depends on the coupling constants. 
This conclusion will not be based on perturbative arguments and it applies to the strong coupling regime as well.

The paper is organized as follows. 
In section \ref{sec:TimeDomainNRSUSY} we construct a family of Lifshitz supersymmetric models that possess a holomorphic structure. We discuss their symmetries and classical properties. We begin in subsection \ref{subsec:ReviewOfN1TDSupersymmetry} by reviewing the $\mathcal{N}=1$ models of Lifshitz  supersymmetry (two real supercharges) which have been previously studied. These theories do not acquire a holomorphic structure. In subsection \ref{sec:TheHolomorphicModel} we present the $\mathcal{N}=2$ holomorphic models of Lifshitz time domain supersymmetry.  
In section \ref{sec:QuantumCorrections} we study the quantum behaviour of these theories. In subsection \ref{subsec:Renormalization} we
discuss renormalization and regularization methods as well as quantum fixed points in Lifshitz field theories. In subsection \ref{subsec:DualScaleRGFlows}, we generalize the study of the renormalization group flow in Lifshitz theories by considering a dual-scale renormalization scheme.  In subsection \ref{subsec:NonRenormalizationTheoremsAGeneralProof} we give a general proof of the non-renormalization theorems based on the symmetries of the models. In subsection \ref{subsec:PerturbativeAnalysis} we provide a perturbative point of view on the quantum behaviour of the theories. In subsection \ref{sec:MarginalCases} we study the marginal cases and show that the theories possess lines of quantum fixed points in which the system has an exact Lifshitz scaling symmetry. In subsection \ref{subsec:GaplessSingularCase} we discuss the gapless singular case. Finally, we conclude in section \ref{sec:Summary}.
Some details are given in the appendices.

\section{Time Domain Supersymmetry}
\label{sec:TimeDomainNRSUSY}
Various types of non-relativistic supersymmetric field theories have been considered in the past from different motivations and points of view (see for example \cite{Orlando:2009az,Dijkgraaf:2009gr,Xue:2010ih,Redigolo:2011bv,Gomes:2014tua,Gomes:2015cia,Chapman:2015wha,Meyer:2017zfg,Gallegos:2018jyg,Auzzi:2019kdd}). Here we restrict our discussion to what we will refer to as ``time domain'' supersymmetry, which corresponds to those cases in which the supersymmetric algebra closes on the Hamiltonian of the system alone (as opposed to other constructions, such as ones in which the supersymmetric algebra follows the relativistic one as in \cite{Gomes:2014tua}). Our focus is on non-relativistic field theories in $d+1$ dimensions which are invariant under space and time translations as well as space rotations (sometimes known as Lifshitz or Aristotelian theories), along with a time domain supersymmetry, without imposing any boost symmetry (either of the Lorentzian or the Galilean types).

In this section we construct and discuss such time domain supersymmetric models. 
We start with a brief review of the minimal non-relativistic $\mathcal{N}=1$\footnote{Note that, in our conventions, $\mathcal{N}=1$ refers to models with 2 real supercharges, which is the minimal number required for an algebra of the form $\{Q,Q\}\sim H$. Accordingly $\mathcal{N}=2$ refers to 4 real supercharges (or 2 complex  ones).} time domain supersymmetric models, which have been studied in various works \cite{Witten:1981nf,Witten:1982df,Witten:1982im,Parisi:1982ud,Sourlas:1985,Orlando:2009az,Dijkgraaf:2009gr,Chapman:2015wha}, and some of their properties. We then construct a family of $\mathcal{N}=2$ models with an $SU(2)$ R-symmetry and a holomorphic structure, which includes both free and interacting theories, and discuss their symmetries and particle content.  

\subsection{A Review of $\mathcal{N}=1$ Time Domain Supersymmetry}
\label{subsec:ReviewOfN1TDSupersymmetry}

We start by reviewing the $\mathcal{N}=1$ time domain supersymmetric models, which have been studied
in various works (see for example \cite{Witten:1981nf,Witten:1982df,Witten:1982im,Parisi:1982ud,Sourlas:1985,Orlando:2009az,Dijkgraaf:2009gr,Chapman:2015wha}).
These are non-relativistic field theories in $d+1$ dimensions, which are invariant under the usual time translations $H$, space translations $P_i$ ($i=1,\ldots,d$) and space rotations $M_{ij}$, as well a complex supercharge $Q$ (or equivalently two real supercharges) and a $U(1)$ R-symmetry charge $R$, satisfying\footnote{Note that the supercharge $Q$ here is a scalar under space rotations. This is not surprising as one does not necessarily expect any specific spin-statistics correspondence in these non-relativistic models.} (see \cite{Chapman:2015wha}): 
\begin{equation}
\begin{split}
\label{eq:Nequal1Algebra}
&\left\{ Q,Q \right\} = 0, \quad
\left\{  Q , Q^\dagger \right\} = 2 H, \\ 
&[M_{ij} , Q ]=0 , \quad 
[P_i, Q] = 0 , \quad
[R,Q] = iQ .
\end{split}
\end{equation}
For models which are additionally invariant under a Lifshitz scaling symmetry $D$ with some dynamical critical exponent $z$ (such as free models), these relations also imply:
\begin{equation}
[ D , Q ] = i  \frac{z}{2} Q .
\end{equation}
As noted in \cite{Chapman:2015wha} (see also e.g. \cite{Sourlas:1985,Damgaard:1987rr,Orlando:2009az,Dijkgraaf:2009gr}), this algebra can be realized in a ($d+1$)-dimensional field theory given by the following action:
\begin{equation}
\label{DetailedBalance3}
\begin{aligned}
S\left[ {\phi ,\psi , \psi^\dagger } \right] &= 
\int d^dx dt \left[ \frac{1}{2} (\partial_t \phi)^2 - \frac{1}{2} \left( \frac{\delta W}{\delta\phi} \right)^2 + i \psi^\dagger\partial_t\psi \right] \\
&  - \int d^dx d^dx' dt \frac{\delta^2 W}{\delta\phi(x)\delta\phi(x')} \psi^\dagger(x) \psi(x') ,
\end{aligned}
\end{equation}
where $\phi$ is a bosonic real field and $\psi$ a fermionic complex field,\footnote{Note that the notation here is different to the one in \cite{Chapman:2015wha}, where $\psi$ was defined as a two component real fermion field.} both of which are scalars with respect to spatial rotations.\footnote{As these are non-relativistic models, and the degrees of freedom involved do not correspond directly to a non-relativistic limit of some relativistic degrees of freedom, standard relativistic spin-statistics relations need not apply here.} The superpotential $W(\phi)$ here is some local functional of the field $\phi(x)$, and will generally contain its spatial derivatives. This action can also be written in superspace formalism as follows:
\begin{equation}
\label{eq:N1ReviewSuperspaceAction}
S = \int dt d^dx d\theta d\theta^\dagger \left[ \frac{1}{2} D\Phi D^\dagger \Phi \right] - \int dt d\theta d\theta^\dagger W(\Phi) ,
\end{equation}
where $\theta, \theta^\dagger$ are Grassmannian superspace coordinates, $\Phi$ is a superfield defined as:
\begin{equation}
\Phi(t,x) \equiv \phi + \theta \psi + \psi^\dagger \theta^\dagger +F \theta^\dagger \theta ,
\end{equation}
$F$ is a real auxiliary field and the covariant derivatives are given by:
\begin{equation}
D = \frac{\partial}{\partial\theta} - i \theta^\dagger \partial_t ,
\qquad
D^\dagger = - \frac{\partial}{\partial\theta^\dagger} + i \theta \partial_t .
\end{equation}
In terms of the fields $\phi, \psi$, the conserved supercharges may be written:
\begin{equation}
\label{eq:N1ReviewSupercharges}
Q = \int d^dx \left[ \partial_t\phi + i \frac{\delta W}{\delta\phi} \right] \psi,
\qquad
Q^\dagger = \int d^dx \left[ \partial_t\phi - i \frac{\delta W}{\delta\phi} \right] \psi^\dagger .
\end{equation}
Of course, one may extend the action \eqref{DetailedBalance3}-\eqref{eq:N1ReviewSuperspaceAction} to any number of superfields.

It is important to mention that these models share many similarities with minimal models of supersymmetric quantum mechanics 
(see \cite{Witten:1982df,Witten:1981nf,Witten:1982im}), and in fact can be viewed as a dimensional extension of it, with the main difference being that the superpotential $W$ is a functional of $\phi(x)$ (rather than a function of a finite number of degrees of freedom). 

A free $\mathcal{N}=1$ model can be obtained by choosing a superpotential of the form: 
\begin{equation}
\label{eq:N1ReviewGeneralFreeSuperpotential}
W_\text{free}(\phi) = \int d^d x \frac{1}{2} \left[ \sum_{l=0}^k h_l\, \phi \nabla^{2l} \phi  \right],
\end{equation}
where $h_l$ are constant parameters. In particular, when only one term of order $k$ is present in the above sum -- that is, when:
\begin{equation}
\label{DetailedBalance4}
W\left( \phi  \right) = \frac{g}{2} \int d^d x \left( \phi \nabla^{2k} \phi \right)  ,
\end{equation} 
one obtains a scale invariant theory with a dynamical critical exponent $z=2k$. The constant $g$ is dimensionless under this scaling symmetry, whereas the 
scaling dimensions of the fields are given by $[\phi]=(d-z)/2$ and $[\psi] = \frac{d}{2}$. 
When more than one term is present in the sum \eqref{eq:N1ReviewGeneralFreeSuperpotential}, the theory is dominated at high energy and momentum scales by the highest derivative term and therefore behaves as a $z=2k$ fixed point in the UV.\footnote{It is for this reason that such theories are often labeled as $z=2k$ Lifshitz theories in the literature, even though, strictly speaking, they are only scale invariant when $h_l = 0$ for $l<k$.} This implies that the perturbative renormalizability properties of the interacting versions of this theory are dictated by the highest derivative terms (see also subsection \ref{subsec:Renormalization} as well as \cite{Anselmi:2007ri,Anselmi:2008ry}). Here we shall restrict the discussion strictly to cases with $k=1$ (that is, where the superpotential contains at most two space derivatives). In this case, the bosonic field is just a free, real $z=2$ Lifshitz scalar, whereas the fermion is a free (spinless) Schr\"odinger fermion (with the possible addition of a chemical potential corresponding to the $l=0$ term), whose $U(1)$ particle number symmetry corresponds to the R-symmetry of \eqref{eq:Nequal1Algebra}.

Interactions that respect the supersymmetric algebra \eqref{eq:Nequal1Algebra} may be introduced to the above free models by adding to the superpotential arbitrary local terms which are polynomial in the superfield $\Phi$ and its spatial derivatives. Depending on the Lifshitz dimension of these deformations, such theories have been shown to be perturbatively renormalizable (see \cite{Anselmi:2007ri,Anselmi:2008ry,Chapman:2015wha}, as well as the discussion in subsection \ref{subsec:Renormalization}). Note, however, that such interactions will generally break the Galilean invariance of the fermionic sector of the free model (with $z=2$).

As an example, a model corresponding to the following superpotential in $2+1$ dimensions was considered in \cite{Chapman:2015wha}:
\begin{equation}
\label{DetailedBalance1}
W\left( \phi  \right) = \frac{1}{2} \int d^2x\left( g\left( \nabla _i \phi  \right)^2 - \sum\limits_{n=1}^\infty g a_n\frac{\phi ^{n + 1} \nabla ^2 \phi }{n + 1} + \sum\limits_{n=1}^\infty c_n \frac{\phi ^{n + 1}}{n + 1}  \right) \ ,
\end{equation}
and it was shown that the action \eqref{DetailedBalance3} indeed represents the most general supersymmetric action one can build out of the fields $\phi, \psi$ (that respects the algebra \eqref{eq:Nequal1Algebra} and does not include interaction terms with time derivatives), 
and that supersymmetry is preserved in these models by quantum corrections (up to first order). Note that in $2+1$ dimensions, the field $\phi$ is dimensionless, and there is therefore an infinite number of marginal and relevant deformations (similar to a relativistic scalar theory in two dimensions). In the following discussion, we will restrict ourselves to cases with $ d\ge 3$.

Similarly to relativistic supersymmetry (and to supersymmetric quantum mechanics), the time domain supersymmetric algebra \eqref{eq:Nequal1Algebra} guarantees that the energy spectrum of the theory is non-negative (regardless of the choice of the superpotential functional $W$ and its properties), and that zero energy states are necessarily invariant under the full $\mathcal{N}=1$ supersymmetry of the theory. Since the classical bosonic potential is given by $ \left|\frac{\delta W}{\delta\phi}\right|^2$, the condition for a (semiclassical) supersymmetric vacuum is given by the equation:
\begin{equation}
\label{eq:N1ReviewClassicalSUSYVacuumCondition}
 \frac{\delta W}{\delta\phi} = 0 .
\end{equation}
Note, however, that unlike the relativistic case, this equation is not an algebraic equation but rather a differential one. For models with a superpotential of the form $W = W_\text{free} + \int d^dx \mathcal{W}_\text{int}$ where $ W_\text{free} $ is given by \eqref{eq:N1ReviewGeneralFreeSuperpotential} and $ \mathcal{W}_\text{int}(\phi) $ is an arbitrary function of $\phi$, if $\mathcal{W}'_\text{int}(\phi_0)=0$ then $\phi=\phi_0$ is certainly a constant solution to equation \eqref{eq:N1ReviewClassicalSUSYVacuumCondition}, however there may also be non-constant solutions to this equation, representing supersymmetric vacua that break the spatial translation symmetry. 

When the functional $W(\phi)$ is positive semi-definite (or at least bounded from below), the model \eqref{DetailedBalance3} is said to satisfy the detailed balance condition. In this case, one can show (see \cite{Witten:1981nf,Parisi:1982ud,Dijkgraaf:2009gr}) that a supersymmetric vacuum state $|0\rangle$ always exists that satisfies the properties:\footnote{An alternative formulation for the property \eqref{eq:N1ReviewDetailedBalanceVacuumProperty} is that any equal-time correlation function of $\phi$ in the vacuum state $|0\rangle$ is given by the following path integral in $d$ dimensions:
\begin{equation}
\left\langle 0 | \phi(t,x_1) \ldots \phi(t,x_k) | 0 \right\rangle \propto \int D\tilde\phi(x)\, \tilde\phi(x_1) \ldots \tilde\phi(x_n) e^{-2 W\{\tilde\phi(x)\}}. 
\end{equation}
}
\begin{align}
\label{eq:N1ReviewDetailedBalanceVacuumProperty}
\left\langle \left. \tilde\phi(x) \right| 0 \right\rangle &= N e^{-W\left\{\tilde\phi(x)\right\}} , \\
\label{eq:N1ReviewDetailedBalanceVacuumFermionCondition}
\psi(x) |0\rangle &= 0 ,
\end{align}
where $N$ is a normalization constant, and for any function $\tilde\phi(x)$, $\left| \tilde\phi(x) \right\rangle$ is a state satisfying $ \phi(x) \left| \tilde\phi(x) \right\rangle = \tilde\phi(x) \left| \tilde\phi(x) \right\rangle $. This can be seen from the requirement $ Q |0\rangle =  Q^\dagger |0\rangle = 0$  and the expressions for the supercharges \eqref{eq:N1ReviewSupercharges}. Alternatively, it can be derived from stochastic quantization arguments: 
The Parisi-Sourlas stochastic quantization procedure (see \cite{Parisi:1982ud,Sourlas:1985,Damgaard:1987rr,Dijkgraaf:2009gr}) famously relates the model \eqref{DetailedBalance3} (and the corresponding quantum correlation functions) to the Langevin equation for a bosonic field $\phi$ in a potential given by $W(\phi)$ and a Gaussian noise source\footnote{Note that in the stochastic quantization approach, the fermions take the role of ghost fields that do not appear on external legs of correlation functions.} (and the corresponding stochastic correlation functions). When the above conditions are satisfied, this equation has a steady state described by a Boltzmann distribution, which corresponds to the supersymmetric vacuum state satisfying \eqref{eq:N1ReviewDetailedBalanceVacuumProperty} of the model \eqref{DetailedBalance3}. This also implies that equal-time correlation functions of $\phi$ in this vacuum are the same as the correlation functions of a scalar boson in a $d$-dimensional Euclidean field theory given by the action $W(\phi)$, and therefore one may deduce many properties of the $(d+1)$-dimensional Lifshitz model from those of the corresponding $d$-dimensional theory. In particular, the renormalization group (RG) flow properties of couplings in $W$ are related to those of the $d$-dimensional theory involving only the bosonic field $\phi$. This might lead one to wonder why the fermions do not contribute to the correlation functions of $\phi$ in the $(d+1)$-dimensional model.

Perturbatively, the answer lies in the quantization of the fermions around the semiclassical vacuum $\phi_0(x)$ that minimizes the functional $W$: Since $ \frac{\delta^2 W}{\delta\phi^2} $ is positive semi-definite at $\phi_0$, the fermions should be quantized such that \eqref{eq:N1ReviewDetailedBalanceVacuumFermionCondition} is satisfied. In fact, when $\phi_0$ is constant, the second-order fermion action around it is just that of a Schr\"odinger fermion (with a non-positive chemical potential), and the semiclassical vacuum corresponds to the standard Galilean vacuum for this fermion. It is well known, however, that upon introducing interactions that preserve the fermion's $U(1)$ particle number symmetry (which is just the $U(1)$ R-symmetry here), particle-number-neutral loops of the Schr\"odinger fermion in Feynman diagrams will vanish in the Galilean vacuum (see for example \cite{Bergman:1991hf}). Therefore fermions do not contribute to Feynman diagrams with only bosons on their external legs.

When the functional $W(\phi)$ is not bounded from below (or from above), the action \eqref{DetailedBalance3} still describes a well-defined model (as the potential is still non-negative), and generally a semiclassical vacuum will still exist. Provided the condition \eqref{eq:N1ReviewClassicalSUSYVacuumCondition} is satisfied, it will be supersymmetric and one can still study the theory perturbatively around this vacuum. When doing so, however, if $\phi_0$ does not minimize $W$, the fermionic modes will not all satisfy the condition \eqref{eq:N1ReviewDetailedBalanceVacuumFermionCondition} (in terms of the free Galilean theory, some of them would represent ``holes'' rather than particles), and as a result may contribute to correlation functions of $\phi$. Non-perturbatively, it is more difficult to tell in this case whether the full quantum theory has a supersymmetric vacuum -- in particular, the equation \eqref{eq:N1ReviewClassicalSUSYVacuumCondition} may have soliton-like vacuum solutions in addition to the constant solutions, and tunneling effects between them may cause the dynamical breaking of supersymmetry in the full quantum theory. We discuss these possibilities more in section \ref{sec:Summary}, but for most of the following discussion we assume the existence of a supersymmetric vacuum.

\subsection{A Holomorphic $\mathcal{N}=2$ Model of Time Domain Supersymmetry}
\label{sec:TheHolomorphicModel}

In this subsection we construct a family of supersymmetric, non-relativistic field theory models in $d+1$ dimensions with $\mathcal{N}=2$ time domain supersymmetry. In addition to time translations, space translations and space rotations, these models are invariant under two complex supercharges (or four real ones) labeled $Q_\alpha$ ($\alpha=1,2$), as well as an $SU(2)$ R-symmetry charge $R^a$ ($a=1,2,3$) satisfying:  
\begin{equation}
\label{eq:AlgebraSusy}
\begin{split}
&\{ Q_\alpha, Q_\beta\} = 0,
\quad \{ Q_\alpha, Q^\dagger_{\dot{\alpha}} \} =2\sigma^0_{\alpha\dot{\alpha}}H, \\
\quad &[M_{ij} , Q_\alpha ]=0 ,
\quad [P_i, Q_\alpha] = 0,
\quad [R^a, Q_\alpha] = \frac{i}{2} (\sigma^a)_{\alpha}{}^\beta Q_\beta,
\end{split}
\end{equation}
where for the fermionic $SU(2)$ indices we use the conventions of \cite{SUSYPrimer},\footnote{Throughout this work the fermions and fermion charges are non-relativistic and scalar under space rotations, though our conventions are inherited from the relativistic structure for convenience. In this sense, the index $\alpha$ does not carry any information about the spin, but rather acts as an index for the representation of the $SU(2)$ R-symmetry.} in which
$\alpha,\dot{\alpha}=\{1,2\}$, and ${\sigma^0}_{\alpha \dot{\alpha}}={\bar{\sigma}}^{0\dot{\alpha}\alpha }=\mathbf{1}_{2\times2}$ (a summary of conventions can be found in appendix \ref{app:NotationaAndConventions}). We also use $(\sigma^a)_{\alpha}{}^\beta \,(a=1,2,3)$ here to denote the Pauli matrices.
In cases with a Lifshitz scaling symmetry $D$ (with some dynamical critical exponent $z$), these relations also imply:
\begin{equation}
 \quad [ D , Q_\alpha ] = i  \frac{z}{2} Q_\alpha  .
\end{equation}

In analogy to the $\mathcal{N}=1$ models and the relativistic Wess-Zumino model, we may construct an off-shell realization of this $\mathcal{N}=2$ algebra in superspace formalism. 
Similarly to the relativistic case, we label superspace coordinates by 
$x^\mu , \theta^\alpha, \theta^\dagger_{\dot{\alpha}}$, where $x^\mu \equiv (t,x^i)$, and $\theta^\alpha, \theta^\dagger_{\dot{\alpha}}$ are anti-commuting two-component coordinates. 
The supersymmetric transformation of these coordinates will be given by:
\begin{equation}
\label{eq:N2HolomorphicSuperspaceTransformationLaws}
\begin{split}
&\delta \theta^\alpha = \epsilon^\alpha, \quad 
\delta \theta^\dagger_{\dot{\alpha}} = \epsilon^\dagger_{\dot{\alpha}}, \\
&\delta t = i \epsilon \sigma_0 \theta^\dagger - i \theta \sigma_0 \epsilon^\dagger, \quad
\delta x^i = 0,
\end{split}
\end{equation}
and therefore in terms of the superspace coordinates, the supercharges are given by:
\begin{equation}
Q_\alpha=i\frac{\partial}{\partial\theta^\alpha}-(\sigma^0\theta^\dagger)_\alpha\partial_t, \qquad Q^\dagger_{\dot{\alpha}}=-i\frac{\partial}{\partial{\theta^\dagger}^{\dot{\alpha}}}+(\theta\sigma^0)_{\dot{\alpha}}\partial_t.
\end{equation}

Continuing the analogy to the relativistic Wess-Zumino model, 
we define a holomorphic superfield $\Phi(x^\mu,\theta,\theta^\dagger)$ as one satisfying the condition:
\begin{equation}
D^\dagger_{\dot{\alpha}}\Phi=0,
\end{equation}
and similarly an anti-holomorphic superfield $\Phi^\dagger$ as one satisfying
\begin{equation}
D_\alpha\Phi^\dagger=0,
\end{equation}
with the supersymmetric covariant derivatives defined as:
\begin{equation}
\label{eq:SuperDerivativesDef}
D_\alpha=\frac{\partial}{\partial\theta^\alpha}-i(\sigma^0\theta^\dagger)_\alpha\partial_t, \qquad D^\dagger_{\dot{\alpha}}=-\frac{\partial}{\partial{\theta^\dagger}^{\dot{\alpha}}}+i(\theta\sigma^0)_{\dot{\alpha}}\partial_t.
\end{equation}

The holomorphic and anti-holomorphic superfields $\Phi, \Phi^\dagger$ can be generally decomposed in terms of component fields as follows:
\begin{equation}
\begin{split}
\label{eq:N2HolomorphicDecomposition}
\Phi&=\phi(y,\vec{x})+\sqrt{2}\theta\psi(y,\vec{x})+\theta\theta F(y,\vec{x}), \\
\Phi^\dagger&=\phi^*(y^*,\vec{x})+\sqrt{2}\theta^\dagger\psi^\dagger(y^*,\vec{x})+\theta^\dagger\theta^\dagger F(y^*,\vec{x}),
\end{split}
\end{equation}
where $\phi$ is a complex bosonic field, $\psi_\alpha$ is a two component complex fermionic field, $F$ is an auxiliary complex bosonic field (needed in order to ensure the closure of the supersymmetric algebra off-shell), and $y,y^*$ are generalized time coordinates defined by:
\begin{equation}
\label{eq:GeneralizedCoordinates}
y\equiv t+i\theta^\dagger\bar{\sigma}^0\theta, \qquad y^*\equiv t-i\theta^\dagger\bar{\sigma}^0\theta.
\end{equation}
From the decomposition \eqref{eq:N2HolomorphicDecomposition} and the transformations \eqref{eq:N2HolomorphicSuperspaceTransformationLaws}, one can readily deduce the supersymmetric transformation laws for the component fields to be:
\begin{equation}
\label{eq:SusyTrans1}
\begin{aligned}
 &\delta\phi=\epsilon\psi, \qquad \delta\phi^*=\epsilon^\dagger\psi^\dagger,\\
&\delta\psi_\alpha=-i(\sigma^0\epsilon^\dagger)_\alpha\partial_t\phi+\epsilon_\alpha F, \qquad \delta\psi^\dagger_{\dot{\alpha}}=i(\epsilon\sigma^0)_{\dot{\alpha}}\partial_t\phi^*+\epsilon^\dagger_{\dot{\alpha}} F^*, \\
&\delta F =-i\epsilon^\dagger\bar{\sigma}^0\partial_t\psi, \qquad \delta F^* =i\partial_t\psi^\dagger\bar{\sigma}^0\epsilon.
\end{aligned}
\end{equation}
In superspace terms, the most general 
action one can build from the holomorphic superfield $\Phi$ which is local and invariant under the supersymmetric algebra \eqref{eq:AlgebraSusy} corresponds to the following Lagrangian:
\begin{equation}
\label{eq:LagrangianGeneral}
L = \int d^2\theta d^2\theta^\dagger\, K(\Phi,\Phi^\dagger) +\int d^2\theta\, W(\Phi) + \int d^2\theta^\dagger\, \bar{W}(\Phi^\dagger),
\end{equation}
where the K\"ahler potential $K(\Phi,\Phi^\dagger) \equiv \int d^dx\, \mathcal{K}(\Phi,\Phi^\dagger)$ is a local, real functional of $\Phi(x)$ (and $\Phi^\dagger(x)$) and the superpotential $W(\Phi) \equiv \int d^dx\, \mathcal{W}(\Phi)$ is a local, holomorphic functional of $\Phi(x)$ (both of which are invariant under spatial translations and rotations, and may contain spatial derivatives of $\Phi$,$\Phi^\dagger$).
Each term in the Lagrangian  \eqref{eq:LagrangianGeneral} is independently invariant under the supersymmetric transformation generated by the supercharges $Q_\alpha$ and $Q^\dagger_{\dot{\alpha}}$ (up to a total derivative). 
Recalling again that the fermions are non-relativistic and do not carry any spin, note that the model \eqref{eq:LagrangianGeneral} can be considered in any number of spacetime dimensions $d+1$. 

If we restrict our discussion to cases which, in the free limit, behave as a $z=2$ Lifshitz fixed point in the UV (that is, cases in which the classical action involves terms with up to 2 time derivatives or 4 space derivatives), $\mathcal{K}(\Phi,\Phi^\dagger)$ will be a general real function of $\Phi$,$\Phi^\dagger$ (with no derivatives), whereas the superpotential density $\mathcal{W}(\Phi)$ will take the general form: 
\begin{equation}
\label{eq:N2GeneralSuperpotnetialWithDerivativeInteractions}
\mathcal{W}(\Phi)=G(\Phi)\partial_i \Phi\partial^i\Phi+ F(\Phi),
\end{equation}
where $F(\Phi)$ and $G(\Phi)\neq 0$ are general holomorphic functions of $\Phi$. 
Further restricting to models which are renormalizable in $d\geq 3$ space dimensions (see the discussion in subsection \ref{subsec:Renormalization}), we shall assume for the majority of the following discussion that $\mathcal{K}(\Phi,\Phi^\dagger) = \Phi^\dagger \Phi$, $G(\Phi)=\text{const.}$ and $F(\Phi)$ is a polynomial of degree $n\leq n^* \equiv \frac{2d}{d-2}$.

Performing the integration over the Grassmannian coordinates $\theta$, $\theta^\dagger$ and eliminating the auxiliary fields $F$, $F^*$ using their equations of motion, one obtains the following expression for the Lagrangian in terms of the component fields:
\begin{equation}
\label{eq:ActionInComponents}
\begin{split}
L =& \int d^dx\,\left[ \partial_t \phi^* \partial_t\phi + i\psi^\dagger\bar{\sigma}^0\psi-\left|\frac{\delta W}{\delta\phi}\right|^2 \right] \\
&-\int d^dx d^dx'\, \frac{1}{2}\frac{\delta^2 W}{\delta\phi(x)\delta\phi(x')}\psi(x)\psi(x') \\
&- \int d^dx d^dx'\, \frac{1}{2} \frac{\delta^2 \bar{W}}{\delta\phi^*(x)\delta\phi^*(x')}\psi^\dagger(x)\psi^\dagger(x'). 
\end{split}
\end{equation}
Much like the $\mathcal{N}=1$ case, this family of models can be viewed as a dimensional extension of the $\mathcal{N}=2$ supersymmetric quantum mechanics models discussed e.g.\ in \cite{Dolgallo1994,Jaffe:1987nx}.
Note also that as these models are a special case of the $\mathcal{N}=1$  models discussed in subsection \ref{subsec:ReviewOfN1TDSupersymmetry}, they can be written in terms of the $\mathcal{N}=1$ action \eqref{eq:N1ReviewSuperspaceAction}, where the $\mathcal{N}=1$ superpotential is related to the $\mathcal{N}=2$ one as follows: 
\begin{equation}
\label{eq:N1SuperpotentialInTermsOfN2}
W^{\mathcal{N}=1}(\phi_1,\phi_2) = 2 \operatorname{Im}\left[ e^{i2\alpha} W^{\mathcal{N}=2} (\phi) \right],
\end{equation}
where $\phi = \frac{1}{\sqrt{2}} (\phi_1 + i \phi_2)$, and $\alpha$ is an arbitrary constant phase\footnote{This can be easily seen by substituting $ \psi_1 \equiv \frac{e^{i\alpha}}{\sqrt{2}}\left( \tilde{\psi_1} + i \tilde{\psi_2} \right)$, $ \psi_2 \equiv \frac{e^{i\alpha}}{\sqrt{2}}\left( -i\tilde{\psi_1}^\dagger + \tilde{\psi_2}^\dagger \right)$ into the action \eqref{eq:ActionInComponents} and comparing with \eqref{DetailedBalance3}, keeping in mind that $W$ is holomorphic in $\Phi$.} (that corresponds to the choice of the $\mathcal{N}=1$ supercharge within the $\mathcal{N}=2$ algebra).

A free $\mathcal{N}=2$ model (with $z=2$ UV scaling) can be obtained by choosing a superpotential density of the form:
\begin{equation}
\label{eq:FreeSuperpotential}
\mathcal{W}_{\text{free}}(\Phi)= \frac{g}{2}\Phi\nabla^2\Phi+\frac{f_2}{2}\Phi^2.
\end{equation}
The space of parameters in the free theory thus consists of the parameter $g$, which we take to be real and positive\footnote{By fixing the arbitrary phase factor in the definition of the superfield $\Phi$, one can always make $g$ real and positive, but this generally leaves $f_2$ complex.} ($g>0,\, g\in \mathbb{R}$) and acts here as a conversion factor between time and space units, as well as the gap parameter $f_2$ which is generally complex and determines the gap in the spectrum. 
Substituting this superpotential into the expression \eqref{eq:ActionInComponents}, the Lagrangian density for the free model reads:
\begin{equation}
\label{eq:LagrangianFreeModels}
\begin{aligned}
\mathcal{L}_{\text{free}}&= \partial_t\phi^*\partial_t\phi-g^2\nabla^2\phi\nabla^2\phi^*-|f_2|^2\phi^*\phi-g(f_2\phi\nabla^2\phi^*+f_2^*\phi^*\nabla^2\phi)\\
&+i\psi^\dagger\bar{\sigma}^0\partial_t\psi-\frac{g}{2}(\psi\nabla^2\psi+\psi^\dagger\nabla^2\psi^\dagger)-\frac{1}{2}(f_2\psi\psi+f_2^*\psi^\dagger\psi^\dagger).
\end{aligned}
\end{equation}

This model consists of a free, complex ($z=2$) Lifshitz scalar field, and two free Schr\"odinger fermion fields (with chemical-potential-like terms). In addition to the symmetries in \eqref{eq:AlgebraSusy}, the free model has several more noteworthy symmetries (see table \ref{TableSymmetries}): 
\begin{itemize}
\item The bosonic sector in invariant under an extra internal $U(1)$ symmetry.
\item When $f_2$ is real, the fermionic sector has a Galilean boost symmetry.
\item In addition to the $SU(2)$ R-symmetry, the fermionic sector has an additional internal $U(1)_M$ symmetry that, when $f_2$ is real, corresponds to the Galilean conserved particle number.
\end{itemize}

Moreover, when $f_2=0$ (or equivalently in the high energy limit), the free model \eqref{eq:LagrangianFreeModels} is invariant under the $z=2$ Lifshitz scaling transformation \eqref{Intro:LifshitzScaling}. Similarly to the $\mathcal{N}=1$ case, the scaling dimensions of the fields are given by $[\phi]=(d-2)/2$ and $[\psi]=d/2$. 
\begin{table} 
\centering
\begin{tabular}{| c | l |} 
	\hline
	Group   & \multicolumn{1}{c|}{Transformation}      \\
	\hline 
	$U\left(1\right)_M$ & $ \begin{pmatrix}\psi_1 \\ \psi_2^* \end{pmatrix} \rightarrow e^{i\theta\sigma^1}\begin{pmatrix}\psi_1 \\ \psi_2^* \end{pmatrix}$.  \\
	\hline
	$U\left(1\right)$   & $\phi\rightarrow e^{i\theta}\phi$.  \\
	\hline
	$SU\left(2\right)$	& $\begin{pmatrix}\psi_1 \\ \psi_2 \end{pmatrix} \rightarrow e^{i\theta\sigma^a/2}\begin{pmatrix}\psi_1 \\ \psi_2 \end{pmatrix}, \qquad a=1,2,3$. \\
	\hline
\end{tabular}\\
\caption{The internal symmetries of the free fields Lagrangian \eqref{eq:LagrangianFreeModels}, for the case of a real $f_2$.}
\label{TableSymmetries}
\end{table}

\begin{figure}[!htb]
 \includegraphics[width=0.8\textwidth]{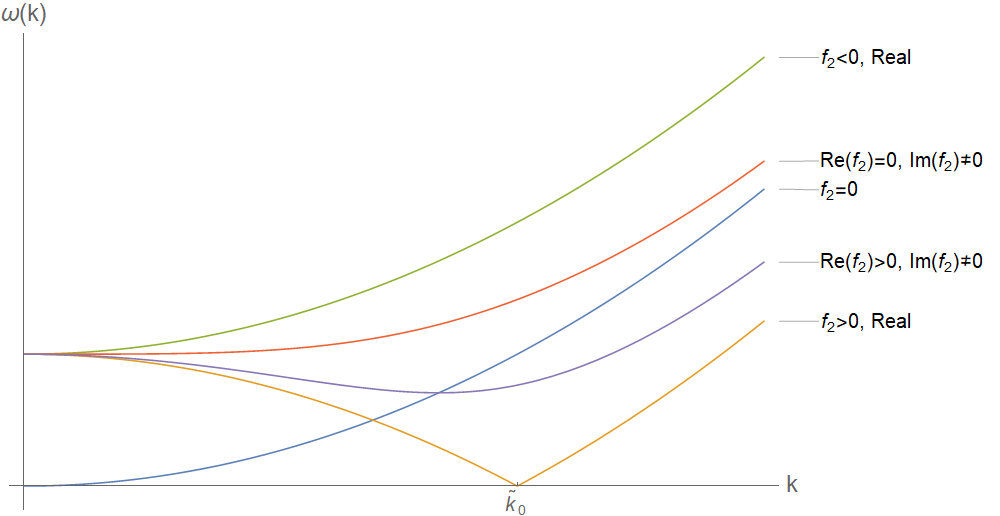}
\centering 
\caption{The free particle dispersion relation.}
\label{fig:HolModelDispRelPlots}
\end{figure}

The free single particle dispersion relation can be easily read off the Lagrangian \eqref{eq:LagrangianFreeModels}, and is given by:\footnote{We denote $k\equiv|\vec{k}|,k^2\equiv k_ik^i$ with $i=1,\ldots,d$.}
\begin{equation}
\omega = \pm | g k^2 - f_2 |.
\end{equation}
Thus both the magnitude and phase of $f_2$ have physical significance to the spectrum: When $\operatorname{Re}(f_2)\leq 0$, the energy is minimal at $k=0$ momentum and the gap is given by $\omega_\text{gap}=|f_2|$. In the case of a purely imaginary $f_2$, for example, the single particle dispersion relation reads $\omega = \pm \sqrt{g^2 k^4 + \operatorname{Im}(f_2)^2} $, and $f_2$ plays a role similar to the relativistic mass.
When $\operatorname{Re}(f_2) > 0$, however, the minimal energy occurs at momenta of magnitude $k=\sqrt{\operatorname{Re}(f_2)/g}$, and the gap is given by $\omega_\text{gap} = \left|\operatorname{Im}(f_2)\right|$. In particular, when $f_2$ is real and positive, the spectrum is gapless and contains a sphere of zero energy states at momenta of magnitude $\tilde{k}_0 = \sqrt{f_2/g} $. As discussed in subsection \ref{subsec:GaplessSingularCase}, with the addition of interactions, this case suffers from IR singularities and is generically strongly coupled at low energies. The various cases are demonstrated in figure \ref{fig:HolModelDispRelPlots}.
For a complete derivation of the particle spectrum and second quantization of the bosons and fermions in \eqref{eq:LagrangianFreeModels}, see appendix \ref{app:Quant}.

One may introduce general (renormalizable) interactions that respect  the supersymmetric algebra \eqref{eq:AlgebraSusy} by adding to the superpotential density a polynomial in $\Phi$, i.e.:
\begin{equation}
\label{eq:N2GeneralSuperpotnetial}
\mathcal{W} = \mathcal{W}_{\text{free}} + \sum_{n=3}^{n^*} \mathcal{W}_{\text{int}}^{(n)},
\end{equation}
where:
\begin{equation}
\label{eq:GeneralSuperInteraction}
\mathcal{W}_{\text{int}}^{(n)} \equiv \frac{f_n}{n}\Phi^n,
\end{equation}
and $f_n$ is a coupling constant. Note that, while these interaction terms are invariant under the $SU(2)$ R-symmetry, they generally break the fermionic Galilean symmetry of the free theory, as well as the $U(1)$ and $U(1)_M$ symmetries (of the bosonic and fermionic sectors respectively) listed in table \ref{TableSymmetries}. One should not expect, therefore, a conservation of the fermionic Galilean particle number in these models.

In terms of the existence of a supersymmetric vacuum state, the $\mathcal{N}=2$ models inherit the properties of the $\mathcal{N}=1$ ones as discussed in subsection \ref{subsec:ReviewOfN1TDSupersymmetry}. The bosonic potential is given by:
\begin{equation}
V=|F|^2 = \left| \frac{\delta W}{\delta\phi} \right|^2\geq0,
\end{equation}
and the condition for a semiclassical supersymmetric vacuum is given by the differential equation:
\begin{equation}
\label{eq:N2SUSYVacuumCondition}
 \frac{\delta W}{\delta\phi} = 0, 
\end{equation}
with the difference being that the superpotential $W$ is now a holomorphic functional. For a superpotential of the form \eqref{eq:N2GeneralSuperpotnetial}, then, the solution $\phi = 0 $ always represents such a supersymmetric vacuum,\footnote{In fact, similarly to the relativistic Wess-Zumino model, since $W$ is holomorphic, as long as the polynomial $F(\phi)$ of \eqref{eq:N2GeneralSuperpotnetialWithDerivativeInteractions} is of degree $n \ge 2$, there is always a constant solution to the equation \eqref{eq:N2SUSYVacuumCondition}, and therefore a supersymmetric semiclassical vacuum always exists. In order to obtain a spontaneous breaking of supersymmetry on this level, one would have to consider a model with multiple interacting holomorphic superfields (as in the O'Raifeartaigh model).} although there may be others $\phi=\phi_0(x)$, either constant or non-constant in space, depending on the form of $W$.

An important distinction in relation to the general $\mathcal{N}=1$ case, however, is the fact that $W$ is holomorphic and therefore the $\mathcal{N}=1$ superpotential \eqref{eq:N1SuperpotentialInTermsOfN2} is never bounded and the detailed balance condition is never satisfied. From the point of view of perturbation theory around the $\phi=0$ vacuum, the interactions preserve the $SU(2)$ R-symmetry, but break the $U(1)_M$ symmetry. Consequently, when $f_2$ is real, the two fermions always represent a particle and ``hole'' pair with interactions that break the Galilean particle number symmetry of the free theory. Therefore unlike the detailed balance case, fermionic closed loops will not vanish, and will contribute to the bosonic correlation functions. In particular, this is required for the cancellations that lead to the non-renormalization discussed in section \ref{sec:QuantumCorrections}. Of course, as in the $\mathcal{N}=1$ case, one must also consider non-perturbative effects which may lead to the dynamical breaking of supersymmetry here (for further discussion see section \ref{sec:Summary}).

To close this section, for later reference we make the following definitions for the above models \eqref{eq:N2GeneralSuperpotnetial}-\eqref{eq:GeneralSuperInteraction}:
\begin{itemize}
\item For the ungapped, IR singular cases with $f_2>0, \quad f_2 \in \mathbb{R}$, we define $f_2 \equiv g\tilde{k}_0^2$, with $\tilde{k}_0 \in \mathbb{R} $. The dispersion relation is then given by $\omega=g|k^2-\tilde{k}_0^2|$, and thus $\tilde{k}_0$ is the momentum of zero energy.
\item For the interaction terms (with $n\ge 3$), we define the coupling constant $ \lambda_n \equiv f_n g^{-n/2}$, which is dimensionless in time (energy) units. When $n=n^*$, $\lambda_n$ is dimensionless in both time and space units. 
\end{itemize}

\section{Quantum Analysis of Lifshitz Field Theories}
\label{sec:QuantumCorrections}

In this section we study the quantum behaviour of the family of $\mathcal{N}=2$ time domain holomorphic supersymmetric models presented in subsection \ref{sec:TheHolomorphicModel} in diverse dimensions and different choices of interactions of the form \eqref{eq:N2GeneralSuperpotnetial}-\eqref{eq:GeneralSuperInteraction}.

In subsection \ref{subsec:Renormalization}, we discuss the renormalization group flow properties of Lifshitz field theories such as the models at hand, review several renormalization and regularization methods for these types of models and study some properties of quantum Lifshitz fixed points.
In subsection \ref{subsec:DualScaleRGFlows}, we make a digression to discuss a dual-scale RG formalism, in which the energy and the momentum scales flow independently, and point out some properties of Lifshitz fixed points in this picture.
In subsection \ref{subsec:NonRenormalizationTheoremsAGeneralProof} we prove non-renormalization theorems for the models at hand, based on the symmetries of the theory. The arguments are similar to the ones made in \cite{Seiberg:1993vc} for the relativistic holomorphic supersymmetry, with a few subtleties (due to the non-boost-invariant nature of the theory). 
In subsection \ref{subsec:PerturbativeAnalysis} we discuss and demonstrate some properties of the perturbative quantum corrections in these models, including a perturbative argument for non-renormalization and some examples of its consequences.

In subsection \ref{sec:MarginalCases}, three different marginal cases are analyzed: $6+1$, $4+1$ and $3+1$ spacetime dimensions with $n=3,4,6$ interactions (respectively) of the form \eqref{eq:GeneralSuperInteraction}. We show that in all three cases, there is a line of quantum critical points, in which the system possesses an exact Lifshitz scaling symmetry with a critical exponent that depends on the coupling.
This conclusion is not based on perturbative arguments, and applies to the strong coupling limit as well. 
Finally, in subsection \ref{subsec:GaplessSingularCase} we discuss the gapless case with $f_2>0$ and its IR properties.

\subsection{Regularization, Renormalization and Fixed Points}
\label{subsec:Renormalization}

We turn to discuss the general procedures of regularization and renormalization, as well as scaling behaviour, in the context of non-boost-invariant field theories. In such a theory, there is no inherent relation between space and time dictated by the symmetry algebra, and therefore one can consider scaling the space and time dimensions separately.
In general, any operator in the theory will carry both time and space dimensions. If an operator  $\hat{O}$ carries dimensions $\left[E\right]^{\Delta_t}\left[p\right]^{\Delta_s}$, where $\left[E\right]$ and $\left[p\right]$ stand for energy (time) and spatial momentum (space) units\footnote{As usual, we use units in which $\hbar=1$.} respectively, 
then one can define its weighted Lifshitz dimension corresponding to a dynamical exponent $z$ as its dimension under a Lifshitz transformation of the form \eqref{Intro:LifshitzScaling}, that is:
\begin{equation}
\Delta_z^{\text{lif}}(\hat{O}) \equiv z \Delta_t (\hat{O}) + \Delta_s (\hat{O}).
\end{equation}
Note that this definition depends on the choice of $z$, which is for now left as an unrestricted parameter for a given theory (for example, for the family of models we consider here, we do not restrict $z$ to be 2 at this point). As we shall see, any specific fixed point will correspond to Lifshitz invariance with respect to a particular value of $z$.

In the free theory \eqref{eq:LagrangianFreeModels} and for a general value of the critical exponent $z$, the parameter $g$ signifying the relative strength of the space and time kinetic terms carries dimensions 
$\left[g\right]=\left[E\right]\left[p\right]^{-2}$. Its weighted Lifshitz dimension is therefore:
\begin{equation}
\Delta_z^{\text{lif}}(g) = z-2.
\end{equation}
Specifically for $z=2$ it is dimensionless $\Delta_{z=2}^{\text{lif}}(g)=0$, aligning with the fact that the free gapless ($f_2=0$) theory is invariant under Lifshitz scaling symmetry with a critical exponent of $z=2$. 

Perturbative regularization and renormalization procedures of non-boost-invariant (Lifshitz) field theories have been previously discussed in e.g. \cite{Anselmi:2007ri,Anselmi:2008ry,Visser:2009fg,Fujimori:2015mea,Arav:2016akx}.  
Generally, they are similar to those of a relativistic theory, with the main difference that in the non-boost-invariant case, the analysis and classification of UV divergences is carried out with respect to the weighted Lifshitz scaling dimension, with the parameter $z=z_\text{uv}$ corresponding to the critical exponent of the free theory at the UV\footnote{Put differently, one chooses the value of $z$ for which the coefficient of the term with the highest number of spatial derivatives in the action of the free theory is dimensionless. The superficial degree of divergence is then defined depending on the weighted Lifshitz dimension corresponding to this value of $z$.} \cite{Anselmi:2007ri}. In analogy with the relativistic case, an operator $\hat{O}$ is called relevant if the corresponding coupling constant $g_{\hat{O}}$ has positive weighted Lifshitz scaling dimension, $\Delta_{z_\text{uv}}^{\text{lif}}(g_{\hat{O}})>0$. Similarly, it is classified as an irrelevant operator in cases where the corresponding coupling constant carries negative weighted Lifshitz scaling dimension $\Delta_{z_\text{uv}}^{\text{lif}}(g_{\hat{O}})<0$, and (classically) marginal when $\Delta_{z_\text{uv}}^{\text{lif}}(g_{\hat{O}})=0$. 
For example, for the family of models discussed in subsection \ref{sec:TheHolomorphicModel}, $z_\text{uv}=2$ and therefore the coupling $f_n$ is relevant when $n<n^*$, (classically) marginal when $n=n^*$ and irrelevant when $n>n^*$.

Various regularization and renormalization methods have been used in the literature for non-boost-invariant field theories.
A subset of regularization methods which are commonly used (see e.g. \cite{Fitzpatrick:2012ww,Alexandre:2011kr,Chapman:2015wha,Alexandre:2013wua}) are \textit{time-first} regularization methods, in which one first performs the integration over energy space and subsequently uses standard relativistic-like regularization procedures to regularize the remaining Euclidean integrals over momentum space. This type of methods can only be used in cases where the integration over energy space converges for all correlation functions one is interested in. 

Consider, for example, the $m$-loop contribution to any $n$-point correlation function in a ($d+1$)-dimensional field theory containing Lifshitz scalar bosons and fermions (of the type considered here) with a UV critical exponent of $z_\text{uv}=2$:
\begin{equation}
I^{(n,m)}(\omega_i,p_i) = \iint \prod_{j=1}^m d^dq_j d\Omega_j \tilde{I}(\omega_i,p_i,q_j,\Omega_j),
\end{equation}
where $(\omega_i, \vec{p}_i)$ ($i=1,\ldots,n$) are the external energies (for time coordinates) and momenta (for space coordinates) respectively which appear in the correlation function, and $(\Omega_j,\vec{q}_j)$ ($j=1,\ldots,m$) are the internal loop energies and momenta. When there are no composite operators in the correlation function, one can start by performing the integration over the energies $\int \prod_{j=1}^m d\Omega_j$ since it is always UV convergent\footnote{This follows from the following arguments: First, note that for almost any possible Feynman diagram or subdiagram, the superficial degree of divergence in energy space alone is negative. The only possible exception is loops containing only a single propagator, when that propagator is first order in time derivatives (such as the fermions in the models discussed in section \ref{sec:TimeDomainNRSUSY}). Such loops can be rendered UV finite via an appropriate choice of regularization or normal ordering. Then the absolute convergence in energy space is guaranteed by the Weinberg-Dyson convergence theorem (see e.g. \cite{Weinberg:1959nj,HahnZimmermann}), applied to the energy space integrals alone.}
(it can be performed, for example, by using contour integration in the complex plane). 
One is then left with an expression of the form:
\begin{equation}
I^{(n,m)}(\omega_i,p_i) = \int \prod_{j=1}^m d^dq_j \hat{I}(\omega_i,p_i,q_j),
\end{equation}
containing only spatial momenta integrations, similar to those of Euclidean field theories. One then proceeds to regularize these remaining $d$-dimensional integrals $\int \prod_{j=1}^m d^d q_j $ using any of the well-known relativistic regularization methods, such as using a spatial UV cutoff $\Lambda_s$, or dimensional regularization by varying the number of space dimensions $d = d_s^\text{phys} - \epsilon_s$. As a more general alternative, one can use a regularization method in which both energy and momentum integrations are regularized separately. For example, one may introduce separate UV cutoffs for spatial momenta $\Lambda_s$ and for energies $\Lambda_t$. Another example is the split dimensional regularization method (introduced in
 \cite{Leibbrandt:1996np,Leibbrandt:1997kh} and used in \cite{Anselmi:2007ri,Arav:2016akx} in the context of Lifshitz field theories), in which one analytically continues both the number of space dimensions $d_s = d_s^\text{phys} - \epsilon_s$ and time dimensions $d_t = 1 - \epsilon_t$ separately.

Any renormalization scheme one chooses to renormalize the theory will inevitably introduce at least one renormalization scale. One may choose a single-scale renormalization scheme, which introduces a scale $\mu_s$ that carries only spatial dimensions $\left[\mu_s\right]=\left[p\right]^1$ (or, alternatively, a scale $\mu_t$ that carries only time dimensions).  This may be, for example, a scale of external (spatial) momenta in the renormalization condition for an ``on-shell'' scheme, a scale introduced as part of a minimal subtraction scheme or, in the Wilsonian approach, a lower bound for spatial Feynman integrals of the form $\int_{\mu_s}^{\Lambda_s} d^dq_j$. The result yields renormalized correlation functions $I^{(n)}_{\text{ren}}(\mu_s,\omega_i,p_i)$ which depend on the external momenta and energies, and the renormalization scale. The time-first regularization methods discussed above clearly lend themselves to such a (spatial) single-scale renormalization scheme.

An alternative and more general approach is to use a dual-scale renormalization scheme, in which one introduces two different renormalization scales: $\mu_s$ for the spatial and $\mu_t$ for the time dimensions, with $\left[\mu_s\right]=\left[p\right]^1$ and $\left[\mu_t\right]=\left[E\right]^1$. 
These can correspond to ``on-shell'' conditions on both the external momenta and energies of the form: $ \omega_i \sim \mu_t$, $p_i \sim \mu_s$. They could appear as part of a minimal subtraction scheme after regularizing both energy and momentum integrations (for example, when using a split dimensional regularization method). In a Wilsonian approach they would appear as the lower bounds on spatial momenta and energy integrals respectively, i.e.\ $\int_{\mu_t}^{\Lambda_t}d\Omega_j \int_{\mu_s}^{\Lambda_s}d^d q_j$. It is important to note that unlike boost invariant theories, there is no natural relation between the two parameters $\mu_s, \mu_t$ that holds at all scales. Although there may be UV and IR Lifshitz fixed points that characterize the RG flow of the quantum theory, those can generally have different values of the dynamical critical exponent $z$ associated with them, and one may not know what they are ahead of time as they can get contributions from quantum corrections (as we demonstrate later). This implies that generally one could consider two-dimensional RG flows in which the momentum and energy scales flow independently.

We now turn to study the RG flow equations in non-boost-invariant (Lifshitz) field theories. For simplicity we first consider the single-scale approach to renormalization, in which only a spatial renormalization scale $\mu_s$ is introduced.  Consider a non-boost-invariant field theory in $d+1$ dimensions, with an action containing a set of parameters (or coupling constants) $c_l,\, (l=1,\ldots,L)$. In the models of the form \eqref{eq:N2GeneralSuperpotnetial} (as discussed in subsection \ref{sec:TheHolomorphicModel}) these are the parameters $c_l=\lbrace g, f_2, f_3,\ldots\rbrace$ representing the kinetic term parameter $g$, the gap term $f_2$ and the coupling constants.

Typically at least one of the parameters $c_l$ has non-vanishing energy dimension. Let us assume then that $c_1$ is such a parameter, that is $\Delta_t(c_1) \neq 0$. Then one can always define dimensionless versions $\tilde{c_l}$ of the other parameters using $c_1$ and $\mu_s$ as follows:
\begin{equation}
\tilde{c_l} \equiv c_l\, c_1^{s_l} \mu_s^{r_l}, \quad l=2,\ldots,L,
\end{equation}
where 
$s_l = -\frac{\Delta_t(c_l)}{\Delta_t(c_1)}$ and
$r_l = -\Delta_s(c_l) - s_l \Delta_s(c_1)$.
For example, for the $\mathcal{N}=2$ supersymmetric family of models discussed in subsection \ref{sec:TheHolomorphicModel}, we have $c_1=g$ and $\tilde{c_n} = f_n g^{-\frac{n}{2}} \mu_s^{-\frac{d-2}{2}(n^*-n)} = \lambda_n \mu_s^{-\frac{d-2}{2}(n^*-n)} $ for $2\leq n \leq n^*$ (it is easy to see that in the marginal case $\lambda_{n^*}$ is indeed dimensionless). $g$ in this case cannot be made dimensionless (as there is no other energy scale). As will be explained in the rest of this subsection, its RG flow properties will be responsible for the value of the critical exponent $z$ associated with a particular fixed point.

Next, consider a renormalized n-point correlation function\footnote{For this discussion, we are considering a correlation function written in momentum and energy space, which does not include the overall delta function factor associated with momentum and energy conservation.} $G^{(n)}_{\text{ren}}(p_i,\omega_i;c_l,\mu_s)$ for some field $\phi$.\footnote{We assume for simplicity that all external fields appearing in the correlation function are identical, but a similar analysis holds in cases where there are various fields and the equations can be easily adjusted.} It will generally depend on the external momenta and energies $(p_i,\omega_i)$, the (spatial) renormalization scale $\mu_s$ and the renormalized coefficients $c_l(\mu_s)$ which run with the scale $\mu_s$ (or alternatively $c_1(\mu_s)$ and $\tilde{c_l}(\mu_s)$).
The Callan-Symanzik RG equation for the n-point correlation function $G^{(n)}_{\text{ren}}$ can be written as follows:
\begin{equation}
\label{eq:CallanSymmanzik1}
\left( \mu_s\frac{\partial}{\partial \mu_s}+ \gamma_{c_1} c_1 \frac{\partial}{\partial c_1} + \sum_{l=2}^L \beta_l\frac{\partial}{\partial \tilde{c_l}}+ n\gamma_\phi \right)G^{(n)}_{\text{ren}}(p_i,\omega_i;\mu_s, c_1, \tilde{c_l})=0,
\end{equation}
(with $\tilde{c_l} = \{\tilde{c_2},\ldots,\tilde{c_L}\}$) where we have defined:
\begin{align}
\gamma_{c_1} (\tilde{c_2},\ldots\tilde{c_L}) &\equiv \frac{\mu_s}{c_1} \frac{\partial c_1}{\partial \mu_s}, \label{eq:AnomalouseDimOfGDef}\\
\beta_l  (\tilde{c_2},\ldots,\tilde{c_L}) &\equiv \mu_s\frac{\partial \tilde{c_l}}{\partial\mu_s}, \\
\gamma_{\phi}  (\tilde{c_2},\ldots,\tilde{c_L}) &\equiv \frac{1}{2}\frac{\mu_s}{Z_{\phi}}\frac{\partial\delta_{Z_\phi}}{\partial\mu_s},\label{eq:GammaFieldDef}
\end{align} 
$Z_\phi$ is the field strength for $\phi$ ($\phi=\sqrt{Z_\phi}\,\phi_{\text{ren}}$) and $\delta_{Z_\phi} \equiv Z_\phi-1$.
Note that, since $c_1$ is the only parameter with a non-vanishing energy dimension, $\gamma_{c_1}$, $\beta_l $ and $\gamma_\phi$ cannot depend on it -- they only depend on the dimensionless couplings $\tilde{c_2},\ldots,\tilde{c_L}$.

At this point we reiterate the fact that since there is no boost invariance in these theories, one can consider two independent scaling transformations: one for space coordinates and another for time coordinates, and therefore each quantity in this analysis, including the $n$-point function $G^{(n)}_\text{ren}$, has two respective dimensions associated with it. The $n$-point function is therefore required to be homogeneous under both of these scaling transformations independently. Put differently, $G^{(n)}_\text{ren}$ is required to be homogeneous under a Lifshitz scaling transformation for \emph{any value of the critical exponent $z$.}
The resulting homogeneity equation for the $n$-point correlation function under a general Lifshitz transformation takes the form:
\begin{equation}
\label{eq:HomogenEq}
\begin{split}
&\left( \mu_s\frac{\partial}{\partial\mu_s}+p_i\frac{\partial}{\partial p_i}+z\omega_i\frac{\partial}{\partial\omega_i}+ \Delta_z^{\text{lif}}(c_1) c_1 \frac{\partial}{\partial c_1} \right. \\ 
&\qquad\qquad\qquad\left. - n\Delta_z^{\text{lif}}(\phi) + (n-1)(d+z) \right) G^{(n)}_{\text{ren}}(p_i,\omega_i;\mu_s, c_1, \tilde{c_l})=0,
\end{split}
\end{equation}  
with $\Delta_z^{\text{lif}}(\tilde{c_l})$ ($\Delta_z^{\text{lif}}(\phi)$) the classical weighted Lifshitz scaling dimension of $\tilde{c_l}$ ($\phi$) for an arbitrary choice of the critical exponent $z$.  
Subtracting the Callan-Symanzik RG equation \eqref{eq:CallanSymmanzik1} from equation \eqref{eq:HomogenEq} we find:  
\begin{equation}
\begin{aligned}
\label{eq:RgMinusHomogen}
&\left(p_i\frac{\partial}{\partial p_i}+z\omega_i\frac{\partial}{\partial\omega_i}
+ \left( \Delta_z^\text{lif}(c_1) - \gamma_{c_1} \right) c_1 \frac{\partial}{\partial c_1}
- \sum_{l=2}^L \beta_l\frac{\partial}{\partial \tilde{c_l}} \right. \\
&\qquad\qquad\qquad \left. 
 - n\left(\Delta_z^{\text{lif}}(\phi) + \gamma_\phi\right)
 + (n-1)(d+z)\right)
 G^{(n)}_{\text{ren}}(p_i,\omega_i;\mu_s, c_1, \tilde{c_l})=0,
\end{aligned}
\end{equation}
again for any value of $z$.

Now, suppose that for specific values of the dimensionless couplings $\tilde{c_l} = \tilde{c_l}^{\text{FP}}$ the beta functions all vanish, i.e.
\begin{equation}
\beta_l \left(\tilde{c_2}^{\text{FP}},\ldots,\tilde{c_L}^{\text{FP}}\right) = 0, \quad 2\leq l \leq L.
\end{equation}
Then at this point in parameter space, we have:
\begin{equation}
\begin{aligned}
\label{eq:RgMinusHomogenAtFixedPoint}
&\left(p_i\frac{\partial}{\partial p_i}+z\omega_i\frac{\partial}{\partial\omega_i}
+ \left( \Delta_z^\text{lif}(c_1) - \gamma_{c_1}^{\text{FP}} \right) c_1 \frac{\partial}{\partial c_1} \right. \\
&\qquad\qquad\qquad \left. 
 - n\left(\Delta_z^{\text{lif}}(\phi) + \gamma_\phi^{\text{FP}} \right)
 + (n-1)(d+z)\right)
 G^{(n)}_{\text{ren}}\left(p_i,\omega_i;\mu_s, c_1, \tilde{c_l}^{\text{FP}}\right)=0,
\end{aligned}
\end{equation}
where $\gamma_{c_1}^{\text{FP}} \equiv \gamma_{c_1}\left(\tilde{c_2}^{\text{FP}},\ldots,\tilde{c_L}^{\text{FP}}\right)$ and $\gamma_\phi^{\text{FP}} \equiv \gamma_\phi\left(\tilde{c_2}^{\text{FP}},\ldots,\tilde{c_L}^{\text{FP}}\right)$. Since \eqref{eq:RgMinusHomogenAtFixedPoint} is true for any choice of $z$, we may choose $z=z^{\text{FP}}$ such that $\Delta_{z^\text{FP}}^\text{lif}(c_1) = \gamma_{c_1}^\text{FP} $, that is:
\begin{equation}
\label{eq:GeneralzAtFixedPoint}
z^\text{FP} = \frac{\gamma_{c_1}^\text{FP} - \Delta_s(c_1)}{\Delta_t(c_1)}.
\end{equation}
For this value of $z$, equation \eqref{eq:RgMinusHomogenAtFixedPoint} takes the form:
\begin{equation}
\begin{aligned}
&\left(p_i\frac{\partial}{\partial p_i}+z^\text{FP} \omega_i\frac{\partial}{\partial\omega_i}
 - n\left(\Delta_{z^\text{FP}}^{\text{lif}}(\phi) + \gamma_\phi^{\text{FP}} \right) \right. \\
&\qquad\qquad\qquad\qquad\qquad \left. 
 + (n-1)(d+z^\text{FP})\right)
 G^{(n)}_{\text{ren}}\left(p_i,\omega_i;\mu_s, c_1, \tilde{c_l}^{\text{FP}}\right)=0.
\end{aligned}
\end{equation}
We therefore conclude that this point in parameter space represents a Lifshitz fixed point with an associated dynamical critical exponent given by $z^\text{FP}$ (which depends on $\gamma_{c_1}^\text{FP}$). The field $\phi$ has a Lifshitz scaling dimension of $ \Delta_{z^\text{FP}}^{\text{lif}}(\phi) + \gamma_\phi^{\text{FP}} $ at this fixed point.

As an example, consider the family of models discussed in subsection \ref{sec:TheHolomorphicModel}. A Lifshitz fixed point will appear at a point in parameter space in which the beta functions for all dimensionless parameters $ \tilde{c_n} = \lambda_n \mu_s^{-\frac{d-2}{2}(n^*-n)} $ vanish. Equation \eqref{eq:GeneralzAtFixedPoint} then implies the following relation between the value of the dynamical critical exponent associated with that fixed point and the anomalous dimension of $g$ at the fixed point:
\begin{equation}
\label{eq:GammaAtFixedPointRelation}
z^\text{FP} = 2 + \gamma_g^\text{FP}.
\end{equation}
Note that $z^\text{FP} \geq 2 $ as long as $\gamma_g^\text{FP} \geq 0$.

\subsection{Dual Scale RG Flows}
\label{subsec:DualScaleRGFlows}

As explained in subsection \ref{subsec:Renormalization}, an alternative approach to the standard, single-scale renormalization of non-boost-invariant field theories is the use of a dual-scale renormalization scheme, utilizing separate scales for momentum ($\mu_s$) and for energy ($\mu_t$). This type of renormalization scheme can prove useful as a tool for analyzing theories flowing between fixed points with different values of the dynamical critical exponent $z$, as it explicitly allows for changing the energy and momentum scales independently, without presupposing a specific relation between them.\footnote{For example, when the physical dispersion relation is unknown, one may consider off-shell renormalization conditions for correlation functions with external propagators having independent values for the momentum and the energy.} It is also a natural fit for regularization methods which treat space and time on an equal footing, such as split dimensional regularization (an example is given in subsection \ref{sec:MarginalCases}). In this subsection we digress to analyze some properties of this dual-scale formalism, and the way RG fixed points are described by it. While the results of this discussion are used for some calculations in later subsections, it is not required for following the rest of this section, and the reader may safely proceed directly to subsection \ref{subsec:NonRenormalizationTheoremsAGeneralProof}. 

We again consider a non-boost-invariant field theory in $d+1$ dimensions, with an action containing a set of parameters $c_l,\,(l=1,\ldots,L)$. We further suppose this theory is renormalized using a dual-scale renormalization scheme, introducing $\mu_s$ as the spatial (momentum) scale and $\mu_t$ as the temporal (energy) scale. We define dimensionless versions of the parameters $\tilde{c_l}$ using these scales, as follows:
\begin{equation}
\tilde{c_l} \equiv c_l\, \mu_t^{-\Delta_t(c_l)} \mu_s^{-\Delta_s(c_l)}.
\end{equation}
For example, for the $\mathcal{N}=2$ supersymmetric family of models discussed in subsection \ref{sec:TheHolomorphicModel}, we may choose $\tilde{c_1} = g \mu_t^{-1} \mu_s^2 $ and $\tilde{c_n} = f_n g^{-\frac{n}{2}} \mu_s^{-\frac{d-2}{2}(n^*-n)} = \lambda_n \mu_s^{-\frac{d-2}{2}(n^*-n)} $ for $2\leq n \leq n^*$.

Given some initial conditions, a dual-scale RG flow for these initial conditions corresponds to a mapping:
\begin{equation}
\mathbb{R}^2 \to \hat{M} \equiv M \times \mathbb{R},
\end{equation}
of the form $ \left(\tilde{c_l}(\mu_s,\mu_t),\ln Z_\phi(\mu_s,\mu_t)  \right) $, where $M$ is the manifold of renormalizable actions parameterized by $\tilde{c_l}$, and $Z_\phi$ is the field strength for the field $\phi$.\footnote{Here we are considering for simplicity the case of a single field $\phi$, but a generalization to any number of fields is straightforward.} The renormalization group action therefore induces a (possibly singular) foliation on the manifold $M$, with leaves of dimension 2 or less. This RG flow may be described by two sets of beta and anomalous dimension functions, defined as follows:
\begin{align}
& \beta^s_l(\tilde{c_k}) \equiv \mu_s \frac{\partial\tilde{c_l}}{\partial\mu_s}, \qquad
\beta^t_l(\tilde{c_k}) \equiv \mu_t \frac{\partial\tilde{c_l}}{\partial\mu_t}, \\
& \gamma^s_\phi(\tilde{c_k}) \equiv \frac{1}{2}\mu_s \frac{\partial\ln Z_\phi}{\partial\mu_s}, \qquad
\gamma^t_\phi(\tilde{c_k}) \equiv \frac{1}{2}\mu_t \frac{\partial\ln Z_\phi}{\partial\mu_t}.
\end{align}
These functions in turn define two vector fields $\hat{\beta}^s, \hat{\beta}^t \in T\hat{M}$ given by:
\begin{equation}
\hat{\beta}^s \equiv \beta^s + 2\gamma^s_\phi Z_\phi \frac{\partial}{\partial Z_\phi}, \qquad
\hat{\beta}^t \equiv \beta^t + 2\gamma^t_\phi Z_\phi \frac{\partial}{\partial Z_\phi},
\end{equation}
with $\beta^s, \beta^t \in TM$ defined as:
\begin{equation}
\beta^s \equiv \sum_{l=1}^L \beta^s_l \frac{\partial}{\partial \tilde{c_l}}, \qquad
\beta^t \equiv \sum_{l=1}^L \beta^t_l \frac{\partial}{\partial \tilde{c_l}}.
\end{equation}

Note that $L_\beta \equiv \operatorname{Span}(\beta^s,\beta^t)$ defines a generalized distribution on $M$. At generic points, this distribution would be two dimensional, but there may be singular points in which $\beta^s$ and $ \beta^t$ become colinear and $L_\beta$ becomes one dimensional.\footnote{Strictly speaking one could also find points with $\beta^s=\beta^t=0$, at which $L_\beta$ is 0-dimensional. These represent more exotic fixed points with independent space and time scale symmetries. We will not consider these cases here.} As will be explained in this subsection, these singular points correspond to RG fixed points in this description. 

From the definition of the RG flow functions, it is clear that $\beta^s, \beta^t$ (and more generally $\hat{\beta}^s, \hat{\beta}^t$) are not arbitrary vector fields. Indeed they must satisfy a constraint: since the distribution $\hat{L}_\beta \equiv \operatorname{Span}(\hat{\beta}^s,\hat{\beta}^t) $ induces a foliation on $\hat{M}$, it must be integrable. Furthermore, since $\hat{\beta}^s,\hat{\beta}^t$ correspond to the coordinate system $\mu_s, \mu_t$ over each leaf of the foliation, they must commute. Put differently, as one flows along a closed curve on the $(\mu_s,\mu_t)$ plane and returns to the initial point, one expects to return to the same physical values of parameters. This translates to the following constraint on these vector fields:
\begin{equation}
\hat{\calL}_{\hat{\beta}^s} \hat{\beta}^t = 0,
\end{equation}
where $\hat{\calL}$ is the Lie derivative on $\hat{M}$. Expressed in terms of the RG functions on $M$, this implies the following two constraints:
\begin{align}
\label{eq:DualScaleRGIntegrabilityCond1}
& \calL_{\beta^s} \beta^t = 0, \\
\label{eq:DualScaleRGIntegrabilityCond2}
& \calL_{\beta^s} \gamma^t_\phi - \calL_{\beta^t} \gamma^s_\phi = 0,
\end{align}
where $\calL$ is the Lie derivative on $M$, and $\gamma^s_\phi, \gamma^t_\phi$ are considered here as scalar functions on $M$.

Consider a renormalized n-point function $G^{(n)}_\text{ren}\left(p_i,\omega_i;\mu_s,\mu_t,\tilde{c_l} \right) $ for the field $\phi$. In the dual-scale description, two Callan-Symanzik equations may be written for $G^{(n)}_\text{ren}$ corresponding to each of the two scales:\footnote{Note that, due to the Frobenius theorem, the constraints \eqref{eq:DualScaleRGIntegrabilityCond1}-\eqref{eq:DualScaleRGIntegrabilityCond2} are necessary and sufficient for this system of equations to be integrable.}
\begin{align}
\label{eq:DualScaleRGCSEq1}
& \left( \mu_s \frac{\partial}{\partial\mu_s} + \sum_{l=1}^L \beta^s_l \frac{\partial}{\partial \tilde{c_l}} + n\gamma^s_\phi \right) G^{(n)}_\text{ren}\left(p_i,\omega_i;\mu_s,\mu_t,\tilde{c_l} \right) = 0, \\
\label{eq:DualScaleRGCSEq2}
& \left( \mu_t \frac{\partial}{\partial\mu_t} + \sum_{l=1}^L \beta^t_l \frac{\partial}{\partial \tilde{c_l}} + n\gamma^t_\phi \right) G^{(n)}_\text{ren}\left(p_i,\omega_i;\mu_s,\mu_t,\tilde{c_l} \right) = 0.
\end{align} 
On the other hand, as in the single scale case (see subsection \ref{subsec:Renormalization}), $G^{(n)}_\text{ren}$ is required to be homogeneous under space and time scaling transformations independently. Thus we have the following homogeneity equations:
\begin{align}
\label{eq:DualScaleRGHomEq1}
& \left( \mu_s \frac{\partial}{\partial\mu_s} + p_i \frac{\partial}{\partial p_i} - n \Delta_s(\phi) + (n-1)d \right) G^{(n)}_\text{ren}\left(p_i,\omega_i;\mu_s,\mu_t,\tilde{c_l} \right) = 0, \\ 
\label{eq:DualScaleRGHomEq2}
& \left( \mu_t \frac{\partial}{\partial\mu_t} + \omega_i \frac{\partial}{\partial \omega_i} - n \Delta_t(\phi) + (n-1) \right) G^{(n)}_\text{ren}\left(p_i,\omega_i;\mu_s,\mu_t,\tilde{c_l} \right) = 0.
\end{align}
Subtracting equations \eqref{eq:DualScaleRGCSEq1}-\eqref{eq:DualScaleRGCSEq2} from equations \eqref{eq:DualScaleRGHomEq1}-\eqref{eq:DualScaleRGHomEq2} respectively and taking a linear combination of the resulting equations, we obtain:
\begin{equation}
\label{eq:DualScaleRGHomMinusCSEq}
\begin{split}
&\left( p_i \frac{\partial}{\partial p_i} + z\omega_i \frac{\partial}{\partial \omega_i} - \sum_{l=1}^L (\beta^s_l + z \beta^t_l) \frac{\partial}{\partial \tilde{c_l}} \right. \\
&\qquad\qquad\qquad 
\left. - n \left(\Delta_z^\text{lif}(\phi) + \gamma^z_\phi\right) + (n-1)(d+z) \right) G^{(n)}_\text{ren}\left(p_i,\omega_i;\mu_s,\mu_t,\tilde{c_l} \right) = 0,
\end{split}
\end{equation}
where $z$ is arbitrary (that is, this equation is satisfied for any value of $z$), and $\gamma^z_\phi \equiv z \gamma^t_\phi + \gamma^s_\phi $.

Suppose that for some point $\tilde{c}^\text{FP} \in M$ and some value $z^\text{FP}$ the RG flow functions satisfy:
\begin{equation}
\label{eq:DualScaleRGFixedPointCondition}
z^\text{FP} \beta^t(\tilde{c}^\text{FP}) + \beta^s(\tilde{c}^\text{FP}) = 0.
\end{equation}   
Then choosing $z=z^\text{FP}$ at this point, equation \eqref{eq:DualScaleRGHomMinusCSEq} takes the form:
\begin{equation}
\begin{split}
&\left( p_i \frac{\partial}{\partial p_i} + z\omega_i \frac{\partial}{\partial \omega_i} - n \left(\Delta_{z^\text{FP}}^\text{lif}(\phi) + \gamma^\text{FP}_\phi \right) \right. \\
&\qquad\qquad\qquad 
\left.  + (n-1)(d+z^\text{FP}) \right) G^{(n)}_\text{ren}\left(p_i,\omega_i;\mu_s,\mu_t,\tilde{c_l}^\text{FP} \right) = 0,
\end{split}
\end{equation}
where $ \gamma^\text{FP}_\phi \equiv \gamma^{z^\text{FP}}_\phi (\tilde{c}^\text{FP}) $.
This implies that $\tilde{c}^\text{FP}$ represents a Lifshitz fixed point with an associated dynamical critical exponent of $z^\text{FP}$, and the field $\phi$ has a Lifshitz scaling dimension of $\Delta_{z^\text{FP}}^\text{lif}(\phi) + \gamma^\text{FP}_\phi$ at this fixed point. 

However, there is an additional subtlety that arises in the dual-scale description. Recall that in this description, the full orbit of the point $\tilde{c}^\text{FP}$ under the RG flow is given by the individual scaling of space and time, and not just by the specific Lifshitz scaling corresponding to $z=z^\text{FP}$. As one does not expect $\beta^s$ and $\beta^t$ to individually vanish at $\tilde{c}^\text{FP}$, this point is clearly not a fixed point of the full RG action. In other words, since $\mu_s$ and $\mu_t$ are individually arbitrary renormalization scales, one is free to change one without changing the other, and the physics should not change (in particular, the system should still be at a Lifshitz fixed point). One is therefore compelled to identify the physical ``fixed point'' with the entire orbit of the point $\tilde{c}^\text{FP}$ in $M$. This naturally raises the question of whether the condition \eqref{eq:DualScaleRGFixedPointCondition} is satisfied over the
entire orbit (with the same value of $z^\text{FP}$), and whether the anomalous dimension $ \gamma^{z^\text{FP}}_\phi (\tilde{c}) $ remains constant over this orbit, as one would expect from physical considerations. Indeed, one can show these properties follow trivially from the constraints \eqref{eq:DualScaleRGIntegrabilityCond1}-\eqref{eq:DualScaleRGIntegrabilityCond2} assumed earlier.

To see this, let $R^\text{FP} \subset M$ be the orbit of the point $\tilde{c}^\text{FP}$. We would like to show that for any point $\tilde{c} \in R^\text{FP}$ the following two conditions are satisfied:
\begin{align}
\label{eq:DualScaleRGFixedPointProp1}
& z^\text{FP} \beta^t (\tilde{c}) + \beta^s(\tilde{c}) = 0, \\
\label{eq:DualScaleRGFixedPointProp2}
& \gamma^{z^\text{FP}}_\phi (\tilde{c}) = \gamma^{z^\text{FP}}_\phi (\tilde{c}^\text{FP}) = \text{const}.
\end{align}
To show property \eqref{eq:DualScaleRGFixedPointProp1}, define a coordinate system $(\tau_1,\ldots,\tau_L)$ in some neighborhood of $\tilde{c}^\text{FP}$ such that $\tilde{c}(0,\ldots,0)=\tilde{c}^\text{FP}$ and $\frac{\partial}{\partial\tau_1} = \beta^s$. Then due to condition \eqref{eq:DualScaleRGFixedPointCondition}, $\beta^t (\tilde{c}^\text{FP}) = (-1/z^\text{FP},0,\ldots,0)$ in this coordinate system. However the constraint \eqref{eq:DualScaleRGIntegrabilityCond1} implies that the components of $\beta^t$ do not depend on $\tau_1$ and therefore $\beta^t \left(\tilde{c}(\tau_1,0,\ldots,0)\right) = (-1/z^\text{FP},0,\ldots,0)$ for any $\tau_1$. That is, \eqref{eq:DualScaleRGFixedPointProp1} is satisfied on the one-dimensional orbit of $\tilde{c}^\text{FP}$ generated by $\beta^s$, which in turn implies that $\beta^t$ generates the same orbit, and it is in fact the full (one-dimensional) leaf induced by the RG flow that contains $\tilde{c}^\text{FP}$. By using property \eqref{eq:DualScaleRGFixedPointProp1} in \eqref{eq:DualScaleRGIntegrabilityCond2} one then obtains $ \calL_{\beta^s}\gamma^{z^\text{FP}}_\phi =  \calL_{\beta^t}\gamma^{z^\text{FP}}_\phi = 0 $ on $R^\text{FP}$, and property \eqref{eq:DualScaleRGFixedPointProp2} follows.

It is important to note, however, that while $z^\text{FP}$ and $ \gamma^{z^\text{FP}}_\phi$ are both constant over the leaf $R^\text{FP}$ corresponding to the fixed point, $\gamma^s_\phi$ and $\gamma^t_\phi$  may not be, and in fact these quantities are renormalization scheme dependent even at the fixed point:

The definitions and assumptions above are covariant with respect to diffeomorphisms of $M$, which correspond to renormalization scheme changes that can be described as redefinition of the parameters $\tilde{c_l}$. It is immediately clear, then, that the properties \eqref{eq:DualScaleRGFixedPointProp1}-\eqref{eq:DualScaleRGFixedPointProp2} are invariant under any diffeomorphism of $M$ that preserves the foliation induced by the RG action. In fact, if a scheme exists in which $\gamma^s_\phi, \gamma^t_\phi$ are constant over $R^\text{FP}$, then they are clearly unchanged under these kinds of scheme changes. However, one can instead consider a larger family of renormalization scheme changes -- those that involve in addition a linear redefinition of the field $\phi$ of the form:
\begin{equation}
\begin{split}
& \phi' = \phi\, h^Z(\tilde{c_k}), \\
& \tilde{c_l}' = h_l(\tilde{c_k}), 
\end{split}
\end{equation}
where $h_l(\tilde{c_k})$ represents a foliation preserving diffeomorphism on $M$. In their infinitesimal form, these are diffeomorphisms of $\hat{M}$ generated by a vector field $\hat{\xi} \in T\hat{M}$ of the form:
\begin{equation}
\label{eq:DualScaleRGDiffVec}
\hat{\xi} \equiv \xi + 2 \xi^Z(\tilde{c}) Z_\phi \frac{\partial}{\partial Z_\phi}, 
\end{equation}
where $\xi \in TM$ is a linear combination of $\beta^s, \beta^t$:
\begin{equation}
\xi \equiv \xi^s(\tilde{c}) \beta^s + \xi^t(\tilde{c}) \beta^t.
\end{equation}
Under this family of diffeomorphisms, the RG flow functions transform as follows:
\begin{align}
& \delta\beta^s = \calL_\xi \beta^s = - (\calL_{\beta^s} \xi^s) \beta^s - (\calL_{\beta^s} \xi^t) \beta^t, \\
& \delta\beta^t = \calL_\xi \beta^t = - (\calL_{\beta^t} \xi^s) \beta^s - (\calL_{\beta^t} \xi^t) \beta^t, \\
& \delta \gamma^s_\phi = \calL_\xi \gamma^s_\phi - \calL_{\beta^s} \xi^Z = \xi^s \calL_{\beta^s} \gamma^s_\phi + \xi^t \calL_{\beta^t} \gamma^s_\phi - \calL_{\beta^s} \xi^Z, \\
& \delta \gamma^t_\phi = \calL_\xi \gamma^t_\phi - \calL_{\beta^t} \xi^Z = \xi^s \calL_{\beta^s} \gamma^t_\phi + \xi^t \calL_{\beta^t} \gamma^t_\phi - \calL_{\beta^t} \xi^Z.
\end{align}
It is easy to check that at a point $\tilde{c} \in R^\text{FP}$, due to properties \eqref{eq:DualScaleRGFixedPointProp1}-\eqref{eq:DualScaleRGFixedPointProp2}, indeed:
\begin{align}
& \delta( z^\text{FP} \beta^t + \beta^s ) = 0, \\
& \delta \gamma^{z^\text{FP}}_\phi  = 0. 
\end{align}
That is, $z^\text{FP}$ and $\gamma^{z^\text{FP}}_\phi$ remain unchanged under such a renormalization scheme change, as one would expect. However, $\delta\gamma^s_\phi$ and $\delta\gamma^t_\phi$ do not vanish separately, even if $\gamma^s_\phi$ and $\gamma^t_\phi$ are separately constant on $R^\text{FP}$. In fact, with an appropriate choice of $\xi^Z$ one may freely change one of them (as the combination $\gamma^{z^\text{FP}}_\phi$ remains fixed). We therefore observe that while for a given fixed point of the dual-scale RG flow $z^\text{FP}$ and $\gamma^{z^\text{FP}}_\phi$ are physical, scheme independent quantities, $\gamma^s_\phi$ and $\gamma^t_\phi$ individually are not. 

In subsection \ref{sec:MarginalCases} we make use of these properties to extract the values of $z^\text{FP}$ and $\gamma^{z^\text{FP}}_\phi$ for the fixed points realized by the marginal cases of the models introduced in subsection \ref{sec:TheHolomorphicModel}.

\subsection{Non-Renormalization Theorem: A General Proof}
\label{subsec:NonRenormalizationTheoremsAGeneralProof}

In this subsection we introduce and prove a non-renormalization theorem for the Lifshitz supersymmetric family of models defined in subsection \ref{sec:TheHolomorphicModel}.

Similarly to the relativistic case (see \cite{Seiberg:1993vc}), one can make a general argument for the non-renormalization of the superpotential in these models, based on its holomorphicity and the symmetries of the theory. Suppose we start with a classical superpotential of the general form:\footnote{For simplicity we assume here a single holomorphic superfield $\Phi$ and a (classical) superpotential with no more than two spatial derivatives, that is with a classical value of $z_{UV}=2$ for the dynamical critical exponent at the UV, in agreement with the previous assumptions in subsection \ref{sec:TheHolomorphicModel}. The following arguments could easily be extended to more general cases as well.}
\begin{equation}
W_\text{tree}(\Phi) = G(\Phi) \partial_i \Phi \partial_i \Phi + F(\Phi),
\end{equation} 
where
$G(\phi)$ and $F(\Phi)$ are holomorphic functions of $\Phi$ with the following expansions: 
\begin{align}
G(\Phi) &= - \sum_{k=0}^{\infty} \frac{1}{k+2} g_{k+2} \Phi^k, \label{eq:GgeneralForm}\\
F(\Phi) &= \sum_{k=2}^{\infty} \frac{1}{k} f_k \Phi^k.\label{eq:FGeneralForm}
\end{align}
The coefficients $g_2, f_2$ correspond to the free part of the superpotential, whereas $g_k, f_k$ for $k \geq 3$ correspond to interactions. Note that, unlike most of this work, we assume here the more general form \eqref{eq:N2GeneralSuperpotnetialWithDerivativeInteractions} for the superpotential, which allows for $d=2$ spatial dimensions as well (for $d=2$, the Lifshitz scaling dimension of the superfield vanishes, and the superpotential may generally contain an infinite number of classically relevant and marginal terms).

As in the relativistic case, we make the following assumptions:
\begin{enumerate}
\item Supersymmetry, and any other relevant global symmetries, are non-anomalous and remain unbroken by quantum corrections,
\item The system is smooth in the weak coupling limit, i.e.\ in the limit $ g_k, f_k \to 0$ for all $k\geq 3$. 
\end{enumerate}
Additionally, we assume that the IR physics of the system can be faithfully described by the microscopic degrees of freedom. It is important to note that the fulfillment of these assumptions is less trivial here than in the analogous relativistic ($3+1$)-dimensional Wess-Zumino model: Whereas the latter model is always IR free, the systems studied here may flow to a finite or strong coupling in the IR (see subsections \ref{sec:MarginalCases}-\ref{subsec:GaplessSingularCase}), and one may have to account for non-perturbative effects and their implications on these assumptions. For instance, as mentioned in section \ref{sec:TimeDomainNRSUSY}, in some cases these systems may have soliton-like semiclassical vacua with a finite tunneling amplitude to the trivial vacuum, and they may change the IR physics. For further details, see the discussion in section \ref{sec:Summary}. 

We consider the \emph{Wilsonian} effective action of the theory associated to some momentum scale $\mu_s$, energy scale $\mu_t$, or both (if one uses a dual-scale renormalization scheme, see discussion in subsection \ref{subsec:Renormalization}). We define this to be the effective action obtained by integrating out a region in momentum and energy space associated with these scales, which does not include any IR singularities of the propagators.\footnote{That is, regions in energy and momentum space which do not include the points $(\omega=0,k=\tilde{k}_0)$ such that $\omega(\tilde{k}_0) = 0$ (where $\omega(k)$ is the single particle dispersion relation).} For the gapped cases (with $ \operatorname{Im}(f_2) \neq 0 $ or $ \operatorname{Re}(f_2)>0$) or the case of $f_2=0$, this corresponds, to integrating out momenta with $k>\mu_s$ (or energies $|\omega| > \mu_t$), similarly to the relativistic case. For the gapless singular case, with a real and positive $f_2$, this requires the integrated-out region to exclude the singular sphere of momenta -- one can choose, for example, to integrate out momenta with $|k-\tilde{k}_0|> \mu_s $ (see discussion in subsection \ref{subsec:GaplessSingularCase}). Unlike the 1PI effective action, the Wilsonian effective action does not suffer from IR divergences as one approaches the gapless limit. For simplicity, for most of this subsection we assume the gapped or $f_2=0$ cases, and return to discuss the gapless singular case in the end.

Due to the assumption that supersymmetry is preserved by the full quantum theory, the effective action will take the general form:
\begin{equation}
L_\text{eff}=\int d^2\theta d^2\theta^\dagger \, K_\text{eff} \left(\Phi,\Phi^\dagger \right) +\int d^2 \theta \, W_\text{eff}\left(\Phi\right) + \text{c.c.} ,
\end{equation}
where $W_\text{eff}\left(\Phi,g_k,f_k\right)$ is a holomorphic functional of the superfield $\Phi$ and depends on the parameters $g_k, f_k$, and similarly $K_\text{eff}\left(\Phi,\Phi^\dagger,g_k,f_k\right)$ is a real functional of $\Phi,\Phi^\dagger$ and the parameters $g_k, f_k$.\footnote{$W_\text{eff}$ and $K_\text{eff}$ can also depend on the renormalization scale $\mu_s$ (or $\mu_t$) as well as UV cutoffs.}

Under the assumptions outlined above, we aim  to show that the effective superpotential $W_\text{eff}(\Phi,g_k,f_k)$ is equal to the classical one $W_\text{tree}$.

As in the relativistic case (see \cite{Seiberg:1993vc}), let us regard the coupling constants $g_k, f_k$ as background superfields. The classical action is then seen to be invariant under global  $U(1) \times U(1)_R$ symmetries, by assigning the following charges to the fields, the superspace coordinates $\theta$ and the superpotential $W$: 
\begin{center}
\begin{tabular}{ c c c }
 & $U(1)$ & $U(1)_R$ \\
 \hline
$\Phi$ & $-1$ & $0$ \\
$ g_k, f_k $ & $k$ & $2$ \\
$\theta$ & $0$ & $-1$ \\
$W$ & $0$ & $2$ 
\end{tabular}
\end{center}

As the parameters $g_k, f_k$ are regarded as background superfields, $W_\text{eff}$ has to be a holomorphic functional of both them and $\Phi$. This holomorphic property and the second assumption above mean that $W_\text{eff}$ can be expanded in non-negative powers of $\Phi$ and its derivatives, as well as the coupling constants $ g_k, f_k $ for $ k \geq 3$ (this also rules out any non-perturbative contributions to $W_\text{eff}$ in terms of these coupling constants -- see section \ref{sec:Summary} for a discussion on non-perturbative considerations). Consider a term in this expansion of degree $n$ in $\Phi$, which has the general form:\footnote{A similar argument will be valid for terms that contain any number of derivatives of $\Phi$. }
\begin{equation}
h(g_2,f_2) \prod_{k=3}^\infty g_k^{l_k}\, f_k^{m_k} \, \Phi^n , 
\end{equation}
where $l_k, m_k \geq 0 $ for all $k \geq 3$, and $h(g_2,f_2)$ is a holomorphic function of $f_2,g_2$ that can also depend on the renormalization and UV cutoff scales. Requiring that $W_\text{eff}$ respects the global symmetries $U(1) \times U(1)_R$ of the original action, we conclude that $h$ must be a homogeneous function of degree $-p$ (that is, $h(\lambda g_2, \lambda f_2) = \lambda^{-p}\, h(g_2,f_2) $) such that the following two conditions are satisfied:
\begin{align}\label{eq:NonRenormCondition1}
&\sum_{k=3}^\infty k (l_k + m_k) - 2p - n = 0, \\
\label{eq:NonRenormCondition2}
&\sum_{k=3}^\infty 2 (l_k + m_k) - 2p = 2 .
\end{align}
Note that by subtracting the second condition from the first we obtain:
\begin{equation}
\sum_{k=3}^\infty (k-2)(l_k + m_k) = n-2 ,
\end{equation}
from which we immediately conclude that $l_k=m_k=0$ for $k>n$. In particular, for $n=2$ the coefficient of $\Phi^2$ does not depend on any of the coupling constants $g_k, f_k$ for $k\geq 3$. It therefore takes the form $ h(g_2,f_2) \Phi^2 $. Restricting to the free case ($g_k = f_k = 0$ for $k\geq 3$) and comparing to the classical action, it is clear that for this term $h(g_2,f_2)=\frac{1}{2} f_2$. Thus we establish non-renormalization for this term, and a similar argument is valid for any term with $n=2$ (such as $-\frac{1}{2} g_2\, \partial_i \Phi \partial_i \Phi $).

For $n\geq 3$, in the weak coupling limit, it is clear that this term corresponds to a Feynman diagram\footnote{In the context of this argument, Feynman diagrams refer to supergraph formalism, or alternatively to diagrams of the theory before integrating out the auxiliary field $F$, so that an interaction of the form $f_k \Phi^k$ always corresponds to a vertex of $k$ propagator lines.} with $n$ external lines, $l_k$ vertices of type $g_k$ and $m_k$ vertices of type $f_k$. If we denote by $I$ the number of internal lines in the diagram, we get from standard counting arguments:
\begin{equation}
\sum_{k=3}^\infty k(l_k+m_k) = n + 2I .
\end{equation}
Comparing with condition \eqref{eq:NonRenormCondition1} we see that $I=p$. Denoting by $V \equiv \sum_{k=3}^\infty (l_k + m_k) $ the total number of vertices in the diagram, condition \eqref{eq:NonRenormCondition2} then implies that $I = V-1$. This equality can only be satisfied in a tree-level diagram. However, the only tree-level diagrams that contribute to the effective action are the 1PI ones, with a single vertex and no internal lines, which correspond to terms of the form $h(g_2,f_2) g_n \Phi^n$ or  $h(g_2,f_2) f_n \Phi^n$. Finally, by comparing to the classical action in the weak coupling limit, the former is excluded, and $h(g_2,f_2)$ is determined to be $\frac{1}{n}$. We are therefore left with the non-renormalized term $\frac{1}{n} f_n \Phi^n$. A similar argument can be used to prove non-renormalization for terms with any number of derivatives of $\Phi$.

The gapless singular case (in which $f_2$ is real and positive) can be handled similarly to the above arguments, except that the effective action is defined by integrating out momenta which are far from the singular sphere in momentum space. It is therefore more convenient to write the effective action in momentum space.\footnote{Note that the parameter $\tilde{k}_0 = \sqrt{f_2/g_2}$ that corresponds to the radius of the singular sphere does not get renormalized itself along the RG flow due to the arguments here, and it is therefore consistent to consider its value to be a fixed parameter in the quantum theory equal to its classical value.} For a small enough value of $\mu_s$, the renormalized fields will be defined inside a shell around the singular sphere, given by the condition on the momenta $|k-\tilde{k}_0|<\mu_s$. A term of degree $n$ in the expansion of the effective superpotential will generally take the form:
\begin{equation}
\begin{split}
\prod_{k=3}^\infty g_k^{l_k}\, f_k^{m_k} \, &\int_{|p_i-\tilde{k}_0|<\mu_s} \frac{d^d p_1\ldots d^d p_n}{(2\pi)^{dn}} \\
&\qquad\qquad h(g_2,f_2;\vec{p}_1,\ldots,\vec{p}_n) \Phi(\vec{p}_1)\ldots\Phi(\vec{p}_n) (2\pi)^d \delta(\vec{p}_1+\ldots+\vec{p}_n) ,
\end{split}
\end{equation}
with $h$ homogeneous in $g_2, f_2$. For $n=2$, it is clear from the arguments above that there is no contribution from $g_k,f_k$ for $k\geq 3$. Then by restricting to the free case we have $h(g_2,f_2;\vec{p}_1,-\vec{p}_1)=-\frac{g_2}{2} p_1^2+\frac{f_2}{2}$ as in the tree level expression. For $n\geq 3$, the only contribution is again from the single vertex diagrams proportional to either $f_n$ or $g_n$, corresponding to the tree-level term with $h(g_2,f_2;\vec{p}_1,\ldots,\vec{p}_n) = \frac{1}{n}$ or $h(g_2,f_2;\vec{p}_1,\ldots,\vec{p}_n) =  \frac{1}{n} \vec{p_1} \cdot \vec{p}_2 $ respectively. 

\subsection{Perturbative Analysis}
\label{subsec:PerturbativeAnalysis}
In this subsection we study the perturbative behaviour of the family of models, and demonstrate some of its properties. In subsection \ref{sec:FeynmanRules} the Feynman rules for these models are given. These are used later on in subsections \ref{subsubsec:OneLoopExample4d} and \ref{sec:MarginalCases}. Subsection \ref{subsec:PerturbativeArgumentforNRTheorems}  briefly presents a general argument that shows that there are no perturbative quantum corrections to the holomorphic superpotential, thus supporting the general proof presented in subsection \ref{subsec:NonRenormalizationTheoremsAGeneralProof}. The perturbative argument is very similar to the relativistic one, which can be found in \cite{Wess:1992cp}. We refer to appendix \ref{app:DetailsPerturbativeAnalysis} for full technical details of this analysis. Subsection \ref{subsubsec:OneLoopExample4d} describes several interesting features of the model with an $n=3$ interaction of the form \eqref{eq:GeneralSuperInteraction} in $3+1$ dimensions, stemming from supersymmetry and the non-renormalization property of the model.

\subsubsection{Feynman Rules}
\label{sec:FeynmanRules}

The expressions for the bosonic and fermionic Feynman propagators may be easily derived from the action \eqref{eq:LagrangianFreeModels}, and are given by: 
\begin{equation}
\label{eq:ScalarProp}
\left\langle\phi(\omega,k)\phi^*(-\omega,-k)\right\rangle=\frac{i}{\omega^2-|gk^2-f_2|^2+i\epsilon},
\end{equation} 
and
\begin{align}
&\left\langle\psi_\alpha(\omega,k)\psi^\dagger_{\dot{\beta}}(-\omega,-k)\right\rangle=\frac{i\omega\sigma^0_{\alpha\dot{\beta}}}{\omega^2-|gk^2-f_2|^2+i\epsilon}, \label{eq:FermionicProp1}\\
&\left\langle\psi^{\dagger\dot{\alpha}}(\omega,k)\psi^{\beta}(-\omega,-k)\right\rangle=\frac{i\omega\bar{\sigma}^{0\,{\dot{\alpha}\beta}}}{\omega^2-|gk^2-f_2|^2+i\epsilon},\label{eq:FermionicProp2}\\
&\left\langle\psi_{\alpha}(\omega,k)\psi^{\beta}(-\omega,-k)\right\rangle=\frac{-i\delta_\alpha^\beta(gk^2-f_2^*)}{\omega^2-|gk^2-f_2|^2+i\epsilon},\\
&\left\langle\psi^{\dagger\dot{\alpha}}(\omega,k)\psi^\dagger_{\dot{\beta}}(-\omega,-k)\right\rangle=\frac{-i\delta^{\dot{\alpha}}_{\dot{\beta}}(gk^2-f_2)}{\omega^2-|gk^2-f_2|^2+i\epsilon}. \label{eq:FermionicProp4}
\end{align}

%%%%%%%%
\begin{figure}[!htb]
 \includegraphics[width=30mm]{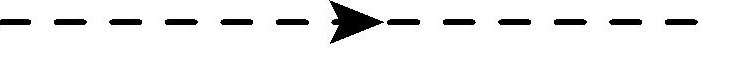}
\centering 
\caption{ Feynman diagram representation for the bosonic propagator, in correspondence with equation \eqref{eq:ScalarProp}.  \label{fig:propagator_s}}
\end{figure}
%%%%%%%%%%%%%%%%%%%%%%%%%%%%%%%
\begin{figure}[!htb]
        \centering
         \begin{subfigure}[b]{0.2\textwidth}
                \includegraphics[width=\textwidth]{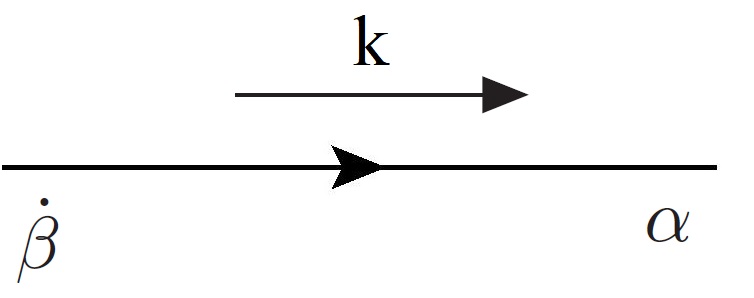}\vspace{0.1cm}
                \caption{$\frac{i\omega\sigma^0_{\alpha\dot{\beta}}}{\omega^2-|gk^2-f_2|^2+i\epsilon}$}
                \label{fig1:propagator_f1}
        \end{subfigure}
~\quad           
        \begin{subfigure}[b]{0.2\textwidth}
                \includegraphics[width=\textwidth]{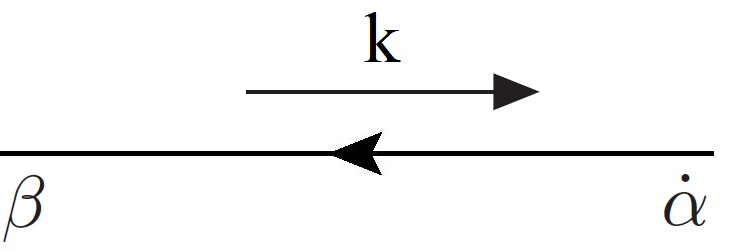}\vspace{0.3cm}
                \caption{$\frac{i\omega\bar{\sigma}^{0\,{\dot{\alpha}\beta}}}{\omega^2-|gk^2-f_2|^2+i\epsilon}$}
                \label{fig1:propagator_f2}
        \end{subfigure}
~\quad     
        \begin{subfigure}[b]{0.2\textwidth}
                \includegraphics[width=\textwidth]{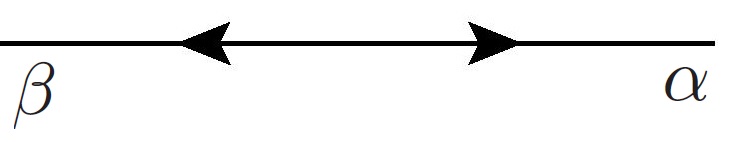}\vspace{0.2cm}
                \caption{$\frac{-i\delta_\alpha^\beta(gk^2-f_2^*)}{\omega^2-|gk^2-f_2|^2+i\epsilon},$}
                \label{fig1:propagator_f3}
        \end{subfigure}
        ~\quad     
        \begin{subfigure}[b]{0.2\textwidth}
                \includegraphics[width=\textwidth]{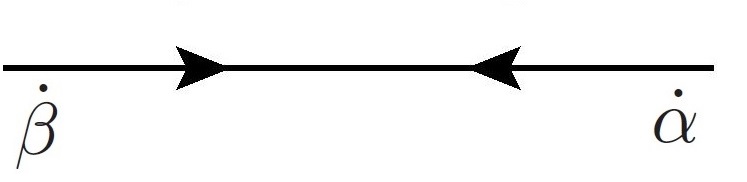}\vspace{0.1cm}
                \caption{$\frac{-i\delta_{\dot{\alpha}}^{\dot{\beta}}(gk^2-f_2)}{\omega^2-|gk^2-f_2|^2+i\epsilon}$}
                \label{fig1:propagator_f4}
        \end{subfigure}
                \caption{Feynman diagram representation for the fermionic propagators. These correspond to equations \eqref{eq:FermionicProp1} - \eqref{eq:FermionicProp4}. In these conventions, arrows on fermionic lines are always directed away from dotted indices or towards undotted indices at a certain vertex.
It should be noted that the choice between figures \ref{fig1:propagator_f1} and \ref{fig1:propagator_f2} is made in accordance with the index contraction order chosen for the corresponding fermionic line in the full diagram, and the direction in which the propagator appears in it. In the diagrams here the index order is assumed to be taken from right to left, matching equations \eqref{eq:FermionicProp1} and \eqref{eq:FermionicProp2}.
 }
        \label{fig1:propagatorsFermions}
\end{figure}
\begin{figure}[!htb]
   \centering
		\subcaptionbox{$-\frac{i\delta_\alpha^\beta f_n(n-1)}{2}$ or $-\frac{i\delta_\alpha^\beta f_n(n-1)}{2}$                 \label{fig2:Int1}}
		{\includegraphics[width=0.45\textwidth]{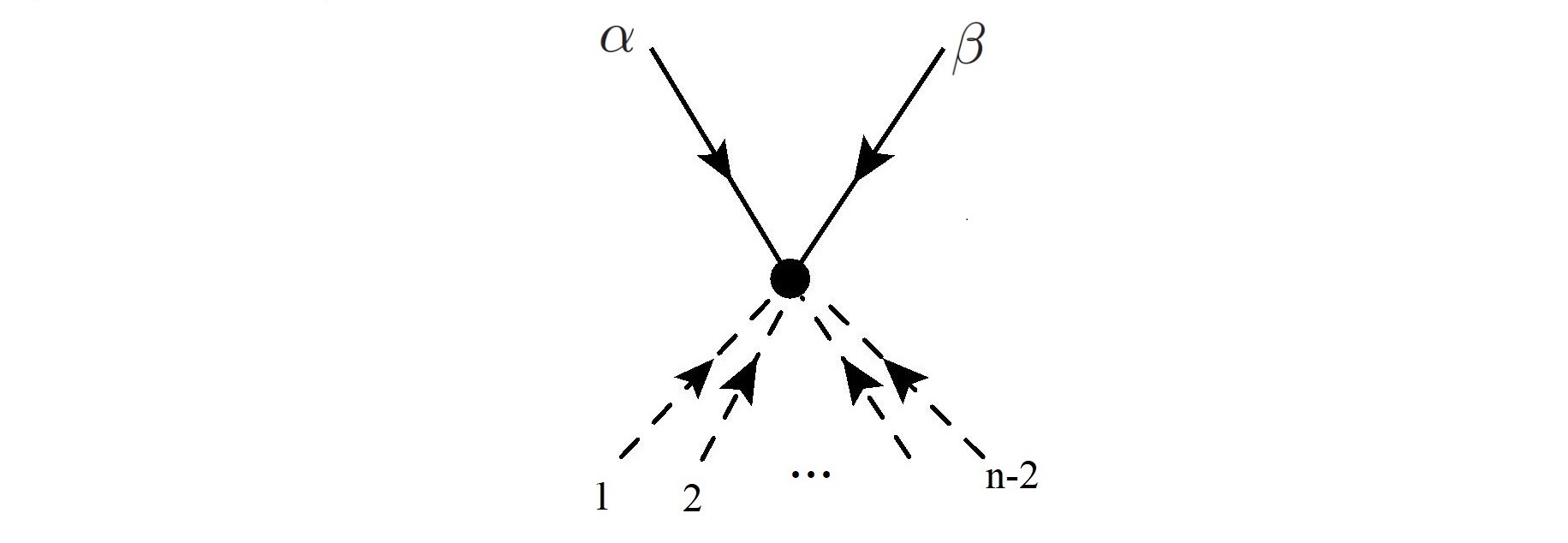}}
	\qquad \qquad
		\subcaptionbox{$-\frac{i\delta^{\dot{\alpha}}_{\dot{\beta}}f_n^*(n-1)}{2}$ or $-\frac{i\delta_{\dot{\alpha}}^{\dot{\beta}}f_n^*(n-1)}{2}$ \label{fig2:Int2Ver}}
		{\includegraphics[width=0.4\textwidth]{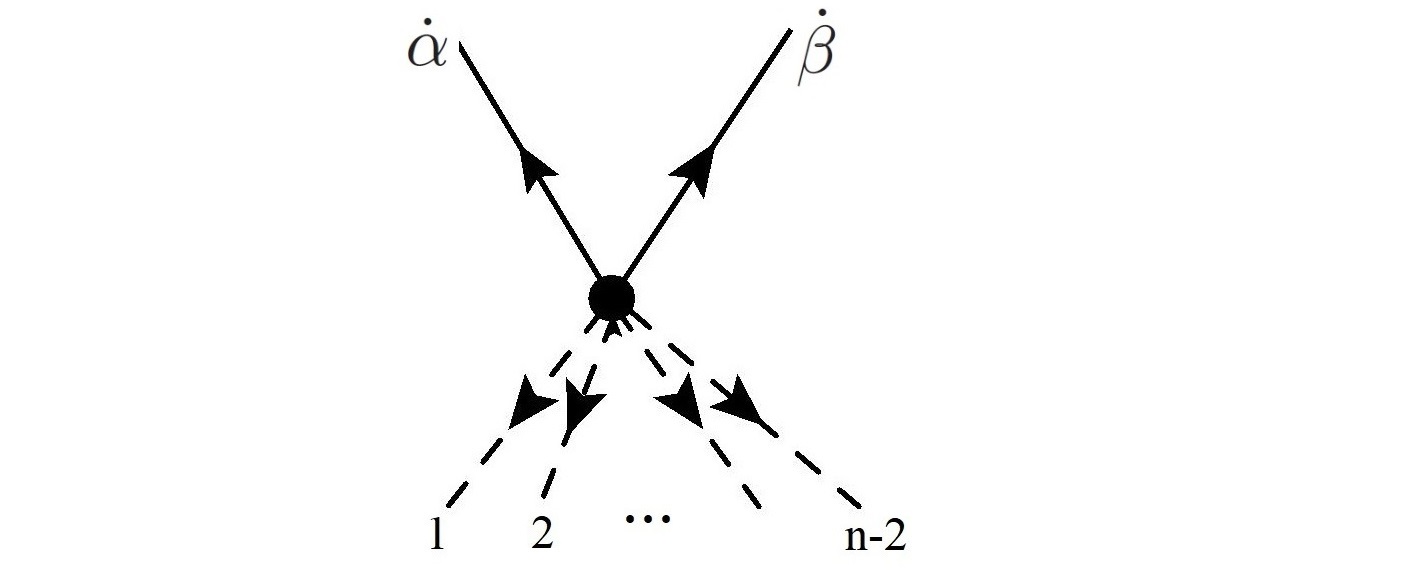}}\qquad \qquad  		\subcaptionbox{$-if_nf_m^* $                \label{fig2:Int3Ver}} {\includegraphics[width=0.19\textwidth]{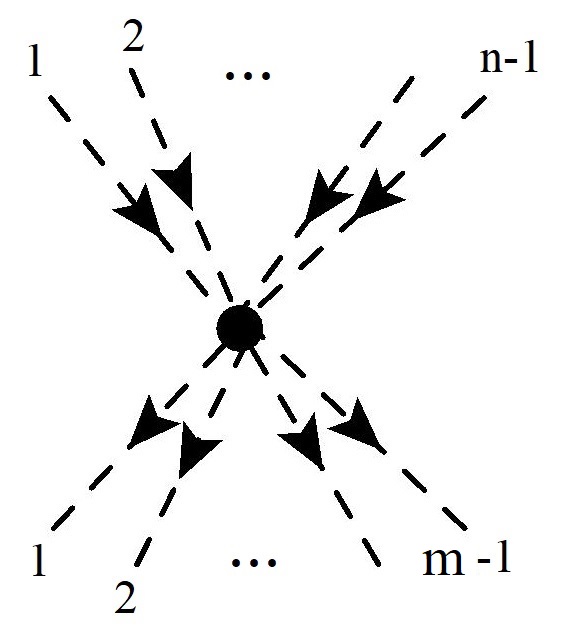}}
	\qquad \qquad               \subcaptionbox{$igf_nq^2-if_2^*f_n$ \label{fig2:Int6Ver}}
                {\includegraphics[width=0.19\textwidth]{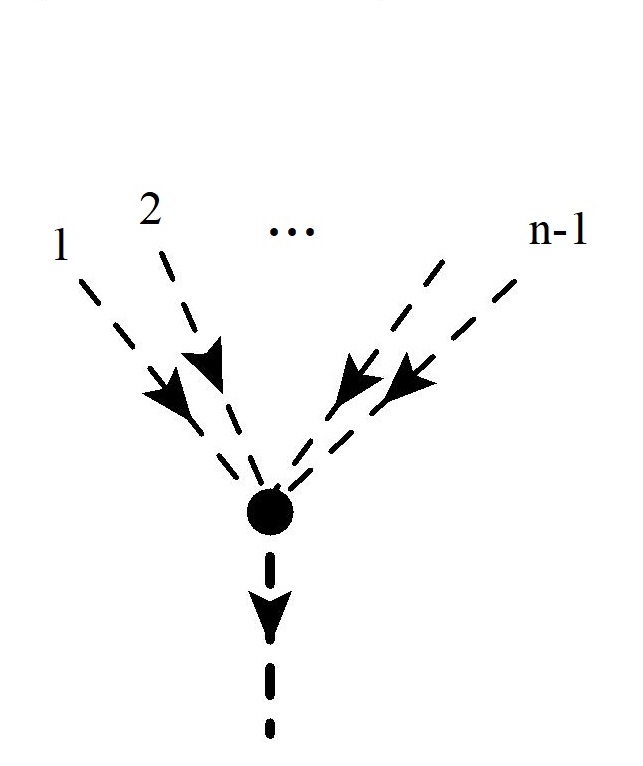}}	
\qquad \qquad                \subcaptionbox{$igf_n^*q^2-if_2f_n^*$\label{fig2:Int7Ver}}
                {\includegraphics[width=0.19\textwidth]{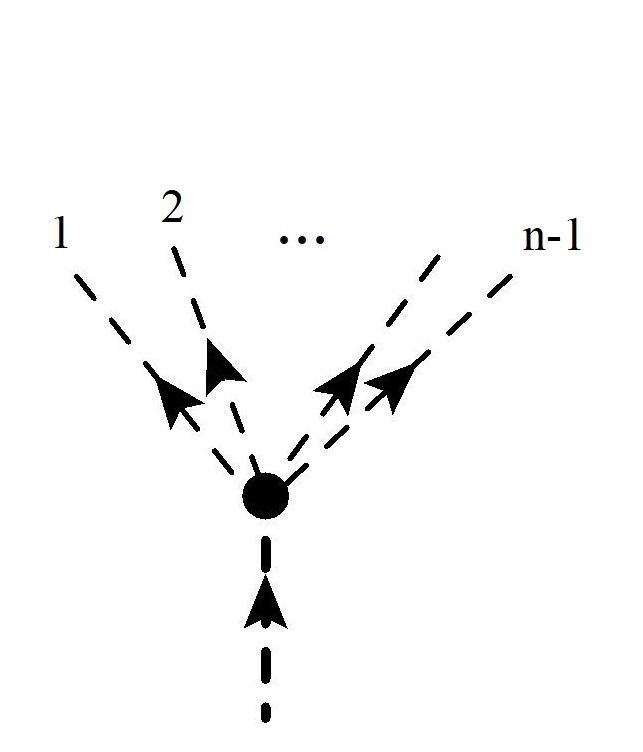}}
        \caption{Feynman rules for a general interaction of the form \eqref{eq:GeneralSuperInteraction}. A thick dashed line represents a boson with a spatial momentum $\vec{q}$ insertion. In figures \ref{fig2:Int1}, \ref{fig2:Int2Ver}, the choice of which rule to use depends on how the vertex connects to the rest of the diagram considered, and on the index contraction order chosen for the corresponding fermionic line.}
\label{fig2:FeynmanRulesVer}
\end{figure}

The visual representations of these propagators in terms of Feynman diagrams are given in figures \ref{fig:propagator_s} and \ref{fig1:propagatorsFermions} respectively. The conventions used here were inherited from those in \cite{Dreiner:2008tw}. 
The Feynman rules for vertices corresponding to a general interaction of the form \eqref{eq:GeneralSuperInteraction} are given in figure \ref{fig2:FeynmanRulesVer}.
Additionally, there is the usual symmetry factor taken into consideration when studying various diagrams, as well as a factor of $-1$ for every closed fermionic loop.

%%%%%%%%%%%%%%%%%%%%%%

\subsubsection{A Perturbative Argument for the Non-Renormalization Theorem}
\label{subsec:PerturbativeArgumentforNRTheorems}

We now present a perturbative argument for the non-renormalization theorem of subsection \ref{subsec:NonRenormalizationTheoremsAGeneralProof}, based on Feynman supergraph considerations. The argument is similar to the one found in \cite{Wess:1992cp} for the relativistic case. We therefore only state the main differences. 
As in the relativistic case, the propagators for the superfields can be constructed from the propagators of the component fields. The details of the calculation, including the propagators in terms of off-shell component fields, are given in appendix \ref{app:DetailsPerturbativeAnalysis}. For example, using \eqref{eq:N2HolomorphicDecomposition} and the definition \eqref{eq:GeneralizedCoordinates} one finds:
\begin{equation}
\label{eq:ScalarProp1}
\begin{aligned}
&\left\langle \Phi(t,x,\theta,\theta^\dagger)\Phi(t',x',\theta',\theta'^\dagger) \right\rangle  =\\ 
&-i(f_2^*+g\nabla^2)\delta(\theta-\theta')e^{-i(\theta{\sigma}^0\theta^\dagger-\theta'{\sigma^0}\theta'^{\dagger})\partial_t}{\mathcal{G}}_{\text{lif}}(t-t',x-x'),
\end{aligned}
\end{equation}
where  
\begin{equation}
\label{eq:PertNRArgumentDeltaLifDefinition}
{\mathcal{G}}_{\text{lif}}(t,x) \equiv {\hat{\mathcal{G}}}_{\text{lif}}(t,x)\delta(t)\delta(x), \qquad {\hat{\mathcal{G}}}_{\text{lif}}(t,x) \equiv\frac{-1}{\partial_t^2+|f_2+g\nabla^2|^2}.
\end{equation}

Similar expressions for the $\langle \Phi^\dagger\Phi^\dagger \rangle$ and $\langle \Phi\Phi^\dagger\rangle$ propagators in superspace are given in appendix \ref{app:DetailsPerturbativeAnalysis}. In analogy to the relativistic case, the propagators of $\left<\Phi\Phi\right>$ and $\left<\Phi^\dagger\Phi^\dagger\right>$ are proportional to $\delta(\theta-\theta')$ and $\delta(\theta^\dagger-\theta'^\dagger)$ respectively. Therefore, any closed loop which contains only  $\left<\Phi\Phi\right>$ (or only $\left<\Phi^\dagger\Phi^\dagger\right>$) propagators clearly vanishes, and thus there are no one-loop contributions, finite or infinite, to the coupling constants,  the gap parameter $f_2$ or the kinetic term parameter $g$.
The generalization of this argument to any loop order follows from the procedure detailed in chapters 9 and 10 of \cite{Wess:1992cp}. The technical adjustments required for the case of the non-boost-invariant, holomorphic time domain supersymmetric model considered here are presented in appendix \ref{app:DetailsPerturbativeAnalysis}, including the free fields super-propagators written in terms of covariant superderivatives of the form \eqref{eq:SuperDerivativesDef} and some useful identities  satisfied by these derivatives.

The Feynman rules for the superfields in a model with a general $n$ interaction of the form \eqref{eq:GeneralSuperInteraction} can be easily deduced in analogy to the relativistic case. This yields the following rules for supergraphs:
\begin{itemize}
\item Each external line represents a holomorphic (or an anti-holomorphic) superfield $\Phi(z)$ ($\Phi^\dagger(z)$). 
\item The propagators $\Phi\Phi$, $\Phi^\dagger\Phi^\dagger$, $\Phi\Phi^\dagger$ correspond to the Lifshitz analogue of the Grisaru-Rocek-Siegel (GRS) propagators:
\begin{equation}
\label{eq:GRSPropagator1}
\left< \Phi(z)\Phi(z')\right>_{GRS}= {\hat{\mathcal{G}}}_{\text{lif}}(f_2^*+g\nabla^2)\frac{D^2}{4\square_t}\delta(z-z'),
\end{equation}
\begin{equation}
\label{eq:GRSPropagator2}
\left< \Phi^\dagger(z)\Phi^\dagger(z')\right>_{GRS}= {\hat{\mathcal{G}}}_{\text{lif}}(f_2+g\nabla^2)\frac{\bar{D}^2}{4\square_t}\delta(z-z'),
\end{equation}
\begin{equation}
\label{eq:GRSPropagator3}
\left< \Phi(z)\Phi^\dagger(z')\right>_{GRS}= {\hat{\mathcal{G}}}_{\text{lif}} \delta(z-z'),
\end{equation}
where we have defined $z\equiv (t,x,\theta,\theta^\dagger)$, $\delta(z)\equiv \delta(t)\delta(x)\delta(\theta)\delta(\theta^\dagger)$ and $\square_t \equiv -\partial_t^2$. 
\item At each $\Phi^n$ vertex with $m$ internal lines, one adds factors of $-\frac{1}{4}\bar{D}^2$ acting on $m-1$ internal propagators. Similar factors of $-\frac{1}{4}D^2$ hold at each $(\Phi^\dagger)^n$ vertex. 
\item \,A factor of $\frac{f_n}{n}$ appears for each vertex accompanied by an integration $\int dt d^dx d^2\theta d^2\theta^\dagger$. 
\item \,In addition one must take into account the usual combinatoric factor that multiplies each diagram. 
\end{itemize}
Using these Feynman rules and the identities presented in appendix \ref{app:DetailsPerturbativeAnalysis} it is easy to follow the relativistic arguments to argue that  any arbitrary closed loop the with a general number of integrations over the whole $\theta,\theta^\dagger$ space can be reduced to an expression containing a
single $d^4\theta$ integral (See appendix \ref{app:DetailsPerturbativeAnalysis} for the full derivation). As in the relativistic case, this leads to the conclusion that the effective action can be written as an expression of the form 
\begin{equation}
\begin{aligned}
\label{eq:EffActionNRPerturb}
&\int d^2\theta d^2\theta^\dagger dt_1d^dx_1\cdots  dt_nd^dx_n \\
& \qquad \left(F_1(t_1,x_1,\theta,\theta^\dagger)\cdots F_n(t_n,x_n,\theta,\theta^\dagger)G(t_1,x_1\cdots,t_n,x_n)\right),
\end{aligned}
\end{equation}
where $G(t_1,x_1\cdots,t_n,x_n)$ is a function which is invariant under translations (both time and space translations) and $F_1,\cdots,F_n$ are functions of superfields and their derivatives. The $F_n$'s do not contain any factors of $\square_t ^{-1}$, and therefore the integration over $d^2\theta d^2\theta^\dagger$ cannot be converted into a $d^2\theta$ integration without adding time-derivatives. One can therefore deduce that the gap parameter $f_2$, the kinetic term parameter $g$ and the coupling constants of the interaction are not renormalized to any order in perturbation theory.

\subsubsection{One Loop Example in $3+1$ Dimensions with an $n=3$ Interaction}

\label{subsubsec:OneLoopExample4d}

In this subsection we demonstrate the consequences of supersymmetry and non-renormalization in the models discussed here, by pointing out some interesting properties for the case of $3+1$ dimensions with an $n=3$ interaction of the form \eqref{eq:GeneralSuperInteraction}. We restrict most of the discussion to the one-loop level in perturbation theory. The Feynman rules for the propagators and vertices are given in subsection \ref{sec:FeynmanRules}. Studying the one-loop Feynman diagrams for this model, we make the following observations: 
\begin{itemize}
\item There are no one-loop quantum corrections to the 1PI (amputated) fermionic amplitudes $\left\langle\psi\psi\right\rangle$ and $\left\langle\psi^\dagger\psi^\dagger\right\rangle$.
This implies there are no one-loop quantum corrections to the energy gap parameter $f_2$ and to the kinetic parameter $g$, aligning with the non-renormalization theorem discussed in previous subsections.

\item There is a cancellation of UV divergences in the one-loop corrections to the 1PI (amputated) bosonic two point function $\left\langle\phi\phi^*\right\rangle$:
The Feynman diagrams corresponding to the one-loop corrections to these correlators are given in figure \ref{fig4:Cancellation2}. 
\begin{figure}[!bt]
   \centering
		\subcaptionbox{                \label{fig4:Cancellation2Diag1}}
		{\includegraphics[width=0.08\textwidth]{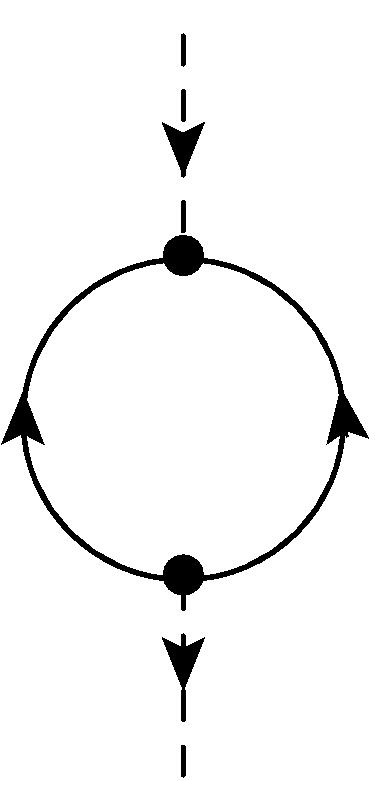}}
	\qquad \qquad
		\subcaptionbox{ \label{fig4:Cancellation2Diag2}}
		{\includegraphics[width=0.085\textwidth]{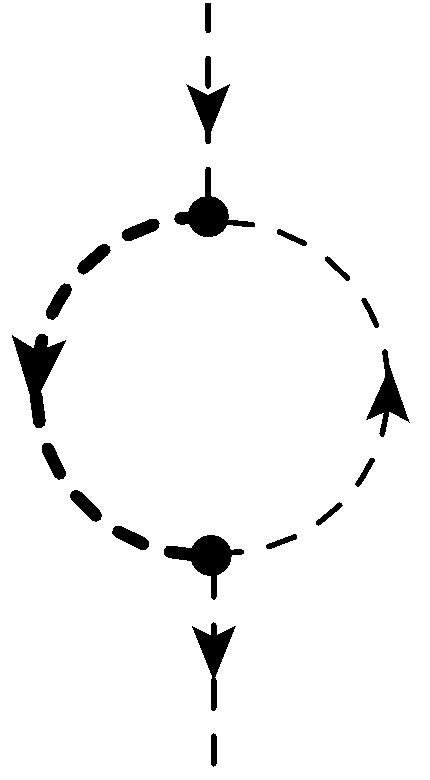}}
			\qquad \qquad
		\subcaptionbox{ \label{fig4:Cancellation2Diag3}}
		{\includegraphics[width=0.08\textwidth]{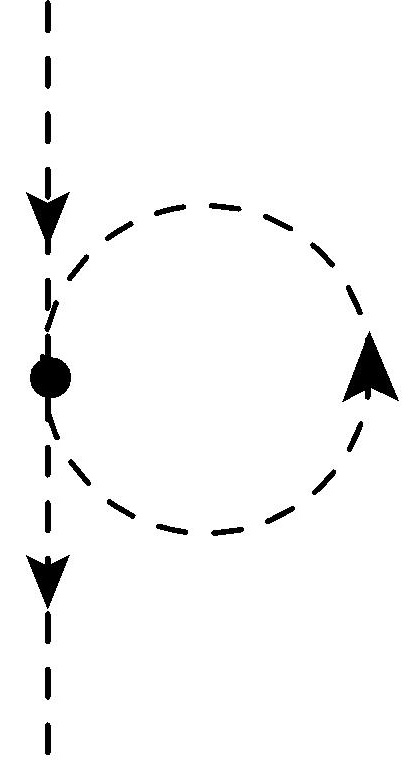}}
		\qquad \qquad 
				\subcaptionbox{ \label{fig4:Cancellation2Diag4}}
		{\includegraphics[width=0.08\textwidth]{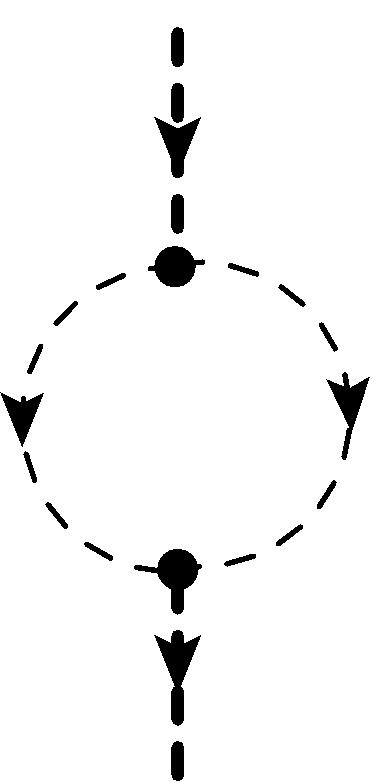}}
        \caption{One-loop corrections for the $\left\langle \phi \phi^* \right\rangle$ propagator in the $n=3$ model. }        \label{fig4:Cancellation2}
\end{figure}
%%%%%%%%%%%%%%%%%%%%%%%%%%
\begin{figure}[!tb]
   \centering
		{\includegraphics[width=0.08\textwidth]{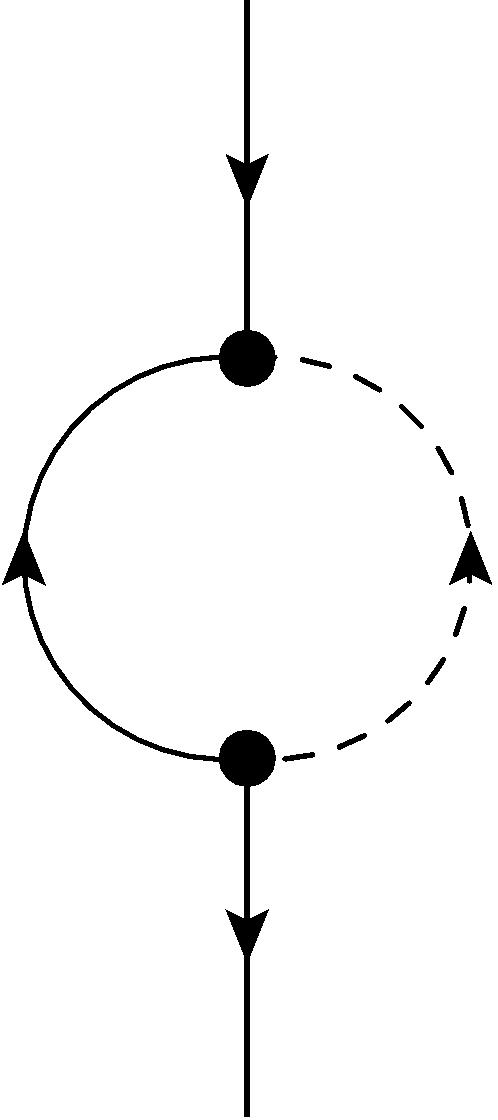}}
        \caption{One-loop corrections for the $\left\langle \psi \psi^{\dagger} \right\rangle$ propagator in the $n=3$ model. }        \label{figfer:Kahler}
\end{figure}
Divergences occur only in the diagrams \ref{fig4:Cancellation2Diag1},\ref{fig4:Cancellation2Diag2} and \ref{fig4:Cancellation2Diag3}. Since the (Lifshitz) degree of divergence here is 1, in order to demonstrate the cancellation of these divergences it is sufficient to show that the sum of these three contributions vanishes for a vanishing external energy (as any terms proportional to positive powers of the external energy will converge by dimensional analysis).

The expression for diagram \ref{fig4:Cancellation2Diag1} reads: 
\begin{equation}
\begin{aligned}
\mathcal{A}& =(-1)(-if_3)(-if_3^*)\\
& \qquad \int \frac{d^3qd\omega_q}{(2\pi)^4}\frac{4(i\omega_q)(i\omega_q)}{\left[\omega_q^2-|gq^2-f_2|^2\right]\left[\omega_q^2-|g(k-q)^2-f_2|^2\right]},
\end{aligned}
\end{equation}
where $k$ is the external momentum and we have omitted the $i\epsilon$ in the denominators for simplicity. 
The diagram \ref{fig4:Cancellation2Diag2} results in: 
\begin{equation}
\mathcal{B}=4(-if_3)(-if_3^*)\int \frac{d^3qd\omega_q}{(2\pi)^4}\frac{(i|f_2-g(k-q)^2|)^2}{\left[\omega_q^2-|gq^2-f_2|^2\right]\left[\omega_q^2-|g(k-q)^2-f_2|^2\right]},
\end{equation}
and finally, the expression for diagram \ref{fig4:Cancellation2Diag3} reads:
\begin{equation}
\mathcal{C}=4(-i|f_3|^2)\int \frac{d^3qd\omega_q}{(2\pi)^4}\frac{i}{\omega_q^2-|gq^2-f_2|^2}.
\end{equation}
It is easy to check that (given an appropriate regularization) the sum of these three contributions vanishes for any value of $k$:
\begin{equation}
\label{eq:CancellationofABC}
\mathcal{A}+\mathcal{B}+\mathcal{C}=0.
\end{equation}
Therefore, in total there are no divergent one-loop corrections to the bosonic two-point function $\left\langle \phi \phi^* \right\rangle$. This is expected due to supersymmetry, since the only one-loop correction to the fermion propagator, given in figure \ref{figfer:Kahler}, is finite, and gives rise to a non-trivial but finite correction to the K\"ahler potential. The remaining bosonic correction described in diagram \ref{fig4:Cancellation2Diag4} is finite and also arises as a result of the corrections to the K\"ahler potential.

\item There is an exact cancellation of the one-loop corrections to the 1PI (amputated) $\left\langle \phi \phi \right\rangle$, $\left\langle \phi^* \phi^* \right\rangle$ correlation functions.  
The relevant diagrams are given in figure \ref{fig3:Cancellation1}.
%%%%%%%%%%%%%%%%%%%%%%%
\begin{figure}[!tb]
   \centering
		\subcaptionbox{                \label{fig3:Diag1}}
		{\includegraphics[width=0.09\textwidth]{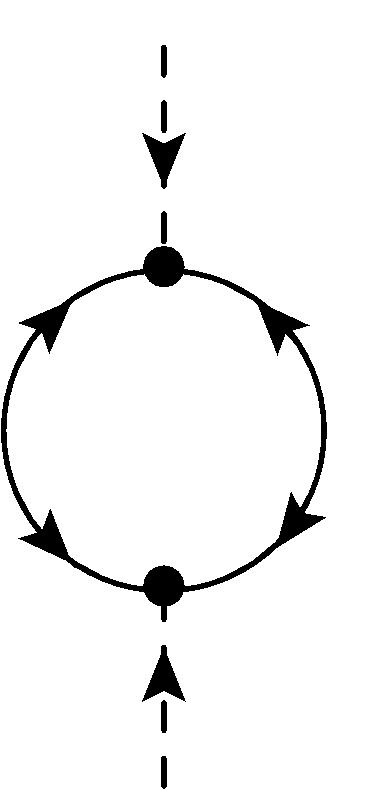}}
	\qquad \qquad
		\subcaptionbox{ \label{fig3:Diag2}}
		{\includegraphics[width=0.095\textwidth]{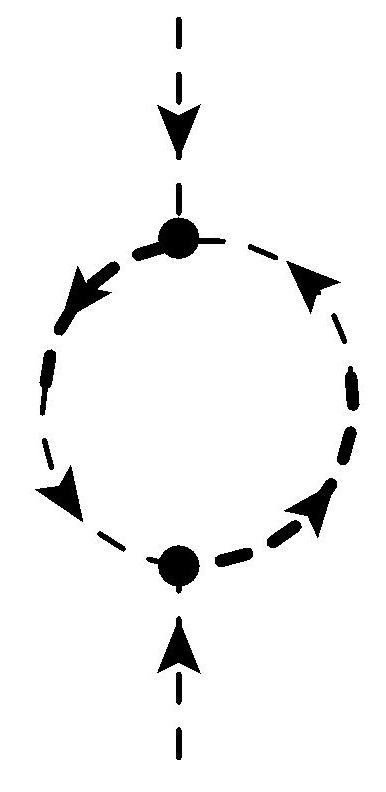}}
        \caption{One-loop corrections for the $\left\langle \phi \phi \right\rangle$ correlation function. }
        \label{fig3:Cancellation1}
\end{figure}

The expression corresponding to diagram \ref{fig3:Diag1} reads: 
\begin{equation}
\begin{aligned}
\mathcal{D}=&4(-if_3)^2(-1)\int \frac{d^3qd\omega_q}{(2\pi)^3} \frac{-i(gq^2-f_2^*)}{\left[\omega_q^2-|gq^2-f_2|^2\right]}\\
&\qquad \qquad\qquad \qquad\qquad\frac{-i(g(k-q)^2-f_2^*)}{\left[(\omega-\omega_q)^2-|g(k-q)^2-f_2|^2\right]},
\end{aligned}
\end{equation}
where $\omega$ and $k$ are the external energy and momentum respectively.
Similarly, the expression for diagram  \ref{fig3:Diag2} reads: 
\begin{equation}
\mathcal{E}=4(if_3)^2\int \frac{d^3qd\omega_q}{(2\pi)^3} \frac{i(gq^2-f_2^*)}{\left[\omega_q^2-|gq^2-f_2|^2\right]}\frac{i(g(k-q)^2-f_2^*)}{\left[(\omega-\omega_q)^2-|g(k-q)^2-f_2|^2\right]}.
\end{equation}
Altogether it is easy to check that the corrections to the correlation function of $\left\langle \phi \phi \right\rangle$ vanish to one-loop order:
\begin{equation}
\label{eq:CancHol11}
D_{(\phi\phi)}(\omega,k) = \mathcal{D} + \mathcal{E} = 0,
\end{equation}
and similarly for $\left\langle \phi^* \phi^* \right\rangle$ corrections.
This cancellation is another indication that the holomorphic structure is indeed preserved to this order in perturbation theory.

\item \, We have shown that all UV divergences in the one-loop corrections to the propagators cancel in this model. In fact, one can check that other than the diagrams in figures \ref{fig4:Cancellation2Diag1},\ref{fig4:Cancellation2Diag2},\ref{fig4:Cancellation2Diag3}, \ref{fig3:Diag1} and \ref{fig3:Diag2}, the only other diagrams (to any perturbative order and with any number of external legs) which have a non-negative superficial degree of divergence\footnote{For an arbitrary 1PI Feynman diagram of order $O(f_3^m)$ in this model with $E_B$ bosonic external legs and $E_F$ fermionic ones, the superficial (Lifshitz) degree of divergence is: $5-\frac{3}{2}m - \frac{1}{2} E_B - \frac{3}{2} E_F$.} are ``tadpole'' diagrams for $\langle\phi\rangle$, which must cancel due to supersymmetry and non-renormalization of the superpotential. Therefore UV divergences in any diagrams for this model will only occur as subdivergences resulting from the appearance of the above set of diagrams (\ref{fig4:Cancellation2Diag1},\ref{fig4:Cancellation2Diag2},\ref{fig4:Cancellation2Diag3}, \ref{fig3:Diag1}, \ref{fig3:Diag2} and the ``tadpole'' diagrams) as subdiagrams. However since these subdiagrams will always appear alongside each other with the same relative signs and relations that led to the cancellation of their divergences in equations \eqref{eq:CancellationofABC} and \eqref{eq:CancHol11}, these subdivergences will similarly cancel. We therefore find this model has the interesting property of being UV finite to all order in perturbation theory. This can also be seen directly from dimensional analysis of supergraphs (see appendix \ref{app:DetailsPerturbativeAnalysis}).

\end{itemize}

\subsection{The Marginal Cases and Exact Lifshitz Scale Symmetry}
\label{sec:MarginalCases}

In this subsection we study the classically marginal cases of the family of supersymmetric models introduced in section \ref{sec:TheHolomorphicModel} (see subsection \ref{subsec:Renormalization} for a definition of marginality in this context). These consist of superpotentials of the form \eqref{eq:N2GeneralSuperpotnetial}-\eqref{eq:GeneralSuperInteraction}, with $f_n\neq 0$ only for $n=n^*\equiv \frac{2d}{d-2}$ (and in particular $f_2=0$). 
Overall, there are three such cases: $n=3$ for $6+1$ dimensions, $n=4$ for $4+1$ dimensions and $n=6$ for $3+1$ dimensions. For all of these cases, the coupling constant $\lambda_n \equiv f_n g^{-n/2}$ is dimensionless in both time and space units.

We would like to argue that each of these three cases realizes a line of fixed points, where the beta function of the marginal coupling constant $\lambda_n$ ($n=3,4,6$) vanishes at the corresponding critical dimension ($d=6,4,3$ respectively). Consider the Wilsonian effective action of these theories associated to some momentum scale $\mu_s$, energy scale $\mu_t$ or both (if one uses a dual-scale renormalization scheme, see discussion in subsections \ref{subsec:Renormalization}-\ref{subsec:DualScaleRGFlows}). As a direct consequence of the non-renormalization theorem proven in subsection \ref{subsec:NonRenormalizationTheoremsAGeneralProof}, the only term in the effective action that transforms non-trivially under the RG flow of the theory is the K\"ahler potential. 
Therefore after canonically normalizing the superfield $\Phi$, the effective Lagrangian takes the form:\footnote{We have omitted here classically irrelevant contributions to the K\"ahler term of the effective action as these are not important for the arguments that follow.}
\begin{equation}
\begin{aligned}
L_{\text{eff}} &= \int d^2\theta d^2 \theta^\dagger d^d x\, \Phi_{\text{cn}}^\dagger\Phi_{\text{cn}}\\
&   + \left( \int d^2\theta d^dx\, \left( \frac{g Z_{\Phi}}{2}\Phi_{\text{cn}}\nabla^2\Phi_{\text{cn}}+\frac{f_n Z_{\Phi}^{\frac{n}{2}}}{n}\Phi_{\text{cn}}^n \right)+ \text{h.c.}\right),
\end{aligned}
\end{equation}
where we have defined $\Phi\equiv\sqrt{Z_{\Phi}}\Phi_{\text{cn}}$, $\Phi_{\text{cn}}$ is the canonically normalized superfield and $Z_{\Phi}$ is its field strength renormalization factor. The canonical effective parameters $g^{\text{cn}}$ and $f_n^\text{cn}$ are therefore given by:
\begin{equation}
g^{\text{cn}}=g Z_{\Phi}, \quad f_n^\text{cn} = f_n Z_{\Phi}^{\frac{n}{2}}.
\end{equation}
However, these are dimensionful parameters. 
The effective dimensionless coupling $\lambda_n^\text{cn}$ is therefore:
\begin{equation}
\label{eq:RenormalizedLambda2}
\lambda_n^\text{cn} = f_n^\text{cn}\,\left(g^{\text{cn}}\right)^{-\frac{n}{2}}= f_n Z_{\Phi}^{\frac{n}{2}} g^{-\frac{n}{2}}Z_{\Phi}^{-\frac{n}{2}}=f_n g^{-\frac{n}{2}}=\lambda_n,
\end{equation}
which implies the beta function identically vanishes for each of the marginal cases ($n=3,4,6$) discussed above, and for any value of the coupling:
\begin{equation}
\label{eq:ZeroBetaFunc}
\beta_n(\lambda_n) = 0.
\end{equation} 
This argument can also be formulated in terms of the RG flow functions of subsection \ref{subsec:Renormalization}: Due to non-renormalization, the beta functions corresponding to the dimensionful parameters $f_n$ and $g$ are both proportional to the anomalous dimension function\footnote{Here are referring to the single-scale RG description.} $\gamma_\Phi$ (as defined in equation \eqref{eq:GammaFieldDef}):
\begin{equation}
\gamma_\Phi \equiv \frac{1}{2}\frac{\mu_s}{Z_\Phi}\frac{\partial \delta_{Z_{\Phi}}}{\partial\mu_s},
\end{equation} 
 with:
\begin{align}
\beta_{f_n} &\equiv \mu_s \frac{\partial f_n}{\partial\mu_s}=n\gamma_\Phi f_n,\\
\label{eq:RenormMarginalCasesBetag}
\beta_g &\equiv \mu_s \frac{\partial g}{\partial\mu_s}=2\gamma_\Phi g.
\end{align}
Equation \eqref{eq:ZeroBetaFunc} for the dimensionless coupling $\lambda_n$ then immediately follows. Note that, due to \eqref{eq:RenormMarginalCasesBetag}, the anomalous dimension corresponding to $g$ (as defined in equation \eqref{eq:AnomalouseDimOfGDef}) is related to $\gamma_\Phi$ via:
\begin{equation}
\gamma_g \equiv \frac{\mu_s}{g}\frac{\partial g}{\partial \mu_s}= 2\gamma_\Phi.
\end{equation}
 
Following the discussion of subsection \ref{subsec:Renormalization}, we therefore conclude that each of these marginal cases realizes a Lifshitz scale invariant theory. Furthermore, in accordance with equation \eqref{eq:GammaAtFixedPointRelation}, the dynamical critical exponent associated with this scale invariance is determined by the anomalous dimension of the field $\Phi$ as follows:  
\begin{equation}
\label{eq:MarginalCaseszGammaRelation}
z=2+\gamma_g(\lambda_n) = 2+2\gamma_\Phi(\lambda_n).
\end{equation}  
That is, the holomorphic structure here implies that (for each of these marginal cases) this family of models describes a line of quantum critical fixed points corresponding to each value of the coupling $\lambda_n$, with the dynamical exponent depending on the coupling.
This is reminiscent of well known families of relativistic superconformal models which realize a set of fixed points for various values of coupling constants, interpolating between weak and strong coupling, such as the $\mathcal{N}=4$ SYM model, although note that unlike those cases (which are relativistic and therefore have $z=1$), here $z$ changes along the marginal directions.
 
It is useful to describe these results from the point of view of the dual-scale RG formalism discussed in subsection \ref{subsec:DualScaleRGFlows}. Recall that in this description we introduce two renormalization scales: a spatial one ($\mu_s$) and a temporal one ($\mu_t$). We then have 2 independent dimensionless parameters in these models, which we may choose to be $\tilde{g} \equiv g \mu_s^2 \mu_t^{-1}$ and $\lambda_n$. Due to non-renormalization (using the same type of arguments as in the single-scale case), we see that the both beta functions of the coupling $\lambda_n$ vanish:
\begin{equation}
\beta^s_n (\tilde{g},\lambda_n) = 0, \quad 
\beta^t_n (\tilde{g},\lambda_n) = 0,
\end{equation}
whereas those of the parameter $\tilde{g}$ are related to the anomalous dimension functions as follows:
\begin{equation}
\label{eq:RenormMarginalCasesDualScaleRGBetag}
\beta^s_{\tilde{g}} (\tilde{g},\lambda_n) = \left(2+2\gamma^s_\Phi(\tilde{g},\lambda_n)\right) \tilde{g},\qquad 
\beta^t_{\tilde{g}} (\tilde{g},\lambda_n) = \left(-1+2\gamma^t_\Phi(\tilde{g},\lambda_n)\right) \tilde{g}. 
\end{equation}
From the discussion in subsection \ref{subsec:DualScaleRGFlows} we conclude that any point $\tilde{g},\lambda_n$ on the parameter space is part of a one-dimensional RG orbit representing a Lifshitz fixed point, and these orbits are just $\lambda_n=\text{const.}$ lines in the parameter space. The dynamical exponent and Lifshitz anomalous dimension of these fixed points are given by:\footnote{Note that $z$ and $\gamma_\Phi$ cannot depend on $\tilde{g}$ as they must remain constant along the fixed point leaves (see subsection \ref{subsec:DualScaleRGFlows}), which in this case are the $\lambda_n=\text{const.}$ lines.}
\begin{align}
\label{eq:RenormMarginalCasesDualScaleRGDynExp}
& z(\lambda_n) = \frac{2+2\gamma^s_\Phi(\tilde{g},\lambda_n)}{1-2\gamma^t_\Phi(\tilde{g},\lambda_n)},\\
\label{eq:RenormMarginalCasesDualScaleRGAnomDim}
& \gamma_\Phi (\lambda_n) = z(\lambda_n) \gamma^t_\Phi(\tilde{g},\lambda_n) + \gamma^s_\Phi(\tilde{g},\lambda_n) .
\end{align}
Viewed as equations for $\gamma^s_\Phi, \gamma^t_\Phi$, \eqref{eq:RenormMarginalCasesDualScaleRGDynExp}-\eqref{eq:RenormMarginalCasesDualScaleRGAnomDim} have a solution only if the condition \eqref{eq:MarginalCaseszGammaRelation} is satisfied, aligning with the single-scale picture. Moreover, these equations then have an infinite set of solutions, corresponding to various possible renormalization schemes.\footnote{In fact, the diffeomorphisms of $\hat{M}$ generated by \eqref{eq:DualScaleRGDiffVec} with the choice $ \xi = 2 \xi^Z(\tilde{g},\lambda_n) \tilde{g} \frac{\partial}{\partial\tilde{g}}$ are examples of a renormalization scheme change that preserves the foliation induced by the dual-scale RG flow, the values of the dynamical exponent $z$ and the Lifshitz anomalous dimension $\gamma_\Phi$ as well as the relations \eqref{eq:RenormMarginalCasesDualScaleRGBetag}, while still changing $\gamma^t_\Phi$ and $\gamma^s_\Phi$ individually.}

It is important to mention here that none of the arguments made so far in this subsection are based on perturbative arguments, and these conclusions should therefore apply to strong coupling as well. However, we did assume the existence of a supersymmetric vacuum state, and that at strong coupling the UV degrees of freedom still correctly describe the physics at lower energies (for a discussion on non-perturbative considerations see section \ref{sec:Summary}).

A natural question which arises in the context of non-boost-invariant theories is whether there are any restrictions on the possible values of the dynamical critical exponent $z$, and in particular whether it can have a value smaller than $z=1$. For the critical cases considered here, it is clear from the relation \eqref{eq:MarginalCaseszGammaRelation} that as long as $\gamma_\Phi (\lambda_n) >0$ we have $z>2$, that is $z$ is larger than its classical value. In the rest of this subsection and in appendix \ref{app:betaFunctions} we show this to be satisfied to the leading order in perturbation theory, for each of the 3 marginal cases. 
Whether this behaviour persists to higher orders in perturbation theory or in strong coupling remains an open question which is left for future work.

In the rest of this subsection we provide an example for the perturbative calculation of $z$ in the marginal cases, by calculating the one-loop quantum corrections to the anomalous dimension for the critical case of the $n=3$ interaction in $6+1$ dimensions. 
As long as supersymmetry is preserved in the quantum theory, the anomalous dimension for the holomorphic field $\Phi$ can be easily calculated from the quantum corrections to the fermionic propagator. 
In this case the leading order non-trivial correction to the fermionic propagator is the one-loop ($\lambda_3^2$) order.  
  
\begin{figure}[!tb]
   \centering
		{\includegraphics[width=0.08\textwidth]{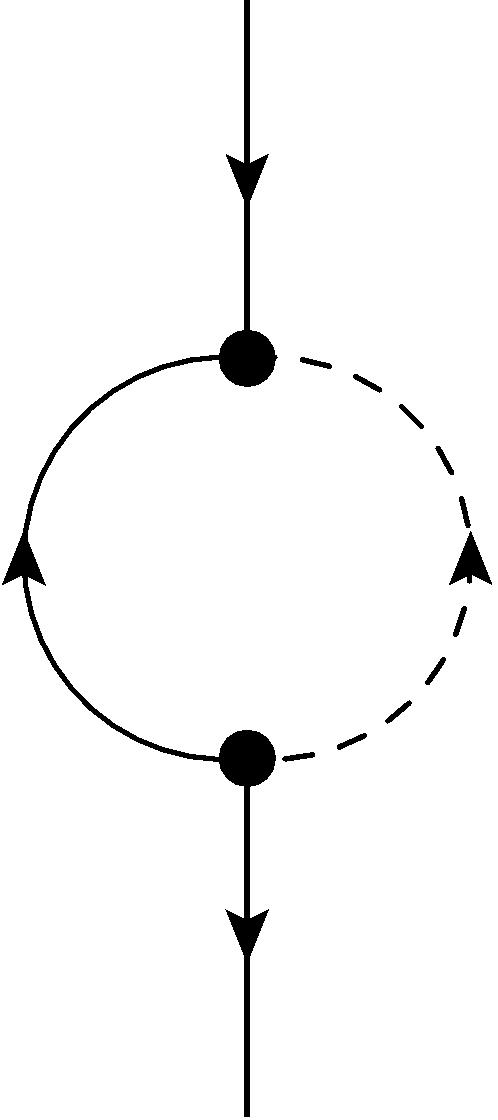}} 
        \caption{The leading order quantum corrections to the fermionic propagator in the critical model of $n=3$ interaction in $6+1$ dimensions. \label{fig:BetaFuncSixSpatialDim} }
\end{figure}

To this order in perturbation theory, using the Feynman rules of subsection \ref{sec:FeynmanRules} it is easy to see that there are no quantum corrections to the $\psi\psi$ or $\psi^\dagger \psi^\dagger$ propagators, as dictated by supersymmetry and the non-renormalization of the parameter $f_2$ (see subsection \ref{subsec:NonRenormalizationTheoremsAGeneralProof}). 
The Feynman diagram for the one-loop correction to the $\psi \psi^\dagger$ fermionic propagator (that is, the self-energy one-loop diagram) is given in figure \ref{fig:BetaFuncSixSpatialDim}, and the corresponding expression reads:
\begin{equation}
\label{eq:MarginalCases6p1dOneLoopProp}
A^{\dot{\beta}\alpha}= 4(-if_3)(-if_3^*) \int \frac{d\Omega}{2\pi}\int \frac{d^6q}{(2\pi)^6}\frac{i\Omega\bar{\sigma}^{0{\dot{\beta}\alpha}}}{\Omega^2-g^2q^4}\frac{i}{(\omega-\Omega)^2-g^2(k-q)^4},
\end{equation}
where $(\omega,k)$ are the external energy and momentum and $(\Omega,q)$ are the internal ones running in the loop (the $i\epsilon$ factors in the denominators have been omitted here for simplicity). 
The Feynman integral in \eqref{eq:MarginalCases6p1dOneLoopProp} is of course divergent and requires regularization and renormalization. To that end, we first extract the UV divergent part of the integral.\footnote{We are using minimal subtraction renormalization schemes here, and it is therefore sufficient to subtract just the divergent part.} This can be done using standard techniques of expansion in external momenta and energies (see e.g.\ \cite{Anselmi:2007ri,Arav:2016akx} for application of these techniques for non-boost-invariant field theories). 
It is easy to see that, due to time reversal invariance ($\Omega \to -\Omega$), the integral \eqref{eq:MarginalCases6p1dOneLoopProp} vanishes for $\omega=0$, and therefore the divergent part is logarithmic and proportional to $\omega$. This is also expected due to supersymmetry (as corrections to the K\"ahler potential involve at least one time derivative). The one-loop correction can therefore be written as follows:
\begin{equation}
\label{eq:Loopfor6dCriticalCase}
A^{\dot{\beta}\alpha} = \omega \left(\frac{\partial A^{\dot{\beta}\alpha}}{\partial \omega}\right)_{\,\omega,k=0} + \ldots
=8|f_3|^2\omega\bar{\sigma}^{0{\dot{\beta}\alpha}}\int \frac{d\Omega}{2\pi}\int \frac{d^6q}{(2\pi)^6}\frac{\Omega^2}{\left[\Omega^2-g^2q^4\right]^3} + \ldots ,
\end{equation}
where ``$\ldots$'' stands for finite terms.
Define the renormalized fermionic field $\psi_{\text{ren}}$ and field strength $Z_\psi$ by the following relation: $\psi = \sqrt{Z_\psi}\psi_{\text{ren}}$, $\delta_{Z_{\psi}}=Z_\psi-1$. 
The counterterm contribution to the self-energy takes the form $i\bar{\sigma}^{0\dot{\beta}\alpha}\omega\delta_{Z_{\psi}}$. Employing a minimal subtraction scheme, we therefore set $\delta_{Z_\psi}$ to cancel the divergent part of the one-loop expression \eqref{eq:Loopfor6dCriticalCase}:
\begin{equation}
\label{eq:DeltaZfor6d}
i\delta_{Z_{\psi}}=-8|f_3|^2\left. \int \frac{d\Omega}{2\pi}\int \frac{d^6q}{(2\pi)^6}\frac{\Omega^2}{\left[\Omega^2-g^2q^4\right]^3} \right|_{\text{div}},
\end{equation} 
(where ``div'' here refers to taking the UV divergent part of the expression after regularization). We demonstrate the regularization and renormalization of these integrals using two different methods. For the first, we use the time-first regularization method, along with a spatial UV cutoff and a single scale renormalization.   
Performing the integral over the running loop energy $\Omega$ using contour integration in the complex plane one finds\footnote{Strictly speaking this integral is IR divergent as well and has to be IR regularized, however such an IR regulator would not change the UV divergences.} 
\begin{equation}
i\delta_{Z_{\psi}}=-i|f_3|^2 \int \left. \frac{d^6q}{(2\pi)^6}\frac{1}{2(gq^2)^3}\right|_{\text{div}}=-\frac{i |\lambda_3|^2}{16(2\pi)^3}\int\left. \frac{dq}{q} \right|_{\text{div}} ,
\end{equation}
where in the second equality we have used the spherical symmetry of the theory. Imposing a spatial UV cutoff $\Lambda_s$, the counterterm coefficient will be chosen to be:\footnote{Alternatively, one may view this as a calculation of the field strength correction in the Wilsonian effective action evaluated at the spatial scale $\mu_s$.} $\delta_{Z_{\psi}} = - \frac{|\lambda_3|^2}{16(2\pi)^3} \log\left(\frac{\Lambda_s}{\mu_s}\right)$,
where $\mu_s$ is a spatial renormalization scale (which carries spatial dimensions $\left[\mu_s\right]=\left[p\right]$). 
This leads to the following result for the anomalous dimension of the theory:
\begin{equation}
\label{eq:GammaFuncFor6d}
\gamma_{\psi} \equiv \frac{1}{2}\frac{\partial\delta_{Z_{\psi}} }{\partial \log(\mu_s)}=\frac{|\lambda_3|^2}{32(2\pi)^3}>0,
\end{equation}
with the dynamical exponent given by the relation \eqref{eq:MarginalCaseszGammaRelation}.

For the second method of regularization and renormalization, we use split dimensional regularization (see subsection \ref{subsec:Renormalization} and \cite{Anselmi:2007ri,Arav:2016akx}) along with a dual scale renormalization (see subsection \ref{subsec:DualScaleRGFlows}). Split dimensional regularization of the integral \eqref{eq:DeltaZfor6d} yields:\footnote{The integral can be found in subsection 3.2.2 of \cite{Arav:2016akx}.}
\begin{equation}
\begin{aligned}
\label{eq:RenormMarginalCasesDualScaleRGDeltaZ}
i\delta_{Z_{\psi}} &= -8|f_3|^2 \int \frac{d^{d_t}\Omega}{(2\pi)^{d_t}}\int \frac{d^{d_s}q}{(2\pi)^{d_s}}\left. \frac{\Omega^2}{\left[\Omega^2-g^2q^4\right]^3} \right|_\text{div}\\
&= \left. -8|f_3|^2 \frac{2 i^{d_t} g^{-d_s/2}}{(4\pi)^{(d_t+d_s)/2}}
\frac{\Gamma\left( \frac{d_s}{4} \right)\Gamma\left( \frac{2+d_t}{2} \right)}{\Gamma\left( \frac{d_t}{2} \right)\Gamma\left( \frac{d_s}{2} \right)\Gamma(3)} \mu^{-\epsilon_\text{lif}/2} \frac{1}{\epsilon_\text{lif}} \right|_\text{div} \\
&= - \frac{i |\lambda_3|^2}{16 (2\pi)^3} \mu^{-\epsilon_\text{lif}/2} \frac{1}{\epsilon_\text{lif}},
\end{aligned}
\end{equation}
where we have defined $d_t \equiv 1-\epsilon_t$, $d_s \equiv 6-\epsilon_s$ and $\epsilon_{\text{lif}} \equiv 2\epsilon_t+\epsilon_s$, and $\mu$ is some scale of dimensions $[\mu]=[E]$ required to make the total dimensions of the expression vanish. A natural choice for $\mu$ would be $\mu=\mu_t$, from which we obtain the following anomalous dimension functions:
\begin{equation}
\gamma^t_{\psi} \equiv \left. \frac{1}{2}\frac{\partial\delta_{Z_{\psi}} }{\partial \log(\mu_t)} \right|_{\epsilon_\text{lif}=0} = \frac{|\lambda_3|^2}{64(2\pi)^3}, \quad
\gamma^s_{\psi} \equiv \left. \frac{1}{2}\frac{\partial\delta_{Z_{\psi}} }{\partial \log(\mu_s)} \right|_{\epsilon_\text{lif}=0} = 0.
\end{equation}
Finally from \eqref{eq:RenormMarginalCasesDualScaleRGDynExp}-\eqref{eq:RenormMarginalCasesDualScaleRGAnomDim} we have, to lowest order in $\lambda_3$:
\begin{equation} 
\gamma_\psi = \left(2+ O(|\lambda_3|^2)\right) \gamma^t_\psi + \gamma^s_\psi = \frac{|\lambda_3|^2}{32(2\pi)^3} + O(|\lambda_3|^4),
\end{equation}
in agreement with the result \eqref{eq:GammaFuncFor6d}. Note that one can also choose $\mu=g \mu_s^2 $ for the arbitrary energy scale in expression \eqref{eq:RenormMarginalCasesDualScaleRGDeltaZ} (or some combination of $\mu_t$ and $ g \mu_s^2$), changing the values of the scheme dependent $\gamma^t_\psi$ and $\gamma^s_\psi$, but leaving the physical $z$ and $\gamma_\psi$ unchanged (see the discussion in subsection \ref{subsec:DualScaleRGFlows}).

In appendix \ref{app:betaFunctions} we give expressions for the leading order corrections to the anomalous dimension in the other two marginal cases ($d=4,3$ with $n=4,6$ respectively). Similarly to the above $d=6$ case, in both cases we find the correction has a positive sign.

\subsection{The Gapless Singular Case } 
\label{subsec:GaplessSingularCase}

In this subsection we make some comments on the gapless singular case of the family of models described in subsection \ref{sec:TheHolomorphicModel}. This is the case in which the gap parameter $f_2$ is real and positive. As stated in subsection \ref{sec:TheHolomorphicModel}, the single particle spectrum of the free model contains a spherical surface of zero energy at momenta of magnitude $\tilde{k}_0 = \sqrt{f_2/g} $. For momenta in the vicinity of this sphere, the dispersion relation takes the approximate form:
\begin{equation}
\omega \approx \pm v |\delta k|,
\end{equation}  
where $ v \equiv 2g \tilde{k}_0 $ and $\delta k \equiv k - \tilde{k}_0 $. The energy is thus approximately linear in the momentum difference from the zero energy surface. This behavior is similar to that of a Fermi liquid near its Fermi surface. In fact, the singular surface here can be thought of as a Fermi surface for the two non-relativistic fermions. The important difference from Fermi liquid theory is that here, due to supersymmetry,  the IR singularity exists in the bosonic sector as well, and is more severe. Near the singular surface, the system can be viewed as a ($1+1$)-dimensional system, with the direction normal to the surface serving as the spatial direction and the directions tangent to it viewed as internal degrees of freedom. The free propagators are then similar to those of a massless ($1+1$)-dimensional relativistic theory, with the bosonic propagator of the form: $\frac{1}{\omega^2-v^2(\delta k)^2}$, and the fermionic one of the form: $\frac{\omega}{\omega^2-v^2 (\delta k)^2}$ or $\frac{v \delta k}{\omega^2-v^2(\delta k)^2}$. Then, similarly to the ($1+1$)-dimensional relativistic case, introducing scalar interactions will generally lead to IR singularities.

To see this more explicitly, consider any Feynman diagram loop containing a bosonic propagator, and let $(\Omega,q)$ be the energy and momentum associated with this propagator respectively, and $(\omega_i, k_i)$ arbitrary energies and momenta external to this loop.\footnote{For simplicity we assume here that the external momenta $k_i$ are chosen such that the singular spheres of the various propagators in the loop do not intersect, and the external energies are chosen such that no two loop propagators are simultaneously on shell when $|\delta q|<\varepsilon$, but are otherwise arbitrary.} The contribution to the loop integral from of region of momentum space where $|\delta q|<\varepsilon$ ($\tilde{k}_0 - \varepsilon < q < \tilde{k}_0 + \varepsilon $) is then given by an expression of the form:
\begin{equation}
\int\limits_{-\infty}^\infty \frac{d\Omega}{2\pi} \int \frac{dS^{d-1}\, \tilde{k}_0^{d-1}}{(2\pi)^{d-1}} \int\limits_{|\delta q|<\varepsilon} \frac{d(\delta q )}{2\pi} \, \frac{i F\left(q,\Omega,k_i,\omega_i \right)}{\Omega^2 - v^2 (\delta q)^2 + O\left((\delta q)^3\right)+ i\epsilon},
\end{equation}
where $d S^{d-1}$ is the measure on the unit sphere associated with the direction of $q$ and $F\left(q,\Omega,k_i,\omega_i \right)$ is some function of of the various energies and momenta which contains the other loop propagators. Performing the integral over $\Omega$ we obtain:
\begin{equation}
\begin{split}
&\int \frac{dS^{d-1}\, \tilde{k}_0^{d-1}}{(2\pi)^{d-1}} \int\limits_{|\delta q|<\varepsilon} \frac{d(\delta q )}{2\pi} \, \frac{ F\left(q,v|\delta q|+O\left((\delta q)^2\right),k_i,\omega_i \right)}{2v |\delta q| \left( 1+ O(\delta q) \right)} + \ldots \\
&\qquad\qquad \approx
\int \frac{dS^{d-1}\, \tilde{k}_0^{d-1}}{(2\pi)^{d-1}} \int\limits_{|\delta q|<\varepsilon} \frac{d(\delta q )}{2\pi} \, \frac{ F\left(\tilde{k}_0 \hat{q},0,k_i,\omega_i \right)}{2v |\delta q|} + \ldots,
\end{split}
\end{equation}
where ``$\ldots$'' stands for finite terms, including the contributions from the poles of other propagators, and $\hat{q}\equiv \vec{q}/|\vec{q}|$. Since $F\left(\tilde{k}_0 \hat{q},0,k_i,\omega_i \right)$ generally does not vanish, it is clear the above term is logarithmically divergent. In order for perturbation theory to be well-defined, then, it requires an IR regulator $\mu_s^\text{IR}$, and as one takes $\mu_s^\text{IR} \to 0$ perturbative quantum corrections to physical observables will diverge.

It is important to note here that the holomorphic supersymmetry of the model ensures the consistency of the assumption that $\operatorname{Im}(f_2)=0$ (and therefore the gaplessness of the model) from a naturalness point of view. In a general, non-supersymmetric theory with these types of bosons and fermions, one would generically expect $\operatorname{Im}(f_2)$ to gain quantum corrections and form a gap along the RG flow so that the gapless singular case discussed here would require fine-tuning (or put differently, renormalization of the theory would require adding such a gap as a counterterm). However here, due to the non-renormalization properties of theory discussed in subsection \ref{subsec:NonRenormalizationTheoremsAGeneralProof}, $\operatorname{Im}(f_2)$ remains vanishing, and furthermore -- the singular sphere radius $\tilde{k}_0$ is not renormalized. 

The fact that these models are strongly coupled in the IR can also be seen from an RG flow analysis. Consider the IR Wilsonian effective action in these gapless, singular cases. Similarly to the RG treatment of Fermi liquid theory (see for example \cite{Shankar:1993pf,Polchinski:1992ed}), since the zero modes are located along the $|k|=\tilde{k}_0$ sphere rather than at $k=0$, for the IR effective action we include only modes within a spherical shell of width $2\mu_s$ around the singular sphere, namely the modes with $|\delta k| < \mu_s $, and integrate out the rest. Another important difference from other cases (where the single particle spectrum is gapped or has $\omega=0$ only at $k=0$) is that as we flow to the IR, we must scale the momenta towards the singular surface rather than $k=0$. We therefore define the scaling procedure of the momenta such that the momentum component of the fields normal to the singular surface $\delta k$ scales along the flow ($\delta k \to s\, \delta k$), whereas the components tangent to the surface do not ($k_\parallel \to k_\parallel$). As in subsections \ref{subsec:Renormalization} and \ref{subsec:DualScaleRGFlows}, the energy may still scale independently. 

Assuming as usual that the IR effective action may still be written in terms of the UV degrees of freedom, due to the arguments outlined in subsection \ref{subsec:NonRenormalizationTheoremsAGeneralProof}, the superpotential will not get renormalized. The (canonically normalized) effective action at scale $\mu_s$ will then take the general form:
\begin{equation}
\begin{split}
\label{eq:GapplessSingularCaseIREffectiveAction}
&S = \int\limits_{|\delta k|<\mu_s} dt d^2\theta d^2\theta^\dagger \frac{d^d k}{(2\pi)^d} \tilde{\Phi}^\dagger(k) \tilde{\Phi}(k)
-  \int\limits_{|\delta k|<\mu_s} dt d^2\theta \frac{d^d k}{(2\pi)^d} \frac{v}{2} \delta k \,\tilde{\Phi}(k) \tilde{\Phi}(-k)
\\
&+ \int\limits_{\substack{|\delta k_i |<\mu_s \\ |\delta(k_1+\ldots+k_{n-1})|<\mu_s}} dt d^2\theta \frac{d^d k_1 \ldots d^d k_{n-1}}{(2\pi)^{d(n-1)}} \frac{f_n}{n} \tilde{\Phi}(k_1)\ldots \tilde{\Phi}(k_{n-1}) \tilde{\Phi}(-k_1-\ldots-k_{n-1}),
\end{split}
\end{equation}
where $d^d k_i \equiv \tilde{k}_0^{d-1} dS^{d-1}_i d(\delta k_i) $, $\,\delta(k_1+\ldots+k_{n-1}) \equiv |k_1+\ldots+k_{n-1}|-\tilde{k}_0$, and we have omitted other terms (higher derivative terms in the superpotential, as well as higher derivative and higher order terms in $\tilde{\Phi}$ in the K\"ahler potential) which are at most (classically) marginal with respect to the free fixed point. Note that, similarly to Fermi liquid theory, the kinematic condition $|\delta(k_1+\ldots+k_{n-1})|<\mu_s$ restricts the interaction to a subset of momenta directions that obey certain geometric relations (rather the entirety of $\left(S^{d-1}\right)^{n-1}$). For example, for $n=3$, this restricts the interaction to momenta with an angle of $2\pi/3$ between them. The interactions between momenta that do not obey this condition become irrelevant in the ``deep'' IR.
Assigning dimensions to the various quantities in \eqref{eq:GapplessSingularCaseIREffectiveAction} according to the scaling prescription outlined above, we have:
\begin{align}
&[\tilde{\Phi}] = [E]^{-\frac{1}{2}} [\delta k]^{-\frac{1}{2}}, \\
\label{eq:GaplessSingularCasevDimensions}
&[v] = [E]^1 [\delta k]^{-1}, \\
&[f_n] = [E]^{\frac{n}{2}} [\delta k]^{-\frac{n}{2}+1}.
\end{align}
In particular, equation \eqref{eq:GaplessSingularCasevDimensions} (along with the discussion of subsection \ref{subsec:Renormalization}) confirms that the free case corresponds to a $z=1$ IR fixed point. 
We can now define the following ``dimensionless'' (in terms of the scaling defined above) coupling:
\begin{equation}
\tilde{\lambda}_n \equiv f_n v^{-\frac{n}{2}} \mu_s^{-1}.
\end{equation}
As usual, the non-renormalization of the superpotential enables us to express the beta functions of $v, f_n$ and $\tilde{\lambda}_n$ in terms of the anomalous dimension of the field $\Phi$:
\begin{align}
&\beta_{v} = 2 \gamma_\Phi v,\\
\label{eq:GaplessSingularCaseBetafn}
&\beta_{f_n} = n \gamma_\Phi f_n,\\
&\beta_{\tilde{\lambda}_n} = - \tilde{\lambda}_n. 
\end{align}
It is therefore clear that $\tilde{\lambda}_n$ diverges as $\mu_s \to 0$, and the interaction is relevant (with respect to the free fixed point), for any value of $n$ and any dimension.\footnote{In fact, if one trusts that the effective action \eqref{eq:GapplessSingularCaseIREffectiveAction} still describes the system in the strongly coupled IR, supersymmetry ensures that $f_n v^{-\frac{n}{2}} $ has dimension of exactly $[\delta k]^1$.} 

While it is difficult to draw conclusions regarding the IR behavior of the models in these gapless, singular cases, we may conjecture that they reach some IR Lifshitz fixed point. If that is indeed the case, and we further assume that the effective action description of \eqref{eq:GapplessSingularCaseIREffectiveAction} is still valid in this regime (and the other assumptions mentioned in \ref{subsec:NonRenormalizationTheoremsAGeneralProof} stand), then the non-renormalization of the superpotential leads us to conclude that the kinetic term in the superpotential will become irrelevant with respect to this fixed point. Furthermore, the discussion of subsection \ref{subsec:Renormalization} then allows us to relate the dynamical critical exponent of this fixed point to the anomalous dimension of $\Phi$: Using equation \eqref{eq:GeneralzAtFixedPoint} for $f_n$ (since in this case $v$ becomes irrelevant with respect to the IR fixed point, the fixed point is determined by $f_n$) along with \eqref{eq:GaplessSingularCaseBetafn} we obtain the relation: $ z = 1 - \frac{2}{n} + 2\gamma_\Phi $. Whether these assumptions are truly satisfied, though, we leave as an open problem.

\section{Discussion and Outlook}
\label{sec:Summary}

In this work we studied the consequences of holomorphic time domain supersymmetry in the context of  Lifshitz (non-boost-invariant) quantum field theories.   
To that end, we constructed a family of such models possessing four real supercharges which satisfy supersymmetric commutation relations closing on the Hamiltonian, endowing these systems with a holomorphic structure.  We found that while these models share some similarities with relativistic models of holomorphic supersymmetry, such as the existence of non-renormalization properties of the superpotential 
(subsection \ref{subsec:NonRenormalizationTheoremsAGeneralProof}), they also yield several new and interesting results. Chief among these is the scale invariance property of the marginal cases studied in subsection \ref{sec:MarginalCases}: We found that each of the three marginal cases realizes a line of interacting quantum critical points with an exact Lifshitz scale invariance (each in a different spatial dimension).
We showed that the dynamical critical exponent $z$ in these cases is related to the anomalous dimension of the superfield, and therefore depends on the coupling constant and changes along the marginal direction. We also calculated the leading order perturbative correction to the anomalous dimension (and therefore the dynamical exponent) in these models and showed that, to this order at least, $z$ is larger than its free limit value ($z>2$).

Another interesting distinction from relativistic supersymmetry lies in the possibility of having supersymmetric vacua with spontaneously broken spatial translation symmetry.  
Whereas in relativistic supersymmetry the moduli space of vacua is represented by an algebraic equation, in the non-boost-invariant (time domain) case it is represented by a differential one (see equation \eqref{eq:N1ReviewClassicalSUSYVacuumCondition} for the $\mathcal{N}=1$ case and equation \eqref{eq:N2SUSYVacuumCondition} for the $\mathcal{N}=2$ holomorphic case). These vacuum equations could therefore have non-homogeneous solutions, representing supersymmetric vacua with broken spatial translations. 
Furthermore, in the holomorphic $\mathcal{N}=2$ case, since the superpotential is not renormalized, the moduli space of the full quantum theory can be studied exactly by solving the semiclassical vacuum equation \eqref{eq:N2SUSYVacuumCondition}. While we haven't focused on these vacua in this work, their existence strongly suggests that these models may serve as an interesting test case for studying spontaneous breaking of translation symmetries in non-boost-invariant field theories.

Most of the discussion of the quantum behaviour of these models in section \ref{sec:QuantumCorrections} relies on perturbative and semiclassical arguments, and does not account for non-perturbative phenomena. While a full analysis of these non-perturbative effects and their consequences seems to be a considerable task beyond the scope of this work, we can make some comments and observations on their possible implications. 

As previously stated, the vacuum equation in these models allows for supersymmetric, non-homogeneous semiclassical vacua, and in particular in some cases there may be soliton-like solutions to the equation (that is, solutions that vanish at spatial infinity in all directions). A simple example (see \cite{Azzollini:2008}) for such a solution is the following for any of the marginal cases discussed in subsection \ref{sec:MarginalCases}:
\begin{equation}
\label{eq:DiscussionMarginalCaseSoliton}
\phi_\text{sol}(x) = \frac{\left[d(d-2)a_n^{-1} B^2 \right]^\frac{d-2}{4}}{\left[ B^2 + \sum\limits_{i=1}^d (x_i-z_i)^2 \right]^\frac{d-2}{2}},
\end{equation}
where $B, z_i \in \mathbb{C} \,(i=1,\ldots,d) $ are arbitrary complex parameters, and $a_n \equiv f_n/g$. These solutions may exist even for cases in which the free single particle spectrum is gapped,\footnote{In fact, it has been shown (see \cite{Berestycki:1983I}) that such a soliton solution always exists in the gapped cases with $d\geq 3$, with a real and negative gap parameter $f_2<0$ and $n<n^*$, and vanishes exponentially at spatial infinity.} and in many cases can belong to $L^2(\mathbb{R}^d)$ (for the marginal case solution \eqref{eq:DiscussionMarginalCaseSoliton} this is true for $d=6$). Moreover, in some cases, an infinite sequence of such soliton solutions may exist (see \cite{Berestycki:1983II}). These classical solutions are non-perturbative in the sense that they ''escape'' to infinity (in field space) in the free limit ($f_n \to 0$).
Any soliton solution of this form spontaneously breaks at least the spatial translation symmetries, and may also break spatial rotations and scale symmetry (in the marginal cases). Therefore one may act on such a solution with any element in the broken spacetime symmetries group to obtain another solution, and in fact, since the vacuum equation is holomorphic here, the same is true for the complex version of these groups. Thus in the IR the soliton may be viewed as a non-relativistic, supersymmetric and gapless quantum mechanics particle, moving on a K\"ahler target space\footnote{Generally this K\"ahler manifold will be non-compact (even if the physical space is compact, as the imaginary directions may still be non-compact), and may become singular as the soliton solution becomes singular for some subset of its parameter space.} parameterized by the complex parameters of the solution (which can be interpreted as the collective coordinates associated with it).

Whenever these soliton solutions are in $L^2(\mathbb{R}^d)$, there is a finite amplitude for tunneling between them and the trivial vacuum (and between each other), even in the infinite volume limit. These configurations therefore generally contribute to any correlation function via instanton solutions that interpolate between them and the trivial vacuum.  The tunneling amplitude from the trivial vacuum to a soliton is proportional to a factor related to the classical Euclidean action of an interpolating instanton, which can be bounded as follows:
\begin{equation}
e^{-S_E} \leq e^{- 2 \left|W\left(\phi_\text{sol}\right)\right|} \sim e^{- C\, |\lambda_n|^{-\frac{2}{n-2}} L_{\text{sol}}^{d-\frac{2n}{n-2}}},
\end{equation}
where $C$ is a constant that depends only on the shape of the soliton, and $L_\text{sol}$ is a length scale related to the size of the soliton (and in the gapped cases controlled by the gap parameter). The soliton contributions are therefore exponentially suppressed at weak coupling, but must be accounted for at strong coupling. 

As in supersymmetric quantum mechanics, the non-perturbative contributions of the soliton vacua may also lead to dynamical breaking of supersymmetry. In the case of (non-degenerate) $\mathcal{N}=2$ supersymmetric quantum mechanics (see \cite{Jaffe:1987nx,Dolgallo1994}) one can use Witten index techniques and the holomorphic structure of the model to show that supersymmetry is not dynamically broken. A generalization of such arguments to the field theory models studied here, however, is complicated by the existence of an infinite number of degrees of freedom, a possible infinite tower of semiclassical vacua and  degeneracies due to the global spacetime symmetries of these models. A full classification of the semiclassical soliton vacua for each of these models, as well as an analysis of their implications for the existence of a stable supersymmetric vacuum, would clearly be desirable 
in order to obtain a non-perturbative understanding of time domain supersymmetry in non-boost-invariant field theories. We leave these subtle issues for future work.

In addition to the supersymmetric family of models introduced in section \ref{sec:TheHolomorphicModel}, we also discussed the properties of the RG flow and Lifshitz fixed points in non-boost-invariant field theories (see subsections \ref{subsec:Renormalization} and \ref{subsec:DualScaleRGFlows}). We obtained a relation between the dynamical critical exponent $z$ and the anomalous dimension of one of the parameters in the theory.\footnote{This is similar to a result in \cite{Aharony:2018mjm}, which studies the RG flow in relativistic systems with quenched disorder where the couplings vary randomly in space. In such systems a Lifshitz fixed point can appear with a dynamical critical exponent which is related to the anomalous dimension corresponding to a source coupled to the energy density.} Additionally, we introduced an alternative approach of a dual-scale RG flow and explained some of the properties of fixed points in this picture. These discussions may prove useful in the larger context of non-boost-invariant field theories.

Several interesting directions for future study follow from this work. First, is the study of spontaneous symmetry breaking in these supersymmetric models -- as previously mentioned, they serve as interesting examples for understanding spontaneous breaking of global symmetries, and in particular space translation symmetries, as well as the associated Goldstone modes, in non-boost-invariant theories. More specifically, analyzing the cases in which space translations are broken to a discrete subgroup (by a periodic solution to the vacuum equation) may help to shed light on the properties of striped phases in certain condensed matter systems (see for example \cite{Gruner:1988zz,Gruner:1994zz,Vojta:2009}).

As mentioned in subsection \ref{sec:MarginalCases}, a question which arises in the study of non-boost-invariant theories is whether there are any restrictions on the value of the dynamical critical exponent $z$, and in particular whether it can have a value smaller than $z=1$. For the scale invariant cases considered in this work, we have shown that $z>2$ to leading order in perturbation theory. It would be interesting to understand whether this persists to higher perturbative orders and non-perturbatively in these models, as well as whether one can find restrictions on $z$ for wider classes of non-boost-invariant theories from general arguments.

Another question left unanswered in subsection \ref{subsec:GaplessSingularCase} is the strongly coupled IR behavior of the gapless singular cases in these models. In particular we would like to understand whether the theory flows to some strongly coupled Lifshitz fixed point in the IR, and whether we can learn anything about this fixed point from the holomorphic supersymmetric structure of the models. 

Another interesting challenge is the construction of vector, matrix and tensor model generalizations of the supersymmetric models introduced here, and studying their large N limit, with the goal of obtaining analytic results for their behavior at weak and strong coupling. The fact that in some cases these models exhibit exact Lifshitz scale symmetry at arbitrary coupling suggests that the large N limits might have a description in terms of a holographic gravity dual, similarly to analogous relativistic systems.

It would be interesting to find a way to gauge these time domain supersymmetric models, or extend them to supersymmetric models with more supercharges.
In the larger context of non-relativistic supersymmetry, unlike their relativistic counterparts, the possible supersymmetry algebras and their representations in non-relativistic (both Galilean invariant and non-boost-invariant) theories have not been classified, as supersymmetry is less restricted in these cases. Such a classification is clearly desirable as a long term goal for understanding non-relativistic field theories.

\acknowledgments

We would like to thank Ofer Aharony, Vladimir Narovlansky, Jacob Sonnenschein, Jerome P.~Gauntlett and Matthew M.~Roberts for valuable discussions. 
IA is supported by the European Research Council under the European Union's Seventh Framework Programme (FP7/2007-2013), ERC Grant agreement ADG 339140.
This work is supported in part by the I-CORE program
of Planning and Budgeting Committee (grant number 1937/12), the US-Israel Binational Science Foundation, GIF and the ISF Center of Excellence. A.R.M gratefully acknowledges the support of the Adams Fellowship Program of the Israel Academy of Sciences and Humanities.

\appendix

\section{Notations and Conventions}
\label{app:NotationaAndConventions}

In this appendix we briefly summarize our notations and conventions.

In this work we consider non-boost-invariant field theories in $d+1$ dimensions, where $d$ is the number of space dimensions. We use Latin letters ($i,j,k\ldots$) for spatial indices, or indices enumerating parameters. 

The fermions and fermion charges in the models we consider are non-relativistic ones. They are scalars under spatial rotations and carry no spin. However, they are charged under an $SU(2)$ global R-symmetry.  
We use Greek letters ($\alpha, \dot{\alpha}, \beta, \dot{\beta} \ldots $) for the $SU(2)$ indices of the fermions.
Although these are not spin indices, we use the relativistic notations and conventions of \cite{SUSYPrimer} for contracting, raising and lowering of the fermionic indices, namely: 
\begin{equation}
\xi \psi = \xi^\alpha\psi_\alpha, \qquad \xi^\dagger \psi^\dagger=\theta^\dagger_{\dot{\alpha}}\psi^{\dagger\dot{\alpha}},
\end{equation}
\begin{equation}
\xi_\alpha=\epsilon_{\alpha\beta}\xi^\beta, \qquad \xi^\alpha=\epsilon^{\alpha\beta}\xi_\beta, \qquad \chi^\dagger_{\dot{\alpha}}= \epsilon_{\dot{\alpha}\dot{\beta}}\chi^{\dagger\dot{\beta}}, \qquad \chi^{\dagger\dot{\alpha}}=\epsilon^{\dot{\alpha}\dot{\beta}}\chi^\dagger_{\dot{\beta}} \,,
\end{equation}
where $\alpha,\dot{\alpha}=\{1,2\}$ and $\epsilon^{12}=-\epsilon^{21}=\epsilon_{21}=-\epsilon_{12}=1 $. We define ${\sigma^0}_{\alpha \dot{\alpha}}={\bar{\sigma}}^{0\dot{\alpha}\alpha }=\mathbf{1}_{2\times2}$ as the unit matrix.
In components, the fermion fields are therefore defined as follows: 
\begin{equation}
{\psi _\alpha } = \left( {\begin{array}{*{20}{c}}
{{\psi _1}}\\
{{\psi _2}}
\end{array}} \right), \qquad {\psi ^\alpha } = \left( {\begin{array}{*{20}{c}}
{{\psi _2}}\\
{{-\psi _1}}
\end{array}} \right), \qquad 
{\psi ^\dagger_{\dot{\alpha}} } = \left( \psi_1^*,\psi_2^* \right), \qquad 
{\psi ^{\dagger\dot{\alpha}} } = \left( {\begin{array}{*{20}{c}}
{{\psi^* _2}}\\
{{-\psi^* _1}}
\end{array}} \right),
\end{equation}
where $\psi_1, \psi_2$ are complex Grassmannian fields.

Throughout this work we generally use the letters $k,p,q$ to denote ($d$-dimensional) momenta and $E,\omega,\Omega$ to denote energies. As usual we use units in which $\hbar=1$.

\section{Free Field Quantization}\label{app:Quant}

In this appendix we detail the canonical second quantization procedure for both the bosonic and fermionic fields described by the free fields Lagrangian density \eqref{eq:LagrangianFreeModels}. 
We start with the bosonic field, whose equation of motion reads:
\begin{equation}
\partial_t^2\phi+g^2\nabla^4\phi+|f_2|^2\phi+g(f_2+f_2^*)\nabla^2\phi=0.
\end{equation}
Its solution can be written in terms of modes expansion as follows:
\begin{equation}
\begin{aligned}
&\phi(t,x) = \int \frac{d^dp}{(2\pi)^d}\frac{1}{\sqrt{2\omega}}\left( a_p e^{ipx-i\omega t} +b_p^\dagger e^{-ipx+i\omega t} \right),\\
& \phi^*(t,x) = \int \frac{d^dp}{(2\pi)^d}\frac{1}{\sqrt{2\omega}}\left( a_p^\dagger e^{-ipx+i\omega t} +b_p e^{ipx-i\omega t} \right),
\end{aligned}
\end{equation}
with 
\begin{equation}
\label{eq:DisersionRelationBosons}
\omega(p) = |gp^2-f_2|.
\end{equation}
The Hamiltonian density that corresponds to the bosonic part of the Lagrangian density \eqref{eq:LagrangianFreeModels} is given by:
\begin{equation}
\mathcal{H}_\text{bos}=\partial_t\phi^*\partial_t\phi+g^2\nabla^2\phi^*\nabla^2\phi+|f_2|^2\phi^*\phi+g(f_2\nabla^2\phi^*\phi+f_2^*\phi^*\nabla^2\phi).
\end{equation} 
By imposing the usual canonical commutation relations:
\begin{equation}
[\phi(x),\partial_t\phi^*(x')] = [\phi^*(x),\partial_t\phi(x')] = i \delta^d (x-x'),
\end{equation}  
one obtains the following commutation relation for $a_p$ and $b_p$:
\begin{equation}
[a_p , a_{p'}^\dagger]=[b_p , b_{p'}^\dagger]=(2\pi)^d\delta^d({p}-{p'}),
\end{equation}
as well as the following expression for the Hamiltonian: 
\begin{equation}
H_{\text{bos}}= \int \frac{d^dp}{(2\pi)^d}\omega(p)\left(a_p^\dagger a_p + b_p^\dagger b_p \right),
\end{equation}
as expected. Note that we have dropped the infinite vacuum energy term.
Next we address the question of quantizing the fermionic fields appearing in \eqref{eq:LagrangianFreeModels}. For convenience we rewrite the fermionic part of the Lagrangian density in terms of the components $\psi_1,\psi_2$:   
\begin{equation}
\label{eq:FermionicFieldsFreeActionInComponents}
\mathcal{L}_\text{ferm}= i\psi_1^*\partial_t\psi_1+ i\psi_2^*\partial_t\psi_2 -g\psi_2\nabla^2\psi_1+g\psi_2^*\nabla^2\psi_1^*-f_2\psi_2\psi_1-f_2^*\psi_1^*\psi_2^*.
\end{equation}
The corresponding equations of motion are given by:
\begin{equation}
\begin{aligned}
&i\partial_t\psi_2+g\nabla^2\psi_1^*+f_2^*\psi_1^*=0,\\
& i\partial_t\psi_1-g\nabla^2\psi_2^*-f_2^*\psi_2^*=0.
\end{aligned} 
\end{equation}
The fields $\psi_1, \psi_2$ can be decomposed in terms of mode expansion as follows:
\begin{equation}
\begin{aligned}
&\psi_1(t,x) =\frac{1}{\sqrt{2}} \int \frac{d^dp}{(2\pi)^d}\left( \tilde{a}_p e^{ipx-i\omega t} +\tilde{b}_p^\dagger e^{-ipx+i\omega t} \right),\\
& \psi_1^*(t,x) = \frac{1}{\sqrt{2}}\int \frac{d^dp}{(2\pi)^d}\left( \tilde{a}_p^\dagger e^{-ipx+i\omega t} +\tilde{b}_p e^{ipx-i\omega t} \right),\\
&\psi_2(t,x) = \frac{1}{\sqrt{2}}\int \frac{d^dp}{(2\pi)^d}\left( \tilde{c}_p e^{ipx-i\omega t} +\tilde{d}_p^\dagger e^{-ipx+i\omega t} \right),\\
& \psi_2^*(t,x) = \frac{1}{\sqrt{2}}\int \frac{d^dp}{(2\pi)^d}\left( \tilde{c}_p^\dagger e^{-ipx+i\omega t} +\tilde{d}_p e^{ipx-i\omega t} \right),
\end{aligned}
\end{equation}
substituting these into the equations of motion one finds the following constraints:
\begin{equation}
\label{eq:EOMConstrainsFreeFermions}
\tilde{c}_p=\frac{(gp^2-f_2^*)}{\omega(p)}\tilde{b}_p, \qquad \tilde{d}_p=\frac{(f_2-gp^2)}{\omega(p)}\tilde{a}_p,
\end{equation} 
with 
\begin{equation}
\omega(p)=|gp^2-f_2|. 
\end{equation}
We therefore have:
\begin{equation}
\begin{aligned}
&\psi_1(t,x) = \frac{1}{\sqrt{2}}\int \frac{d^dp}{(2\pi)^d}\left( \tilde{a}_p e^{ipx-i\omega t} +\tilde{b}_p^\dagger e^{-ipx+i\omega t} \right),\\
& \psi_1^*(t,x) =\frac{1}{\sqrt{2}} \int \frac{d^dp}{(2\pi)^d}\left( \tilde{a}_p^\dagger e^{-ipx+i\omega t} +\tilde{b}_p e^{ipx-i\omega t} \right),\\
&\psi_2(t,x) =\frac{1}{\sqrt{2}} \int \frac{d^dp}{(2\pi)^d}  \frac{(gp^2-f_2^*)}{\omega} \left(\tilde{b}_p e^{ipx-i\omega t} -\tilde{a}_p^\dagger e^{-ipx+i\omega t} \right),\\
& \psi_2^*(t,x) = \frac{1}{\sqrt{2}}\int \frac{d^dp}{(2\pi)^d}\frac{(gp^2-f_2)}{\omega}\left( \tilde{b}_p^\dagger e^{-ipx+i\omega t} -\tilde{a}_p e^{ipx-i\omega t} \right).
\end{aligned}
\end{equation}
The Hamiltonian density which corresponds to the Lagrangian density \eqref{eq:FermionicFieldsFreeActionInComponents} reads: 
\begin{equation}
\mathcal{H}_\text{ferm}=g\psi_2\nabla^2\psi_1-g\psi_2^*\nabla^2\psi_1^*+f_2\psi_2\psi_1+f_2^*\psi_1^*\psi_2^*.
\end{equation}
Imposing the canonical anti-commutation relations:
\begin{equation}
\{\psi_1(x),\psi_1^*(x')\} = \{\psi_2(x),\psi_2^*(x')\} = \delta^d(x-x'),
\end{equation}
one finds the relations for $\tilde{a}_p, \tilde{b}_p$:
\begin{equation}
\{\tilde{a}_p , \tilde{a}_{p'}^\dagger\}=\{\tilde{b}_p , \tilde{b}_{p'}^\dagger\}=(2\pi)^d\delta^d({p}-{p'}),
\end{equation}
as well as the following expression for the Hamiltonian in terms of creation and annihilation operators: 
\begin{equation}
\begin{aligned}
H_{\text{ferm}} &= \int \frac{d^dp}{(2\pi)^d} \omega(p) \left( \tilde{a}_p^\dagger \tilde{a}_p+\tilde{b}_p^\dagger \tilde{b}_p \right),
\end{aligned}
\end{equation}
where we have again dropped the infinite vacuum energy term.

\section{More Details for Perturbative Superspace Analysis}
\label{app:DetailsPerturbativeAnalysis}

In this appendix we elaborate on the technical details involved in the supergraph formulation of Feynman diagrams for the holomorphic, time domain supersymmetric family of models introduced in subsection \ref{sec:TheHolomorphicModel}, as well as the perturbative argument for non-renormalization as discussed in subsection \ref{subsec:PerturbativeArgumentforNRTheorems}. As most of the details are similar to those of the relativistic Wess-Zumino model (see \cite{Wess:1992cp}), we mostly highlight the differences.

The Lagrangian density of the free model (in $d+1$ dimensions) given in \eqref{eq:LagrangianGeneral}, \eqref{eq:FreeSuperpotential} can be written in the following form:\footnote{We use the conventions of \cite{Wess:1992cp}, in which
$\int d^2\theta \delta(\theta) = \int d^2 \theta^\dagger \delta(\theta^\dagger) = 1$.}
\begin{equation}
\label{eq:SuperspacePropLag}
\mathcal{L}=\int d^2\theta d^2\theta^\dagger \left( \Phi\Phi^* +\left(\frac{g}{2}\Phi\nabla^2\Phi+\frac{f_2}{2}\Phi^2\right)\delta(\theta^\dagger)+ \left(\frac{g}{2}\Phi^*\nabla^2\Phi^*+\frac{f_2^*}{2}{\Phi^*}^2\right)\delta(\theta)\right).
\end{equation}
In terms of component fields, the Lagrangian \eqref{eq:SuperspacePropLag} yields the following propagators: 
\begin{align}
& \left< \phi(t,x) \phi^*(t',x') \right> =i{\mathcal{G}}_{\text{lif}}(t-t',x-x'), \\
& \left< \phi^*(t,x) F^*(t',x') \right> = -i(f_2+g\nabla^2){\mathcal{G}}_{\text{lif}}(t-t',x-x'),\\
& \left< \phi(t,x) F(t',x') \right> = -i(f_2^*+g\nabla^2){\mathcal{G}}_{\text{lif}}(t-t',x-x'), \\
&\left< F(t,x) F^*(t',x') \right> =-i\partial_t^2{\mathcal{G}}_{\text{lif}}(t-t',x-x'),\\
&\left< \psi_\alpha(x,t) \psi^\beta(t',x') \right>= i\delta_\alpha^\beta(f_2^*+g\nabla^2){\mathcal{G}}_{\text{lif}}(t-t',x-x'), \\
&\left< \psi^{\dagger\dot{\alpha}}(x,t) \psi^\dagger_{\dot{\beta}}(t',x') \right>= i\delta^{\dot{\alpha}}_{\dot{\beta}}(f_2+g\nabla^2){\mathcal{G}}_{\text{lif}}(t-t',x-x') ,\\
&\left< \psi_\alpha(x,t) \psi^\dagger_{\dot{\beta}}(t',x') \right>=-\sigma^0_{\alpha\dot{\beta}}\partial_t{\mathcal{G}}_{\text{lif}}(t-t',x-x'),
\end{align}
where ${\mathcal{G}}_{\text{lif}}(t,x)$ is defined in \eqref{eq:PertNRArgumentDeltaLifDefinition}. 

In addition to the expression \eqref{eq:ScalarProp1}, using the definitions in equation  \eqref{eq:GeneralizedCoordinates} one finds the following expressions for the super-propagators: 
\begin{equation}
\label{eq:ScalarProp2}
\begin{aligned}
&\left\langle \Phi^\dagger(t,x,\theta,\theta^\dagger)\Phi^\dagger(t',x',\theta',\theta'^\dagger) \right\rangle  \\ 
&=-i(f_2+g\nabla^2)\delta(\theta^\dagger-\theta'^\dagger)e^{i(\theta{\sigma}^0\theta^\dagger-\theta'{\sigma^0}\theta'^{\dagger})\partial_t}{\mathcal{G}}_{\text{lif}}(t-t',x-x'),
\end{aligned}
\end{equation}
and:
\begin{equation}
\label{eq:ScalarCompProp3}
\begin{aligned}
\left< \Phi(t,x,\theta,\theta^\dagger)\Phi^\dagger(t',x',\theta',\theta'^\dagger)\right>=
i e^{-i(\theta\sigma^0\theta^\dagger+\theta'\sigma^0\theta'^\dagger-2\theta'\sigma^0\theta^\dagger)\partial_t}{\mathcal{G}}_{\text{lif}}(t-t',x-x').
\end{aligned}
\end{equation}

These free fields super-propagators can be written in terms of covariant superderivatives as follows:
\begin{align}
&\left< \Phi(z)\Phi(z')\right>=  \frac{i}{4}(f_2^*+g\nabla^2){\hat{\mathcal{G}}}_{\text{lif}}\bar{D}^2\delta(z-z')     ,\label{eq:ScalarPropSuperspace1}\\
&\left< \Phi^\dagger(z)\Phi^\dagger(z')\right>=  \frac{i}{4}(f_2+g\nabla^2){\hat{\mathcal{G}}}_{\text{lif}}D^2\delta(z-z') \label{eq:ScalarPropSuperspace2}  , \\
& \left< \Phi(z)\Phi^\dagger(z')\right>= \frac{i}{16}{\hat{\mathcal{G}}}_{\text{lif}}\bar{D}^2D^2\delta(z-z'),\label{eq:ScalarPropSuperspace3} \\
&\left< \Phi^\dagger(z)\Phi(z')\right>= \frac{i}{16}{\hat{\mathcal{G}}}_{\text{lif}}D^2\bar{D}^2\delta(z-z'),
\end{align}
where we have defined $z\equiv (t,x,\theta,\theta^\dagger)$ and $\delta(z)\equiv \delta(t)\delta(x)\delta(\theta)\delta(\theta^\dagger)$, ${\hat{\mathcal{G}}}_{\text{lif}}$ is defined in \eqref{eq:PertNRArgumentDeltaLifDefinition} and the superderivatives are defined in equation \eqref{eq:SuperDerivativesDef}. Note that, to derive these expressions, we have used the identity:
\begin{equation}
\frac{1}{16}\frac{\bar{D}\bar{D}DD}{\square_t}\Phi = \Phi, \qquad \text{if}\quad \bar{D}_{\dot{\alpha}}\Phi=0,
\end{equation}
where $\square_t \equiv -\partial_t^2$.

As in the relativistic case, based on the Feynman rules for the supergraph formalism (as detailed in section \ref{subsec:PerturbativeArgumentforNRTheorems}), the form of the GRS propagators \eqref{eq:GRSPropagator1}-\eqref{eq:GRSPropagator3} and using the identities:
\begin{equation}
\label{eq:Identity1}
\bar{D}^2D^2\bar{D}^2=16\square_t\bar{D}^2,
\end{equation}
\begin{equation}
\label{eq:Identity2}
D^2\bar{D}^2D^2=16\square_t D^2,
\end{equation}
which hold for the time domain superderivatives \eqref{eq:SuperDerivativesDef},  the expression for any arbitrary closed loop the with an integration over the whole superspace for each interaction vertex can be reduced to an expression containing a single $d^4\theta$ integral.

We briefly summarize the argument for this:
Suppose that $\theta_i, \theta^\dagger_i$ denote the Grassmannian coordinates corresponding to the $i$-th vertex in the loop.
Each propagator in the loop will contribute an expression of the form: $ D_i^2 \bar{D}_i^2 D_i^2 \ldots \delta(\theta_i-\theta_j)\delta(\theta^\dagger_i-\theta^\dagger_j)$ or $ \bar{D}_i^2 D_i^2 \bar{D}_i^2 \ldots \delta(\theta_i-\theta_j)\delta(\theta^\dagger_i-\theta^\dagger_j) $, with overall $l_i$ instances of $D_i^2$ and $k_i$ instances of $\bar{D}_i^2$ ($l_j,k_i \in \mathbb{N}$). 
For any $l_i>1$ or $k_i>1$ one can use the identities \eqref{eq:Identity1}, \eqref{eq:Identity2} to reduce the number of superderivatives until one remains with $l_i, k_i = \lbrace 0, 1 \rbrace$ (leaving non-negative powers of the $\square_t$ operator).

The expression can then be further simplified by repeatedly integrating the superderivative factors by parts and subsequently performing the remaining $\delta$-function integrations over the Grassmannian coordinates. Eventually either the expression vanishes or one is left with a single $\int d^4 \theta$ integration, with no further factors of $\delta(\theta_i-\theta_j)$ or $\delta(\theta^\dagger_i-\theta^\dagger_j)$, and all remaining $D^2$ or $\bar{D}^2$  factors operating on lines external to the loop.   
This process can be repeated over all loops, and thus the integration over all $\theta_i$ space is reduced to a single $d^4\theta$ integral, as promised. 
We are then led to the conclusion that the effective action can be written in the form given in equation \eqref{eq:EffActionNRPerturb} (with no negative powers of $\square_t$ appearing). Therefore the holomorphic superpotential is not renormalized quantum mechanically in perturbation theory.

\section{The Anomalous Dimension in Other Marginal Cases}
\label{app:betaFunctions}

In subsection \ref{sec:MarginalCases} we discussed the marginal cases for the family of time domain supersymmetric models introduced in subsection \ref{sec:TheHolomorphicModel}, and showed they exhibit exact Lifshitz scaling invariance for any value of the dimensionless coupling. We related the dynamical critical exponent $z$ to the anomalous dimension of the superfield, and calculated it to the leading non-trivial order in perturbation theory for the case of $6+1$ dimensions with an $n=3$ interaction.
In this appendix we study the leading non-trivial order perturbative corrections to the anomalous dimension in the other marginal cases -- $4+1$ dimensions with an $n=4$ interaction and $3+1$ dimensions with an $n=6$ interaction -- with the goal of determining the sign of the anomalous dimension (and thus whether $z>2$ in these cases). We use a time-first regularization method along with single-scale renormalization.

In both cases we find that the sign of the anomalous dimension is indeed positive in the leading non-trivial order in perturbation theory.

\subsection{$4+1$ Dimensions with a $\Phi^4$ Interaction}

Consider first the marginal case in $4+1$ dimensions with an $n=4$ interaction of the form \eqref{eq:GeneralSuperInteraction}.
As in subsection \ref{sec:MarginalCases}, the anomalous dimension of the field $\Phi$ can be calculated from the corrections to the fermionic propagator. 
The leading order ones are described in figure \ref{fig:BetaFuncfourSpatialDim}, and the corresponding expression reads: 
\begin{figure}[!htb]
   \centering
		{\includegraphics[width=0.1\textwidth]{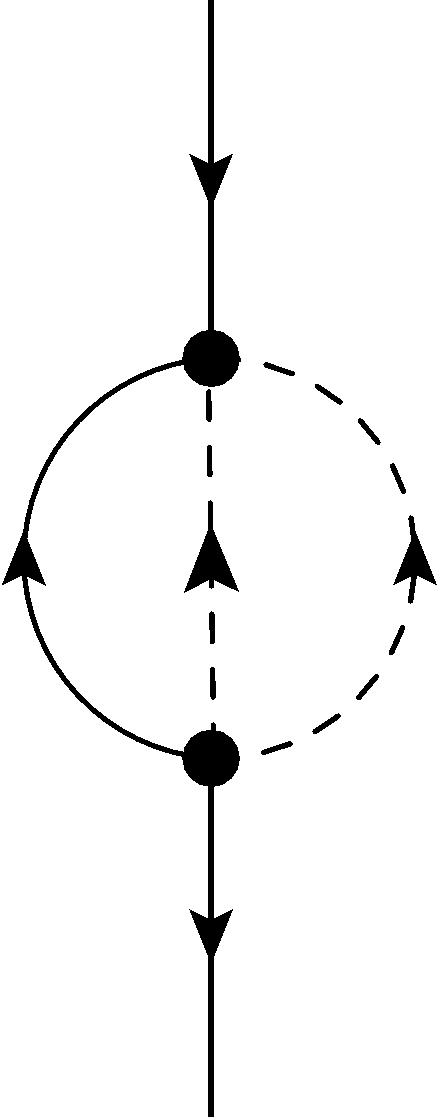}}
        \caption{\label{fig:BetaFuncfourSpatialDim}Leading order quantum corrections to the fermionic propagator in $4+1$ dimensions with an $n=4$ interaction. }
\end{figure}
\begin{equation}
\begin{aligned}
B^{\dot{\beta}\alpha} &=8\left(-\frac{3if_4}{2}\right)\left(-\frac{3if_4^*}{2}\right) \int \frac{d\Omega_{1}}{(2\pi)}\int \frac{d^4q_1}{(2\pi)^4}\int \frac{d\Omega_{2}}{(2\pi)}\int \frac{d^4{q_2}}{(2\pi)^4}\\
& \frac{i(\omega-\Omega_{1}-\Omega_{2})\bar{\sigma}^{0{\dot{\beta}\alpha}}}{\left[(\omega-\Omega_{1}-\Omega_{2})^2-g^2(k-q_1-q_2)^4\right]}\frac{i}{\left[\Omega_{1}^2-g^2q_1^4\right]}\frac{i}{\left[\Omega_{2}^2-g^2q_2^4\right]},
\end{aligned}
\end{equation}
where $(\omega,\vec{k})$ are the external energy and momentum respectively and  $(\Omega_{l},\vec{q}_l)\, (l=1,2)$ are the loop energies and momenta.
Extracting the UV divergent part  of the integral similarly to subsection \ref{sec:MarginalCases}, one finds: 
\begin{equation}
\begin{aligned}
B^{\dot{\beta}\alpha} &= -18i|f_4|^2\bar{\sigma}^{0{\dot{\beta}\alpha}}\omega\int \frac{d\Omega_{1}}{(2\pi)}\int \frac{d^4q_1}{(2\pi)^4}\int \frac{d\Omega_{2}}{(2\pi)}\int \frac{d^4{q_2}}{(2\pi)^4}\\
& \frac{(\Omega_{1}+\Omega_{2})^2+g^2(q_1+q_2)^4}{\left[(\Omega_{1}+\Omega_{2})^2-g^2(q_1+q_2)^4\right]^2}\frac{1}{\left[\Omega_{1}^2-g^2q_1^4\right]}\frac{1}{\left[\Omega_{2}^2-g^2q_2^4\right]} + \ldots, 
\end{aligned}
\end{equation}
where ``$\ldots$'' again stands for UV finite terms. Then setting the counterterm contribution to the self-energy to cancel the divergent part, we get:
\begin{equation}
\begin{aligned}
i\delta_{Z_\psi} &=18i|f_4|^2\int \frac{d\Omega_{1}}{(2\pi)}\int \frac{d^4q_1}{(2\pi)^4}\int \frac{d\Omega_{2}}{(2\pi)}\int \frac{d^4{q_2}}{(2\pi)^4}\\
&\left. \frac{(\Omega_{1}+\Omega_{2})^2+g^2(q_1+q_2)^4}{\left[(\Omega_{1}+\Omega_{2})^2-g^2(q_1+q_2)^4\right]^2}\frac{1}{\left[\Omega_{1}^2-g^2q_1^4\right]}\frac{1}{\left[\Omega_{2}^2-g^2q_2^4\right]} \right|_\text{div} ,
\end{aligned}
\end{equation}
where we have used the same definitions for $\delta_{Z_{\psi}}$ as those in section \ref{sec:MarginalCases}. Performing the two integrals over the energies using contour integration (both integrals converge, one after the other, in either order) one finds:
\begin{equation}
\delta_{Z_\psi}=-\frac{9|f_4|^2}{8g^4}\int \frac{d^4q_1}{(2\pi)^4} \int \frac{d^4q_2}{(2\pi)^4} \left. \frac{1}{q_1^2q_2^2(q_1^2+\vec{q}_1\cdot\vec{q}_2+q_2^2)^2} \right|_\text{div}.
\end{equation} 

Since the integrand in the above expression is everywhere negative, it is clear that after imposing a spatial UV cutoff and introducing a spatial renormalization scale $\mu_s$, the anomalous dimension $\gamma_\psi$ will be positive (in this perturbative order):
\begin{equation}
\gamma_{\psi} \equiv \frac{1}{2}\frac{\partial\delta_{Z_{\psi}} }{\partial \log(\mu_s)}>0,
\end{equation}
and therefore $z>2$ due to \eqref{eq:MarginalCaseszGammaRelation}.

\subsection{$3+1$ Dimensions with a $\Phi^6$ Interaction}

\begin{figure}[!htb]
   \centering
		{\includegraphics[width=0.1\textwidth]{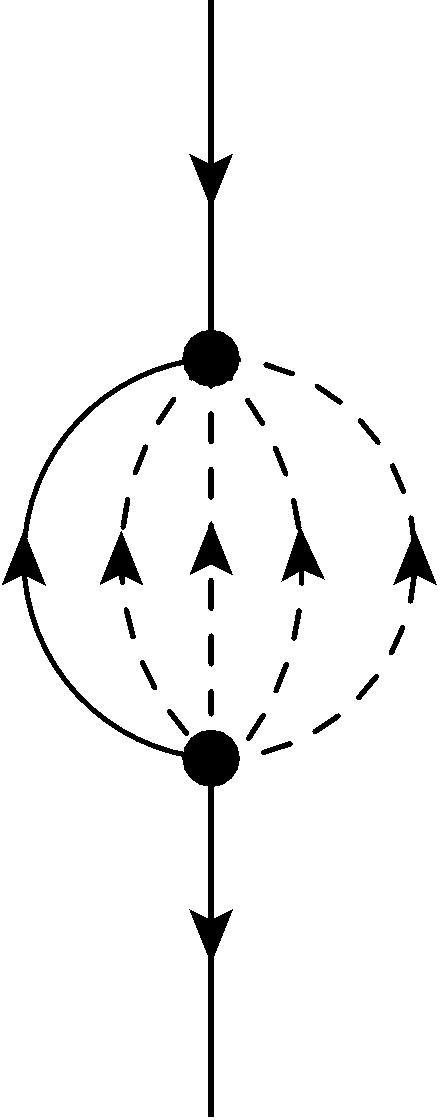}}
        \caption{\label{fig:BetaFuncThreeSpatialDim}Leading order quantum corrections to the fermionic propagator in a model of $f_6\Phi^6$ interaction in $3+1$ dimensions. }
\end{figure}

Next consider the marginal case 
in $3+1$ dimensions with an $n=6$ interaction of the form \eqref{eq:GeneralSuperInteraction}. The leading order quantum corrections to the fermionic propagator contain four loops and are described in figure \ref{fig:BetaFuncThreeSpatialDim}.
The corresponding expression reads:
\begin{equation}
\begin{aligned}
C^{\dot{\beta}\alpha} &=-\left|\frac{5f_6}{2}\right|^2 96 \prod_{i=1}^4 \left( \int\frac{d\Omega_{i}}{(2\pi)}\int\frac{d^3q_i}{(2\pi)^3}\frac{i}{\left[\Omega_{i}^2-g^2q_i^4\right] }\right)\\
& \qquad \frac{i\bar{\sigma}^{0\dot{\beta}\alpha}\left(\omega-\sum\limits_{i=1}^4 \Omega_{i}\right)}{\left(\omega-\sum\limits_{i=1}^4 \Omega_{i}\right)^2-g^2\left(k-\sum\limits_{i=1}^4 q_i\right)^4},
\end{aligned}
\end{equation}
where $(\omega,\vec{k})$ are the external energy and momentum respectively and $(\Omega_l, \vec{q}_l) \,(l=1,\ldots,4) $ are the loop energies and momenta. 
Extracting the UV divergent part as in the previous cases and setting the self-energy counterterm to cancel it, we get:
\begin{equation}
\begin{aligned}
\delta_{Z_\psi}& = -600|f_6|^2\prod_{i=1}^4 \left( \int\frac{d\Omega_{q_i}}{(2\pi)}\int\frac{d^3q_i}{(2\pi)^3} \right)\left(\frac{1}{\left[\Omega_{i}^2-g^2q_i^4\right]}\right)\\
&\qquad \qquad\left.\frac{(\sum\limits_{j=1}^4\Omega_j)^2+g^2(\sum\limits_{j=1}^4q_j)^4}{\left[(\sum\limits_{j=1}^4\Omega_j)^2-g^2(\sum\limits_{j=1}^4q_j)^4\right]^2}\right|_\text{div}.\\
\end{aligned}
\end{equation}
Performing the integrals over the energies one finds:
\begin{equation}
\begin{aligned}
\delta_{Z_\psi}& = -\frac{75|f_6|^2 }{8 g^6}\prod_{i=1}^4 \left( \int\frac{d^3q_i}{(2\pi)^3} \right)  \left. \frac{1}{ q_1^2 q_2^2 q_3^2 q_4^2 \left(\sum\limits_{1\leq i\leq j \leq 4} \vec{q}_i\cdot\vec{q}_j\right)^2}\right|_\text{div},
\end{aligned}
\end{equation}
which again leads to the conclusion that the anomalous dimension is positive, and $z>2$, to leading order in perturbation theory.

\end{document}